\documentclass[twocolumn,aps,prb,superscriptaddress,longbibliography]{revtex4-2}
\usepackage{graphicx}
\usepackage{color}
\usepackage{amsmath}
\usepackage[export]{adjustbox}
\usepackage{enumitem}
\usepackage{amssymb}
\usepackage{bbm}
\usepackage{hyperref}
\usepackage[normalem]{ulem}
\usepackage{adjustbox}

\hypersetup{
	colorlinks = true,
	linkcolor = blue,
	citecolor = blue,
	urlcolor  = blue,
}

\newcommand{\be}{\begin{equation}}
	\newcommand{\ee}{\end{equation}}

\newcommand{\bea}{\begin{eqnarray}}
	\newcommand{\eea}{\end{eqnarray}}

\renewcommand{\epsilon}{\varepsilon}

\def\nn{\nonumber\\}

\begin{document}
 \title{Topological metamaterial platform for bosonic qubits}

	\date{\today}

\author{D. O. Oriekhov}
\altaffiliation{These authors contributed equally to this work.}
\affiliation{Kavli Institute of Nanoscience and QuTech, Delft University of Technology, 2628 CJ Delft, the Netherlands}
    
    \author{Mykola Luhanko }
\altaffiliation{These authors contributed equally to this work.}
\affiliation{V.N. Karazin Kharkiv National University, Kharkiv, Ukraine}
\affiliation{Haiqu, Inc., 95 Third Street, San Francisco, CA 94103, USA}
    \affiliation{Department of Physics, University of Zürich, 8057 Zürich, Switzerland}

    \author{Genya Crossman}
    \affiliation{Kavli Institute of Nanoscience and QuTech, Delft University of Technology, 2628 CJ Delft, the Netherlands}

    \author{Guliuxin Jin} 
    \affiliation{Kavli Institute of Nanoscience and QuTech, Delft University of Technology, 2628 CJ Delft, the Netherlands}

    \author{Radoica Dra\v{s}ki\'{c}}
    \affiliation{Kavli Institute of Nanoscience and QuTech, Delft University of Technology, 2628 CJ Delft, the Netherlands}

	\author{Eliska Greplova}%
	\email{e.greplova@tudelft.nl}
	\affiliation{Kavli Institute of Nanoscience and QuTech, Delft University of Technology, 2628 CJ Delft, the Netherlands}

	\begin{abstract}
    Topologically protected qubits promise easier scaling of quantum devices through lower error rates and resilience to specific types of disorder. At the same time, they are currently harder to engineer in comparison to unprotected qubits. Here, we propose a quantum chip design solution that builds the topological protection bottom up from building blocks that are more accessible to engineer. Specifically, we encode the qubit into the chiral edge modes of a Chern insulator. We propose a quantum-spin-Hall analog built from two complex-conjugate copies of the Hofstadter model in an array of superconducting resonators, encode the logical states in counter-circulating wave packets in the two arrays, and construct a universal gate set -- $X$, $T$, $Z$, Hadamard and a controlled phase -- by optimizing the couplings at the scale of the lattice. The single-qubit gates reach process fidelities of $98.8\%$ to $99.99\%$ and the two-qubit
    controlled-phase gate $82.0\%$. In each case the error is dominated by leakage out of the computational subspace: conditioned on the excitation remaining in the computational subspace, the gates are accurate to better than $99.99\%$ for single-qubit gates and to $98.6\%$ for the two-qubit gate,
    respectively. We analyze behavior of these gates under disorder and outline pathways for mitigating it in an experimental setting.
	\end{abstract}
	\maketitle

\section{Introduction}

Quantum computing has advanced rapidly across a range of platforms, yet scaling these
platforms to large system sizes remains a major challenge. Generally, two directions are
pursued. The first is full error correction, where millions of physical qubits encode a much
smaller number of logical qubits whose states can be corrected. The second is topological
quantum computing, where the qubit states and the operations acting on them are protected by a
symmetry of the system, so that higher fidelities are reached and far fewer qubits are needed.
The trade-off between these two approaches is the ease of implementation. Qubits without
topological protection, such as the transmon~\cite{Koch2007transmon} or the Loss-DiVincenzo
spin qubit~\cite{spin-qubits}, are relatively easy to fabricate, and superconducting chips
with hundreds of qubits are already available. Topological qubits, such as Majorana
qubits~\cite{majorana}, are in practice far more challenging to realize. While the field has
progressed significantly~\cite{Wang2022poormans,Dvir2023Kitaev,Zatelli2024,Bordin2025,Haaf2025,Loo2026},
a topologically protected Majorana qubit remains to be demonstrated experimentally. 

A middle ground between these two directions is currently emerging: quantum systems are
engineered such that protection against the leading source of noise arises naturally, and
ideally in a way that is straightforward to implement on each platform. While such protection
cannot match that of a Majorana qubit, it can still increase state and operation fidelities and
therefore decrease the overhead of fault-tolerant scaling. Examples of this approach include cat
qubits, where encoding the qubit states as a superposition of two coherent states protects
against bit flips~\cite{Reglade2024, Guillaud2023}, and fluxonium qubits, which are protected
from flux noise~\cite{Manucharyan2009science}. More recently, this idea has expanded to
implementing topological models on quantum devices, with the intention of leveraging aspects of
topological protection in the operation of quantum chips. Examples include the implementation of
the Su-Schrieffer-Heeger model~\cite{Su1979} in superconducting resonator
arrays~\cite{Splitthoff2024PRR, Jouanny2025natcomm}, the proposal to leverage the symmetries of
that model to protect quantum entanglement~\cite{JinGreplova2023}, and protocols in which
topological edge states mediate the transfer of a state between distant sites of a dimerized
chain~\cite{Bello2016}.

In this work, we put forward an architecture that bridges the research on
unprotected qubits and topological quantum computing. Specifically, we show how to engineer a
two-dimensional model of the integer quantum Hall
system~\cite{Halperin1982PhysicalReviewQuantizedHallConductance,BTtiker1988PhysicalReviewAbsenceBackscatteringQuantum}
with dynamically moving edge states in a superconducting resonator lattice. Additionally, we
show how to encode the qubit degrees of freedom into spin-Hall edge states, and how to implement
a universal set of quantum gates on them.

\begin{figure}
    \centering
    \includegraphics[scale=1.0]{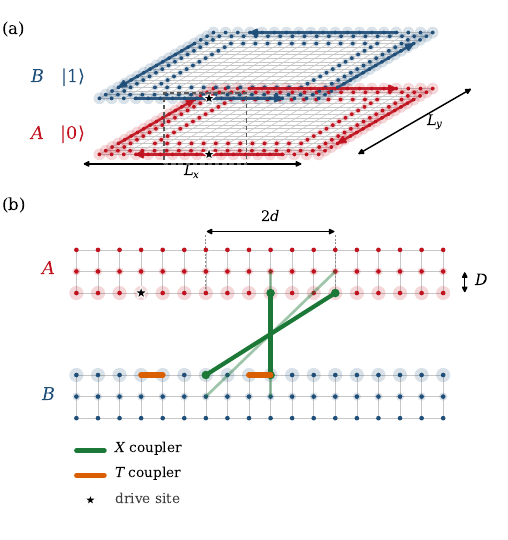}
    \caption{\label{fig1:system-gates}%
The platform and the structure of a gate coupler.
(a) Oblique view of the two resonator arrays. Array $A$ (red, lower) and array
$B$ (blue, upper) each carry a single chiral edge mode; the arrows give the direction of propagation.
The logical states are $|0\rangle$, a
wave packet circulating in $A$, and $|1\rangle$, a packet circulating in $B$. The star marks the resonator
where the packet is injected and read out. Shading of the sites indicates the weight of the edge
mode, and the dashed box marks the region drawn in (b).
(b) The coupler region at lattice resolution. The $X$-gate coupler consists of a
straight rung and a long diagonal, spanning $2d$ along the edge and repeated over $D$ rows inward from the edge.
The $T$-gate coupler consists instead of two
modified bonds lying within array $B$. }
\end{figure}

The main obstacle to realizing topologically protected edge states in native materials,
and to leveraging them for quantum computing, lies in the fabrication and control of devices
made from materials that are topological insulators and host the relevant
modes~\cite{Bordone2019SemiconductorScienceQuantumComputingQuantum,Koenig2007science}. The
quantum spin Hall state itself has so far been realized in only a small number of crystalline
systems, such as the HgTe quantum well~\cite{Koenig2007science}. More recently, spin-polarized
helical edge transport has been detected at zero external magnetic field in monolayer graphene
proximitized by the interlayer antiferromagnet CrPS$_4$~\cite{Ghiasi2025}. Neither platform,
however, removes the difficulty that motivates us here. In graphene the topological gap is not
intrinsic but induced: it requires a van der Waals heterostructure in which an adjacent magnetic
layer supplies the necessary spin-orbit and exchange coupling, so that the gap is set by the
quality of an interface instead of chosen by design. In both platforms the operating
frequencies, the size of the gap and the degree of disorder are properties of the material and not something that we can choose by design. This challenge is the main motivation for the present
work. We propose to use an artificial metamaterial topological insulator to realize an analog
of the quantum spin Hall state, and to encode a qubit architecture into that state.
Specifically, we propose to build an array of superconducting resonators coupled such that they
realize the Hamiltonian of the Hofstadter model, where spin Hall states appear naturally and
their chirality is protected by a symmetry of the Hamiltonian. 

Comparable lattices built from superconducting resonators have been realized before, for
example quantum hyperbolic lattices~\cite{Kollar2019Nature} and a one-dimensional SSH chain
topological insulator~\cite{Splitthoff2024PRR}. In this work, we move beyond proving the
existence of a protected mode, or realizing a fixed Hamiltonian, toward a quantum computing
modality: we define the qubit states and the operations acting on them, and assess their
fidelity. Our architecture is motivated by the metamaterial-based classical topological
insulator built from the pendula array of Ref.~\cite{Huber2015Science}. That approach has since
been carried further, to a phononic quadrupole topological insulator in a mechanical
metamaterial~\cite{SerraGarcia2018} and, closer to the present proposal, to an $LC$ circuit
network in which both the quadrupole transitions and the edge mode topology were measured
directly~\cite{SerraGarcia2019}. Related results have been established in artificial electronic
lattices~\cite{Slot2017,Kempkes2018}. The advantage of this approach, which the field of
classical metamaterials has already benefited from, is the ability to build topological
protection bottom-up and to operate it within a predefined range of frequencies, without having
to combat the disorder and variability of native material samples.

This paper is organized as follows. In Sec.~\ref{sec:model-shape} we introduce the model,
its lattice Hamiltonian and the qubit encoding, and we describe how the injection
process is optimized by engineering the shape of the wave packet. In Sec.~\ref{sec:gates} we
describe the implemented gate set, first the single-qubit gates and then the two-qubit gate,
giving for each gate the optimization strategy and the evaluation of its performance. In
Sec.~\ref{sec:disorder} we evaluate the performance of these gates in the disordered system,
benchmarking both a single application and several sequential applications of the same gate.
Finally, in Sec.~\ref{sec:conclusions} we summarize the results and the routes for improving the gates. 

\section{Platform and encoding}
\label{sec:model-shape}

\subsection{Model}
\label{sec:properties}

We base our design on one of the best-known topological insulator Hamiltonians, the
Hofstadter model~\cite{Hofstadter1976,Thouless1982,Dana1985}. A single copy of this model lacks
time-reversal symmetry, so that only one chirality of edge mode arises. In order to obtain a
system in which both chiralities coexist, it is convenient to take two complex-conjugate copies,
as was done in Ref.~\cite{Huber2015Science}:
\begin{align}
    \label{eq:model-no-gates}
    &H_{meta}=\left(\begin{array}{cc}
\mathcal{H}_{\Phi} & 0 \\
0 & \mathcal{H}_{\Phi}^*
\end{array}\right),\\
&\mathcal{H}_{\Phi}=\\
&\sum\limits_{ x,y} [\omega_{0}\hat{a}_{x,y}^{\dagger}\hat{a}_{x,y}+f_y \hat{a}_{x,y}^{\dagger}\hat{a}_{x,y\pm l,} + f_x e^{\pm   i\Phi y/l} \hat{a}_{x,y}^{\dagger} \hat{a}_{x\pm l,y}].\nonumber
\end{align}
Here the block $\mathcal{H}_{\Phi}$ is a single Hofstadter Hamiltonian on a square
lattice, and $\Phi=2\pi/3$ is the magnetic flux parameter that breaks time-reversal symmetry.
The lattice has three sites per unit cell and an inter-site distance $l$, and the operators
$\hat{a}_{x,y}^{\dagger},\hat{a}_{x,y}$ create and annihilate a boson on the node with
coordinates $x,y$.  
In Fig.~\ref{fig1:system-gates} we show the two arrays that realize the upper and lower
blocks of \eqref{eq:model-no-gates}. Each array carries the complex hoppings
$f_x e^{\pm i\Phi y/l}$ of a single Hofstadter model, with the sign of the flux opposite in
the two, and the two are not connected to one another except through the gate couplers
introduced in Sec.~\ref{sec:gates}. Throughout this work we set $f_x=f_y=f$, the choice that
maximizes the bulk gap, as shown in Appendix~\ref{app:spectral}.

Two energy scales appear. The onsite term $\omega_0$ is the resonator frequency, and the
engineered hopping $f$ is much smaller, $\omega_0\gg f$: the band structure of
\eqref{eq:model-no-gates} is not the whole spectrum of the metamaterial but a narrow window
engineered inside it. The onsite term $\omega_0$ is a
uniform offset and drops out of the dynamics. This is exact rather than an
approximation: a term $\omega_0\mathbbm{1}$ commutes with every other term in
\eqref{eq:model-no-gates}, so it factors out of the evolution operator as the global
phase $e^{-i\omega_0 t}$. All simulations are performed in the frame rotating at $\omega_0$ and all frequencies quoted in this work are measured from $\omega_0$ in units of $f$; the ratio $\omega_0/f$ enters only when converting the dimensionless results into laboratory units. Throughout this work we set $\hbar=1$.

The key implementation challenge in translating this model onto a metamaterial lattice is
the presence of complex couplings. In the representation~\eqref{eq:model-no-gates} the lattice
bonds along $y$ are real and those along $x$ carry the flux phase $e^{\pm i\Phi y/l}$, so that a
third of the lattice bonds are complex, while every gate coupling we introduce in
Sec.~\ref{sec:gates} is real. Ref.~\cite{Huber2015Science} gives a unitary transformation that
renders the lattice entirely real, but the same mechanism does not apply here: as we show in
Appendix~\ref{app:real-form}, it would leave half the bonds of every gate coupler complex. One
may therefore have either a real lattice or real gate couplers, but not both. We
keep~\eqref{eq:model-no-gates}, since the complex bonds are then confined to the bulk, where
they take a fixed value and repeat a single regular pattern, while the gate couplings remain
real. This is the more useful arrangement for two reasons. First, we want the gate couplings to
be tunable. Second, schemes for synthesizing complex couplings limit their strength relative to
the bare coupling~\cite{Rosen2024}, whereas the gate couplings we need are of order $f$ or
larger, as listed in Table~\ref{tab:params_fidelity}. 

The block-diagonal form $\mathcal{H}=\mathrm{diag}(\mathcal{H}_\Phi,\mathcal{H}_\Phi^*)$ of Eq.~\eqref{eq:model-no-gates} admits two antiunitary symmetries built from the complex conjugation operator $\hat{\mathcal{K}}$ and the Pauli matrices $\sigma_{x,y,z}$ acting in the array index. The first, $\hat{\mathcal{K}}\sigma_x$, squares to $+1$ and carries no topological index. The relevant one is
\begin{align}
\label{eq:AII-symmetry}
    \hat{\mathcal{T}} = i\sigma_y \hat{\mathcal{K}},\qquad \hat{\mathcal{T}}^2 = -1,
\end{align}
which commutes with $\mathcal{H}$ and places the model in symmetry class AII, the class of the quantum spin Hall effect~\cite{Kane2005PRL,Bernevig2006Science}, with a $\mathbb{Z}_2$ index; the same construction underlies the mechanical analog of Ref.~\cite{Huber2015Science}. Both statements can be verified directly on the matrix~\eqref{eq:model-no-gates}. Since $\hat{\mathcal{T}}$ exchanges the two arrays, it is a symmetry of the pair and not of either array separately, and independent disorder in the two arrays breaks it. The spectrum of a single block consists of three bands that differ in their first Chern
number, $1$, $-2$ and $1$. Consequently, a single edge of a single array carries exactly one
chiral mode in each gap between two bands~\cite{Hatsugai1993}. This is also why we build the
system out of two flux-inverted copies of the Hofstadter model: we want two counter-propagating
chiral modes to pass the same local coupler structure. We emphasize that the gates of
Sec.~\ref{sec:gates} necessarily \emph{break} the symmetry~\eqref{eq:AII-symmetry}, since an
inter-array coupling block commutes with $\hat{\mathcal{T}}$ only if it is antisymmetric, and
the coupler that implements the $X$ gate is not.

The band structure of a ribbon of finite width, translation invariant along $x$ or along $y$, is shown in Fig.~\ref{fig2:signal-vs-bandstructure}(a) and (b). The horizontal axis is the momentum along the invariant direction, $k_x$ or $k_y$, in units of the inverse lattice constant $1/l$, and the vertical axis is the frequency $\omega$ measured from the onsite term $\omega_0$ in units of $f$. Three bulk bands are separated by two gaps, whose centers we denote $\omega_{\pm}$; a single edge branch crosses each gap, and the dash-dotted line marks the working frequency $\omega_+$ used throughout. The crossing points of the edge branches lie either at the edge or at the center of the Brillouin zone, and are protected by the combination of inversion and time-reversal symmetry. Each chiral branch corresponds to a mode living on one side of the ribbon only, as drawn in Fig.~\ref{fig1:system-gates}(a).

Finally, the transverse shape of the chiral mode near the edge sets how deep a gate coupler must reach. %
The mode decays by roughly a factor of eleven per magnetic cell, a decay length of $1.24\,l$, so that the two outermost rows already carry $91.2\%$ of its weight. 

\subsection{Encoding}
\label{sec:computational-basis-def}

At a fixed frequency, the two-array structure of Fig.~\ref{fig1:system-gates} hosts
exactly one chiral mode per array in each gap. We can therefore choose the qubit basis within
the upper gap, marked by the dash-dotted line in Fig.~\ref{fig2:signal-vs-bandstructure}, as
follows: the mode of the lower array $A$, which circulates clockwise, is the state
$|0\rangle$, and the mode of the upper array $B$, which circulates counterclockwise, is the state
$|1\rangle$. 

Then, the basis states at a fixed time $t=T_{M}$ are defined as 
\begin{align}\label{eq:basis-definition}
	&|0\rangle (t=T_M)\equiv |\leftarrow\rangle = \sum_{i,j} \Psi^{a}_{i,j}(T_M) \hat{a}_{i,j}^{\dagger} |A;0_{i,j}\rangle,\nn	&|1\rangle (t=T_M)\equiv|\rightarrow\rangle  = \sum_{i,j} \Psi^{b}_{i,j}(T_M) \hat{b}_{i,j}^{\dagger} |B;0_{i,j}\rangle.
\end{align}
Here $i$ and $j$ label the resonator at lattice position $(i,j)$ within an array, $\hat{a}^{\dagger}_{i,j}$ and $\hat{b}^{\dagger}_{i,j}$ create a single excitation on that resonator in arrays $A$ and $B$ respectively, and $\Psi^{a}_{i,j}(T_M)$ and $\Psi^{b}_{i,j}(T_M)$ are the complex amplitudes of the propagating wave packet on that resonator at the measurement time. The sums run over all sites of the corresponding array, so each logical state is a single excitation coherently spread over the whole edge, not an excitation of one resonator.
The two arrays $A$ and $B$ that carry these states are the ones drawn in Fig.~\ref{fig1:system-gates}(a), $A$ carrying the clockwise and $B$ the counterclockwise chiral mode. The gray lines there are the hoppings of Eq.~\eqref{eq:model-no-gates}; the colored links in Fig.~\ref{fig1:system-gates}(b) are the additional couplings that implement the gates, described in Sec.~\ref{sec:gates}.

Both logical states contain exactly one excitation, delocalized over the array, with
every resonator otherwise empty; $|A,B;0_{i,j}\rangle$ denotes that empty state. We work
throughout in this single-excitation sector. Because the Hamiltonian conserves the number
of excitations, the sector is closed: an initial state with one photon stays a state with
one photon, and no other filling ever enters the dynamics or the gate matrix elements.
Without inter-array couplings between resonators, this setting allows for a natural protection against bit-flip error ($|\leftarrow\rangle$ to $|\rightarrow\rangle$ and vice versa). Our model is quadratic, so the excitations do not interact.

A physical resonator array retains a residual Kerr nonlinearity,
$\sum_x\tfrac{K}{2}\hat{n}_x(\hat{n}_x-1)$ within an array and $\sum_x\chi\,\hat{n}^{a}_{x}
\hat{n}^{c}_{x}$ between two of them. Both annihilate any state that carries a single
excitation: the self-Kerr term requires two photons in the same mode, and the cross-Kerr term
one photon in each of two modes. The single-qubit gates of Sec.~\ref{sec:gates} are therefore
exact however large $K$ and $\chi$ may be, and nothing we report for them depends on the
circuit being linear.
The two-qubit gate of Sec.~\ref{sec:cphase} places one photon in each qubit, which is
precisely the configuration on which the cross-Kerr term stops being zero; what a device must deliver is therefore not the absence of
the nonlinearity but its localization, large at the coupler and small along the rest of the
edge; the strength required there is given in Sec.~\ref{sec:cphase}.

The platform offers a further option: the modes of opposite chirality in the lower gap
can be used to host a second qubit, which would allow for dual-rail encoding with error
correction. We denote the frequencies of the modes in the two gaps by $\omega_{\pm}$, and
indicate the frequency $\omega_+$ of the mode used in our simulations in
Fig.~\ref{fig2:signal-vs-bandstructure}. Finally, we note that the particle-hole symmetry
of a single Hofstadter model about the band center $\omega_0$ ensures that the chiral modes in
the upper and the lower gap have equal group velocities.

The protection of an idle qubit relies on the following: First, within one array the mode is chiral and the bulk is gapped with Chern numbers $1$, $-2$, $1$, so intra-array disorder has no counter-propagating state to back-scatter into and cannot mix the two logical states. Second, the arrays are physically disconnected, so no perturbation local to either array can transfer weight between them at all.

The residual error channel is therefore not a logical bit flip but leakage out of the computational subspace: weight that remains in the edge mode but is no longer in the wave packet that encodes the qubit, which is what we quantify throughout this work. This bears on how that error should be assessed. First, because the weight stays in the edge mode and the dynamics is unitary, the distortion that a given coupler imprints on the packet is a fixed and known transformation, and it can in principle be reduced by design: by a longer packet, which is narrower in frequency, by couplers whose action varies less across the bandwidth of the packet, or by shaping the packets injected in the two arrays so as to compensate the distortion of a given circuit. For the $X$ gate, doubling the duration of the envelope to $T_{\rm drive}=100\,f^{-1}$ at fixed couplings reduces the leakage from $1.2\%$ to $0.30\%$, at the cost of a packet twice as long on the perimeter. Second, the part of a packet that is distorted in this way does not come back as a wrong answer (in a sense that wrong gate was performed), it just fails to arrive in the mode that is read out. A readout matched to that mode and able to register the absence of a photon would therefore herald the event, making leakage an erasure rather than a Pauli error, and error-correcting schemes tolerate erasures far more readily than Pauli errors~\cite{Kubica2023erasure,Levine2024dualrail}. We do not simulate the readout in this work, so we only note this as a property the encoding may permit.

In addition, the mode is coherently distributed over many resonators, so that a perturbation of any single one shifts its phase only slightly, which suppresses dephasing. Finally, the edge modes disperse almost linearly over a wide range of in-gap frequencies, and keeping the weight of the wave packet within that linear region prevents it from spreading as it propagates.

The unitary dynamics simulated in this work must be compared with the loss rate of the resonators themselves, which is not part of the simulation. The quality factors of individual superconducting resonators reach $10^4$ in typical devices and $10^6$--$10^7$ in the best reported cases \cite{Romanenko2020PRA,Ory2025}. A packet completes one lap of the perimeter in a time $T_{\rm lap}=2(L_x+L_y)/v$, with $v$ the group velocity of the edge mode given in Sec.~\ref{sec:injection}, which is
$143.6\,f^{-1}$ for the $56\times56$ arrays, and in laboratory units the photon lifetime is $T_1 = Q/\omega_0$. The number of laps before the photon is lost is therefore
\begin{align}
\label{eq:laps-before-loss}
    N_{\rm laps}=\frac{Q}{(\omega_0/f)\,(f\,T_{\rm lap})},
\end{align}
which depends on the ratio $\omega_0/f$ between the resonator frequency and the engineered hopping. This ratio is set by the achievable coupling strength: $\omega_0/f=100$ corresponds to $50$~MHz. With $Q=10^6$ one obtains roughly $700$ laps at $\omega_0/f=10$ and $70$ laps at $\omega_0/f=100$ on the $56\times56$ arrays; with $Q=10^4$ the corresponding numbers are $7$ and below one. Since only about two gates fit along one perimeter (the packet is $vT_{\rm drive}=78\,l$ long on a perimeter of $224\,l$), resonator quality, rather than the topological protection, limits the achievable circuit depth, and the platform requires resonators in the high-$Q$ regime.

\subsection{Injection}
\label{sec:injection}
\begin{figure}
    \centering
    \includegraphics[scale=1.0]{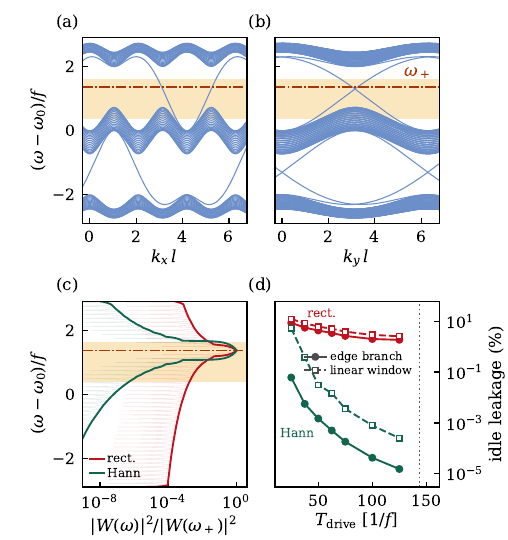}
    \caption{\label{fig2:signal-vs-bandstructure}%
Band structure, spectrum of the injected signal and leakage of an idle qubit.
(a), (b) Band structure of the model~\eqref{eq:model-no-gates} for ribbons of
width $56\,l$ that are translation invariant along $x$ and along $y$,
respectively. The dash-dotted line marks the working frequency $\omega_+$. In the shaded band,
$(\omega-\omega_0)/f \in [0.38, 1.62]$, the edge-mode group velocity stays within $10\%$ of its value at $\omega_+$.
(c) Normalized power spectrum $|W(\omega)|^2$ of the injected signal at
$T_{\rm drive} = 50/f$ for a rectangular (red) and a Hann (green) envelope. Heavy lines are the upper envelope of the spectrum and faint
lines the spectrum itself.
(d) Leakage of the identity operation against $T_{\rm drive}$:
the fraction of the injected weight lying outside the in-gap edge branch
(filled circles) and outside the shaded window (open squares).
The dotted gray line marks one lap of the perimeter, $T_{\rm drive} = 143.6/f$. }
\end{figure}

The basis states~\eqref{eq:basis-definition} at the time of measurement are defined
through the vector of distributed wave packets $(\Psi_{i,j}^{a}(t),\Psi_{i,j}^{b}(t))$. In what
follows we write the equations in the Heisenberg picture, in order to align the notation with
the tight-binding Hamiltonian in second quantization; in this format the time-dependent
coefficients $\Psi$ are absorbed into the corresponding creation operators. In this section we
discuss the injection procedure and the optimization of the injected signal, with the aim of
performing quantum computation at the highest possible fidelity. We assume that the drive is
applied at one of the edge sites, and in the simulations we choose the central site $x=L/2$
along the $x$ axis, at the $y=0$ edge. Here and in the gate Hamiltonians of Sec.~\ref{sec:gates}, $x$ denotes the position along the edge on which the packet is injected and the gates act, and $y$ the depth into the array. In the gauge of Eq.~\eqref{eq:model-no-gates} this edge is the one along which the hoppings are real, so that it runs along the $y$ direction of that equation. The drive is a periodic function applied for a finite
time, and the equations of motion for the two driven arrays read
\begin{align}
&\frac{d \hat{a}_{x,y}^{\dagger}(t)}{d t}=\frac{i}{\hbar}\left[H_{\Phi}\left(\hat{a}^{\dagger}, \hat{a}\right), \hat{a}_{x,y}^{\dagger}(t)\right]+\delta_{x,\frac{L}{2}}\delta_{y,0}A(t)e^{i\omega_+ t},\\
&\frac{d \hat{b}_{x,y}^{\dagger}(t)}{d t}=\frac{i}{\hbar}\left[H_{\Phi}^{*}\left(\hat{b}^{\dagger}, \hat{b}\right), \hat{b}_{x,y}^{\dagger}(t)\right]+\delta_{x,\frac{L}{2}}\delta_{y,0}B(t)e^{i\omega_+ t}.
\end{align}
Here the weight functions $A(t)$ and $B(t)$ describe the injection of the wave packet
during the finite interval $0<t<T_{\rm drive}$, and the time $t$ is measured in units of the inverse hopping
frequency $1/f$. We set $A(t)=B(t)=0$ for $t<0$ and $t>T_{\rm drive}$. In order to make the two injections
equivalent, we choose $A(t)=A\,W(t)$ and $B(t)=B\,W(t)$ with the same window function $W(t)$. 

The drive in these equations is a classical amplitude, and a classical drive prepares a
coherent state and not a single photon. For the wave-packet dynamics this makes no difference:
the system is linear, so the mean-field amplitude obeys the same equation as the
single-particle wavefunction, and the simulation gives the latter exactly. The encoding of
Sec.~\ref{sec:computational-basis-def}, however, requires the edge mode to hold a single-photon
Fock state. In a
device the role of the drive would therefore be played by a single-photon source, for example
a transmon coupled to the drive site that releases its excitation with a shaped envelope~\cite{Pechal2014PRX}. The
simulation with a classical drive gives the temporal mode that this photon must occupy.

Next, we address the challenge posed by the spreading of the injected wave packet in
time. Since the state is injected through a single edge resonator, part of the signal is immediately lost into the bulk modes (a fraction quantified by $L_{\rm idle}$ below), while the remaining part travels along the edge as a
superposition of edge modes of different frequency and momentum. In a perfectly fabricated
system without disorder, the only mechanism by which the wave packet can spread in space and
change its shape is the nonlinearity of the edge-mode dispersion. In
Fig.~\ref{fig2:signal-vs-bandstructure} we therefore compare the band structure of
model~\eqref{eq:model-no-gates}, with periodic boundary conditions along $x$ in panel (a) and
along $y$ in panel (b), with the Fourier transform of the injected signal $W(t)\exp(i\omega_{+}t)$ in panel (c). The shaded band in Fig.~\ref{fig2:signal-vs-bandstructure} shows the limit at which the
group velocity $v=d\omega(k)/dk$ starts to deviate more than $10\%$ from its value at the
band gap center $\omega_+$, which is $v=1.56\,f l$ and which we use throughout as the
propagation speed of the packet.

We use the Hann window
\begin{align}
    \label{eq:hann-window}
    W_{\rm H}(t)=\tfrac{1}{2}\left[1-\cos\left(2\pi t/T_{\rm drive}\right)\right],\qquad 0<t<T_{\rm drive},
\end{align}
whose first derivative also vanishes at the end points, so that its power decays as $(\omega-\omega_+)^{-6}$~\cite{Harris1978}. Its spectrum therefore stays inside the linear part of the edge branch,
Fig.~\ref{fig2:signal-vs-bandstructure}(c), whereas the discontinuities of a rectangular window
of the same length leave tails that decay only as $(\omega-\omega_+)^{-2}$ and fall outside
it; the comparison is given in Appendix~\ref{app:envelope}.

Panel (d) of Fig.~\ref{fig2:signal-vs-bandstructure} gives the baseline for the
gates of Sec.~\ref{sec:gates}: the leakage of the identity operation, that is,
of a packet that circles the array with no gate acting on it. We measure it against a fixed
target, the chiral edge branch itself. The idle leakage $L_{\rm idle}$ is the fraction of the
injected weight that does not lie in the eigenstates of the finite array within the upper bulk
gap. These are stationary states, so $L_{\rm idle}$ is fixed once the drive is over and does
not grow as the packet propagates. For the Hann envelope at $T_{\rm drive}=50\,f^{-1}$ it is
$1.5\times10^{-3}\,\%$, against $4.3\%$ for a rectangular envelope of the same length.
Restricting the target to the part of the branch inside the linear window gives $0.031\%$ and
$6.1\%$, respectively. The basis states~\eqref{eq:basis-definition} are the full injected
packets; since these lie in the edge branch to within $L_{\rm idle}$, restricting them to the
branch would change the numbers quoted below by no more than that. All results reported in
the remainder of this paper use the Hann envelope~\eqref{eq:hann-window} with
$T_{\rm drive}=50\,f^{-1}$.

The Hann window is a simple fixed choice and not the outcome of an optimization; the temporal shape of the control signal could be refined further using quantum optimal-control methods, for example gradient-based approaches such as GRAPE~\cite{Khaneja2005grape} or smooth pulse optimization for leakage reduction~\cite{Motzoi2009}, and we leave it to future work.

\section{Gate Implementation}
\label{sec:gates}

Having established the computational basis, we now introduce a way to perform gate
operations. In the following we show how to implement the gate set
\begin{align}
    \{X, Z, T,H, \textit{CPhase}\}.
\end{align}
which is one of several possible building blocks for universal quantum
computation~\cite{Barenco1995,Boykin2000,Kitaev1997}. In this section we first show how to
perform the single-qubit gates, which are implemented by turning on additional hopping terms in
the system. We then add a two-qubit controlled-phase gate, implemented with a density--density
(Kerr) nonlinearity between two arrays. For every gate in this section we take the system size $L_x=L_y=56$. 

\subsection{Design method and figures of merit}
\label{sec:design-method}

The single-qubit gates between the chiral modes of our metamaterial system are performed
by turning on a coupling either between the two arrays, or within one of them. The idea behind every gate is the same: we introduce a local term that
breaks time-reversal or inversion symmetry near the edge, such that only the
exponentially localized edge modes are affected; the corresponding couplings are shown schematically in
Fig.~\ref{fig1:system-gates}. The $X$ gate requires the mixing of modes from the two arrays,
whereas the $Z$ and $T$ gates modify the phase of a mode within a single array only, and the
Hadamard gate is composed of an $X$-type coupler and $T$-type bonds. For the
two-qubit controlled-phase gate we instead propose to use a density--density nonlinearity. Each gate acts as an edge-located scatterer that transforms one or several modes
coherently. Key technical novelty of this scheme
lies in where the optimization takes place. In a conventional superconducting processor the
device is fixed and the gate is shaped by a time-dependent control pulse. Here the reverse
holds: the couplings are static, and the gate is designed by choosing which lattice bonds to
modify and by how much. A coupler is specified by a handful of discrete choices, namely which
bonds are altered, how many rows $D$ from the edge they extend over, and how far apart ($d$) they sit
along the edge, together with one continuous strength per modified bond. Since a working coupler
needs only a handful of such strengths, the continuous part of the problem is small enough for a direct numerical search against the target unitary, while the discrete choices are scanned
exhaustively.

We first fix the figures of merit, since the same set is used for every gate in this
section as well as for the disorder study of
Sec.~\ref{sec:disorder}. A gate here is not a unitary acting on an isolated two-level
system: the packet can also leave the computational subspace altogether when a coupler distorts it, and any useful fidelity measure has to separate how much weight survived
from how accurately the surviving weight was rotated. We therefore work with the
projected gate matrix, which retains both the amplitude and the phase information. Let
$|\psi^{\rm free}_i(t)\rangle$ be the trajectory obtained with all gate couplings
switched off for the input $i\in\{A,B\}$, and $|\psi_j(t)\rangle$ the trajectory
with the gate on. The gate matrix is
\begin{align}
\label{eq:gate-matrix-M}
    M_{ij}(t)=\frac{\langle\psi^{\rm free}_i(t)|\psi_j(t)\rangle}{W^2},
\end{align}
with $W^2$ the norm of the gate-free trajectories at the measurement time. Using
the gate-free trajectories as the reference basis fixes the phases of $M$ by
construction. From $M$ we
form, with $N=2$ the dimension of the computational subspace,
\begin{align}
\label{eq:fidelity-1X-metrics}
    &L = 1-\frac{\mathrm{Tr}\,(M^{\dagger}M)}{N},\qquad
    P_{\rm surv} = 1 - L, \nn
    &F_{\rm pro}=\frac{|\mathrm{Tr}\,(U^{\dagger}M)|^{2}}{N^{2}},\qquad
    F_{\rm cond}=\frac{F_{\rm pro}}{P_{\rm surv}},
\end{align}
where $U$ is the target gate. Here $L$ is the leakage out of the computational
subspace and $P_{\rm surv}$ the survival (in the sense of Ref.~\cite{Wood2018PRA}) and
$F_{\rm pro}$ is the process fidelity of the gate including that loss. The
conditional fidelity $F_{\rm cond}$ divides the loss out and measures how
accurately the intended rotation was performed on the weight that did survive.
By construction $F_{\rm pro}=P_{\rm surv}\,F_{\rm cond}$, so the two numbers factorize the gate
error into weight lost from the computational subspace, $1-P_{\rm surv}=L$, and misalignment of the
weight that remains, $1-F_{\rm cond}$. We quote both because the two respond to different design choices: the former
depends on how the coupler acts on the spectral content of the injected packet, the latter on
how accurately the coupling strengths are set. The average gate
fidelity $F_{\rm avg}=(NF_{\rm pro}+1)/(N+1)$~\cite{Horodecki1999PRA,Nielsen2002PLA}
is listed in Appendix~\ref{sec:table-of-optimized-params} for reference only, since that relation is exact only for trace-preserving maps and our projected map is not one.
The reverse process, seepage from its complement back into the computational subspace,
is fixed by unitarity to be $(N-\mathrm{Tr}\,M^{\dagger}M)/N_L$ with
$N_L=6270$ the dimension of the complement, giving values of order $4\times10^{-6}$
against leakage of order $10^{-2}$.

The optimization objective for every single-qubit gate is the process fidelity
$F_{\rm pro}$ of Eq.~\eqref{eq:fidelity-1X-metrics}, evaluated at a single application of the gate to a freshly injected packet. Couplers with more than two free strengths, namely the $X$ couplers deeper than two rows and the Hadamard coupler, are optimized with the evolutionary, non-gradient algorithm
CMA-ES~\cite{Hansen2001,Hansen2016}, which suits the problem because each evaluation requires
integrating the equations of motion of the full lattice and no gradient
is available in closed form.  We use a population of $12$ to $14$ per generation over $40$ to $45$ generations, and seed each
run with the best couplings previously found for that geometry. The $X$ strengths are bounded
to $[0,5]f$; for the Hadamard coupler the inter-array strengths are bounded to $[-5,0]f$ and
the intra-array ones to $[-0.9,5]f$.  The time evolution is obtained with the \texttt{scipy}
integrators~\cite{Virtanen2020}. For a description of the algorithm and modern
implementations see Ref.~\cite{katiraeefar2025cmaes}. Couplers with one or two strengths, which include the adopted $X$ gate and the $T$ and $Z$
gates, are optimized with the Nelder--Mead simplex method, started from the best point of a
coarse grid or from an earlier estimate, with each coupling bounded to $[0,5]f$ for the $X$
gate and to $[0,4]f$ for the $T$ and $Z$ gates.

Reconstructing the gate matrix, $M$, requires the gate-free reference trajectory for each input, which is
natural in simulation but not something an experiment has access to. If one wanted to fine-tune the $X$ gate on a device, it would be done instead by monitoring the energy that
arrives in the far array. Two such ratios are available. The first is the energy in
subsystem $B$ after a mode injected into $A$ has crossed the coupler, relative to what
was injected,
\begin{align}
\label{eq:fidelity-1X}
    \mathcal{E}_{\rm trans}(T_{M})& = \frac{|E_{B}[t=T_{M}]|}{|E_{A}[t=T_{\rm drive}]|}=\frac{|\langle\rightarrow |H_{\Phi}^{*}|\rightarrow\rangle|_{T_{M}}}{|\langle\leftarrow |H_{\Phi}|\leftarrow\rangle|_{T_{\rm drive}}},\\
    \label{eq:measurement-time-T1X}T_M&=\frac{2(L_x+L_y)}{v} + \frac{T_{\rm drive}}{2},
\end{align}
and the second compares the energy in $B$ after three applications with that after one,
\begin{align}
\label{eq:fidelity-3X}
    &\mathcal{E}_{\rm ret} =\frac{E_B[t=T_{M,3X}]}{E_B[t=T_{M,1X}]},\nn
    &T_{M,3X}=\frac{6\left(L_x+L_y\right)}{v}+\frac{T}{2},\, T_{M,1X}=\frac{2\left(L_x+L_y\right)}{v}+\frac{T}{2}.
\end{align}
For the $T$ and $Z$ gates the corresponding quantity is the accumulated phase,
\begin{align}
    \label{eq:fidlity-T-gate}
    \Theta_{T}(T_M)=\frac{\arg\left(\langle \rightarrow | U_T | \rightarrow\rangle_{t=T_M}\right)}{\pi/4},
\end{align}
measured at the time $T_M$ of Eq.~\eqref{eq:measurement-time-T1X}. Equation~\eqref{eq:measurement-time-T1X} is the nominal arrival time, $168.6\,f^{-1}$ on the $56\times56$ arrays. A coupler shifts the arrival of the packet, so in the simulations $T_M$ is calibrated for each gate and each application as the time at which $F_{\rm pro}$ is largest; for a single passage the calibrated values lie between $161\,f^{-1}$ and $188\,f^{-1}$. For the Hadamard gate it is the phase imbalance $\Theta_H$
\begin{align}
\label{eq:H-gate-phase-fidelity-definition}
    \Theta_{H}=\frac{1}{\pi}\left\langle \left|\phi_{A, \mathrm{in\,A}}-\phi_{B, \mathrm{in\,B}}\right| \right\rangle .
\end{align}
It is the difference between the phases of the ``out $A$ from in $A$'' and ``out $B$ from in $B$''
elements, normalized to $\pi$. All of these are measurable without a reference trajectory, which is their virtue, but they all carry the same limitation: they
are blind to any weight lost from the computational subspace. We therefore optimize $F_{\rm pro}$ in
simulation and quote these observables only as the quantities an experiment would monitor.

\subsection{X-gate}
\begin{figure}
    \centering
    \includegraphics[scale=1.0]{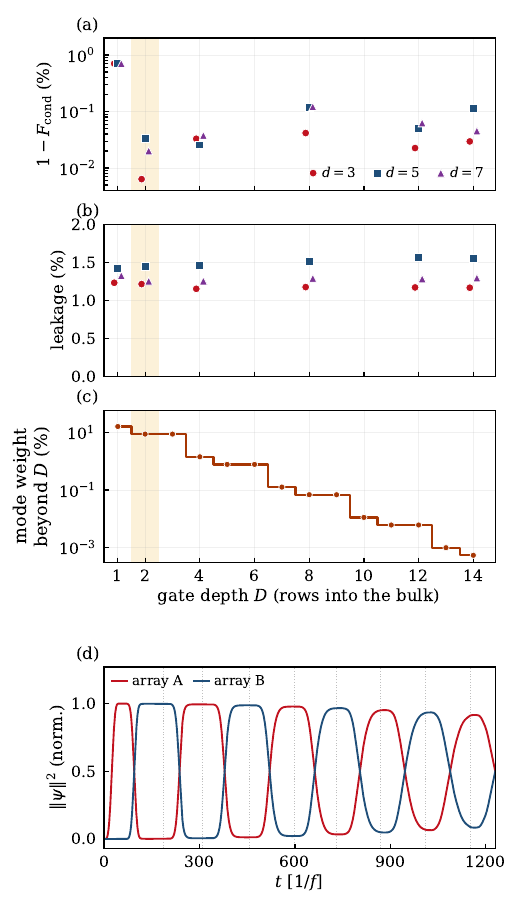}
    \caption{\label{fig:X-gate}%
The $X$ gate: dependence on the depth $D$ of the coupler and on the separation
$d$ of its attachment points along the edge, and repeated application.
(a) Deviation of the conditional fidelity from unity, $1-F_{\rm cond}$,
for $d=3$ (red circles), $d=5$ (blue squares) and $d=7$
(purple triangles), with the coupling strengths re-optimized at
every pair $(d,D)$.
(b) Leakage out of the computational subspace for the same
configurations.
(c) Fraction of the edge-mode weight lying beyond depth $D$, at the gap center $\omega_+$.
The shaded band in panels (a)--(c) marks $D=2$, the depth adopted for the $X$ gate.
(d) Weight $\|\psi\|^2$ in array $A$ (red) and array $B$ (blue), normalized
to the gate-free trajectory, for the adopted coupler applied once per lap to a packet
injected into array $A$. Dotted vertical lines mark the measurement times $T_M$. }
\end{figure}

In the Hamiltonian formulation, we define the X-gate as
\begin{align}
\label{eq:hamiltonian-X}
H_{X}=\sum_{y}g^X_{y}\left(\hat{a}^{\dagger}_{x_0,y}\hat{b}_{x_0,y}+\hat{a}^{\dagger}_{x_0+d,y}\hat{b}_{x_0-d,y}\right)+\mathrm{h.c.} 
\end{align}
Here the operators $\hat{a}$ act in the lower array $A$ of
Fig.~\ref{fig1:system-gates}, that is, in the first block of the
Hamiltonian~\eqref{eq:model-no-gates}, and the operators $\hat{b}$ act in the second array $B$. The coordinates $x_0$ and $x_0\pm d$ are the attachment points of the inter-array coupling. As drawn in Fig.~\ref{fig1:system-gates}(b), the structure consists of a straight rung joining $(x_0,y,A)$ to $(x_0,y,B)$, together with a single long diagonal joining $(x_0+d,y,A)$ to $(x_0-d,y,B)$; it therefore spans $2d$ along the edge. The summation over $y$ runs over the $D$ rows inward from the edge that the coupler occupies, which is the gate depth. The two links at a given depth $y$ are given the same strength $g^X_y$, and the strengths at different depths are optimized independently. 

Two further discrete choices remain: the number of inter-array links and their separation
$d$ along the edge. We use two links throughout: two separated attachment points can be switched independently, which is what allows a single mode to be addressed within a sequence of gates.
The separation is a weak parameter, and unlike the depth it shows no threshold. Optimizing
the two couplings independently at every separation from $3l$ to $9l$, at fixed depth $D=2$,
the process fidelity spans $98.52\%$ to $98.78\%$. Majority of this range is leakage: the coherent error $1-F_{\rm cond}$ stays below $0.04\%$ at every separation. The variation is not monotonic in $d$, and it does
not follow the period of the magnetic unit cell. We therefore adopt
$d=3l$, which is the best of the seven and also the most compact, spanning $6l$ along the edge, so that the long diagonal link is as short as possible. The separations $d=5$ and $7$ are shown in Table~\ref{tab:params_fidelity} for comparison, the others in the data repository~\cite{Supplement-code}.
The second discrete parameter is the depth $D$ of the gate, that is, how many rows of resonators inward from the edge the modified couplings extend. Ideally the edge mode would live entirely on the outermost row; in practice it decays exponentially into the bulk, so a coupler of finite depth captures only part of it and a deeper coupler also reaches sites where the mode is negligible. Figure~\ref{fig:X-gate}(a) and (b) resolve the two effects, using couplings re-optimized independently at every depth $D$ and separation $d$. Depth behaves as a threshold: A single-row coupler is poor, with $1-F_{\rm cond}\simeq0.71\%$, and the three separations agree there to within $0.003$ percentage points. Adding the second row improves the coherent error by up to two orders of magnitude, to $0.006$--$0.033\%$. Beyond that nothing is gained: $D=4$ is level with $D=2$. Interestingly $D=8$ is consistently a factor of two to six \emph{worse}. The explanation is in Fig.~\ref{fig:X-gate}(c). The first two rows already carry $91.2\%$ of the mode, and the third row carries none at all, because every third row of the three-site magnetic cell is the sublattice on which the edge mode amplitude vanishes identically. A coupler deeper than two rows therefore achieves very little additional overlap while introducing additional couplings that must themselves be tuned, and the optimization problem becomes artificially harder.
The leakage tells the analogous story. Across the whole grid of eighteen configurations it varies only between $1.15\%$ and $1.56\%$, with no trend in $D$, and what little variation there is follows $d$. We accordingly fix $D=2$ for the $X$, $T$ and $Z$ gates reported below; the inter-array coupler of the Hadamard gate extends one row further (Sec.~\ref{sec:hgate}).

We simulate the $X$ gate application by integrating the equations of motion of $H_{meta}+H_X$.
The value of gate coupling values $g_y^X$ would be experimentally implemented via capacitive or inductive couplings~\cite{Splitthoff2024PRR}.

We show the resulting performance in Fig.~\ref{fig:X-gate}(d) over eight applications, by the last of which the leakage has grown to $22\%$ (Fig.~\ref{fig:phase-H-gates}(e)). The results are presented for the optimized coupler of Fig.~\ref{fig1:system-gates}(b), at depth $D=2$ rows from the edge, with the strengths
\begin{align}
    \label{eq:optimized-g_X}
    g^{X}/f=\{1.7474,\, 0.6692\}.
\end{align}
The optimization is given in Table~\ref{tab:params_fidelity}.
For a single application at $d=3$, across the depths listed there, the conditional fidelity reaches $99.958\%$ to $99.994\%$, while the leakage stays between $1.15\%$ and $1.21\%$. Between $97\%$ and $99\%$ of the total gate error is therefore leakage, and almost none of it
is a misaligned gate rotation. The coupler does its job essentially exactly, and
what limits the gate is the weight that the coupler removes from the reference packets. This weight stays on the edge: the idle packet leaks only $1.5\times10^{-3}\,\%$
(Sec.~\ref{sec:injection}), and the gate adds nothing to that. Leakage depends on how the
coupler acts on the spectral content of the packet, and for a fixed envelope, weaker and
intra-array couplers leak least.

\subsection{Phase gates: $T$ and $Z$}\label{sec:zgate}

\begin{figure}
    \centering
    \includegraphics[scale=1.0]{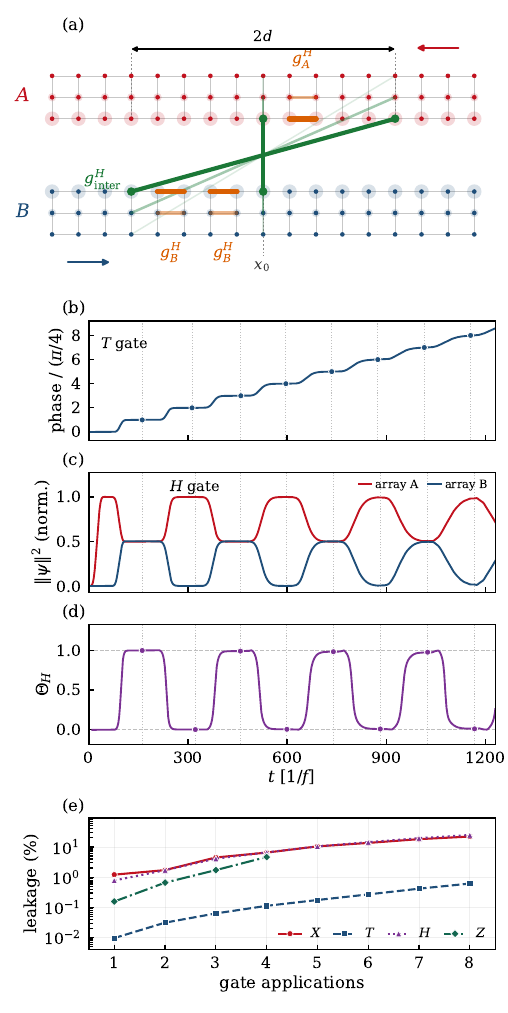}
    \caption{\label{fig:phase-H-gates}%
The phase gates and the Hadamard gate in the clean system.
(a) Geometry of the Hadamard coupler, drawn at lattice resolution for the adopted
$d=5$. Green: the inter-array coupler. Orange: the three intra-array bonds.
The shading of the sites indicates the
edge-mode weight and the arrows the direction of circulation.
(b) Phase accumulated on $|1\rangle$ by the $T$ gate, in units of $\pi/4$; circles mark the ideal value
$n$ at the $n$-th application.
(c) Weight $\|\psi\|^2$ in array $A$ (red) and array $B$ (blue), normalized
to the gate-free trajectory, for the $H$ gate and a packet injected into array $A$.
(d) Phase imbalance $\Theta_H$ of
Eq.~\eqref{eq:H-gate-phase-fidelity-definition}, in units of $\pi$.
(e) Leakage $L$ against the number of gate
applications, for the $X$ (circles), $T$ (squares), $H$ (triangles) and $Z$ (diamonds) gates.
Dotted vertical lines in (b)--(d) mark the measurement times $T_M$. }
\end{figure}

Next, we turn to the implementation of the $T$ gate. This gate applies a phase $\pi/4$ to
a qubit state and is essential for the realization of non-Clifford operations. Here we
illustrate how to implement it on the state $|1\rangle=|\rightarrow\rangle$, which resides in array $B$. In our setup, the gate is realized by strengthening two bonds of array $B$ near its edge, as shown schematically in Fig.~\ref{fig1:system-gates}(b). The geometry is described by the same two parameters as the $X$ gate. The depth $D$ is the number of rows inward from the edge over which the bonds are modified, and we again take $D=2$; the separation $d$ is the distance along the edge between the two modified bonds, which occupy the columns $x_0$ and $x_0+d$ with $d=5$. Note that here the two intra-array bonds lie on the same side of $x_0$ and the structure spans $d$. Each bond is horizontal, joining neighboring resonators along the edge within array $B$, and the two bonds at a given depth carry the same strength, so the gate has $D$ free couplings rather than $2D$. The Hamiltonian term for this gate has the form:
\begin{align}
\label{eq:hamiltonian-T}
   H_{T}=\sum_{y}g^T_{y}\left(\hat{b}^{\dagger}_{x_0,y}\hat{b}_{x_0+1,y}
   +\hat{b}^{\dagger}_{x_0+d,y}\hat{b}_{x_0+d+1,y}\right)+\mathrm{h.c.},
\end{align}
where the sum over $y$ runs over the $D$ rows inward from the edge and
$g^T_y$ is the strength on both bonds at depth $y$.
The $Z$ gate uses the same mechanism as the $T$ gate, acting on the same pair of
modified bonds of array $B$ shown in Fig.~\ref{fig1:system-gates}(b), with the target phase
$\pi$ in place of $\pi/4$.

We fix the two coupling strengths by optimizing the same objective used for the $X$
gate, the process fidelity at a single application, and obtain
\begin{align}
    \label{eq:optimized-g_T-depth-2}
    g^{T}/f=\{0.738,\, 3.991\},
\end{align}
the second of which lies at the upper bound of the search range, which gives $F_{\rm pro}=99.990\%$ and $F_{\rm cond}=100.000\%$ with a leakage of
$0.0097\%$. This is the best of the single-qubit gates on every measure for the following structural reason: the $T$ gate modifies two bonds within a single array and opens no channel between the arrays, so the perturbation it applies to the
propagating mode is far weaker than that of the $X$ or $H$ couplers.
We compose the $Z$ gate from two passages through one coupler that imprints a phase of
$\pi/2$ per passage, $Z=S^2$. Re-optimizing the two strengths on $F_{\rm pro}$ of the
two-passage gate gives
\begin{align}
    \label{eq:optimal-Z-couplings}
    g^{Z}/f=\{2.064,\, 0.863\},
\end{align}
with $F_{\rm pro}=99.840\%$, $F_{\rm cond}=99.997\%$ and a leakage of $0.157\%$, listed
alongside the other gates in Table~\ref{tab:params_fidelity}. A single passage through
this coupler is an $S$ gate, with $F_{\rm pro}=99.959\%$, $F_{\rm cond}=99.999\%$ and a leakage
of $0.040\%$. We use two passages because a $Z$ gate completed in one passage on the same bonds
reaches only $F_{\rm pro}=96.7\%$ within the bound on the couplings. The larger phase requires a stronger coupling on the edge row than the $T$ gate and the leakage is correspondingly higher, but it
remains well below the $1.2\%$ of the $X$ gate.

The performance of individual and sequential $T$ gates is shown in Fig.~\ref{fig:phase-H-gates}(b).
In Fig.~\ref{fig:phase-H-gates}, panel (c) is for injection into array $A$, while panels (b), (d) and (e) use $M(t)$, which is built from both inputs.
The accuracy of the phase is best seen over repeated applications. The phase accumulated
after $n$ gates is $n\pi/4$ to within $0.25\%$ for $n$ up to eight, and the conditional
fidelity stays above $99.99\%$ throughout, so that the only accumulating error is the leakage,
which reaches $0.62\%$ by the eighth application. The leakage of all gates over repeated applications is compared in Fig.~\ref{fig:phase-H-gates}(e).

\subsection{H (Hadamard)-gate}
\label{sec:hgate}

Finally, we describe a Hadamard gate realization. It combines an $X$-type coupler between the
arrays with $T$-type bonds inside them, so the Hamiltonian is built from the terms of
Eqs.~\eqref{eq:hamiltonian-X} and \eqref{eq:hamiltonian-T}, and the target is
$U_H = (\sigma_x+\sigma_z)/\sqrt{2}$ in the computational basis. The inter-array coupler has separation $d=5$ and extends three rows inward. Three intra-array
bonds, each two rows deep, complete it. In array $B$ a pair of bonds two columns apart sits at
$x_0-4$ and $x_0-2$, both between that array's two attachment points; the two share one
strength per row. In array $A$ a single bond sits at
$x_0+1$, between that array's attachment points, with strengths of its own; the
arrangement is drawn in Fig.~\ref{fig:phase-H-gates}(a). This leaves seven
coupling strengths, three inter-array and four intra-array.

We fix the seven coupling strengths by optimizing the process fidelity
$F_{\rm pro}$ of Eq.~\eqref{eq:fidelity-1X-metrics} at a single gate application, with the
CMA-ES procedure of Sec.~\ref{sec:design-method}, and obtain
\begin{align}
    \label{eq:optimal-H-couplings}
    &g^{H}_{\rm inter}/f=\{-1.9420,\, -0.3616,\, -0.0034\},\nn
    &g^{H}_{B}/f=\{1.2037,\, -0.4626\},\nn
    &g^{H}_{A}/f=\{-0.1588,\, 1.5677\},
\end{align}
giving $F_{\rm pro}=99.22\%$ and $F_{\rm cond}=99.996\%$ with a leakage of $0.78\%$, the lowest
of the inter-array gates and below the $X$ gate's $1.21\%$. 

Repeating the optimization at the other separations gives $98.68\%$, $98.93\%$ and $99.25\%$ for
$d=3$, $7$ and $9$, an irregular dependence like the $X$ gate's, and we adopt $d=5$ because it
ties the best of them to within the spread between optimizer seeds while being the most compact,
spanning $10\,l$ along the edge. Fig.~\ref{fig:phase-H-gates}(c)--(e) show the gate over eight consecutive
applications. The remaining coherent error is $0.004\%$, meaning
the gate is almost entirely limited by leakage: $99.5\%$ of the gate error is weight that left the
computational subspace. The $H$ gate now leaks less than the $X$ gate, $0.78\%$ against
$1.21\%$, despite modifying bonds within the arrays as well as between them, because it transfers only half of the weight across. It also stands up well to repetition, reaching $98.29\%$ after two
applications and $75.06\%$ after eight (Appendix~\ref{sec:deep-circuit-appendix}).

\subsection{Controlled-phase gate via density nonlinearity}
\label{sec:cphase}
In this section we present a model for a controlled-phase gate $C\!-\!R(\phi)$. Applied repeatedly, and combined with the single-qubit gates of the preceding sections, it yields a controlled-NOT; the number of repetitions required is quantified below. The main reason to focus on the phase $\phi\ll \pi$ is to avoid creation of strong scatterer in the system, that would lead to higher leakage out of the edge mode during the gate operation. Below we show the proposed protocol and discuss its fidelity in terms of that leakage in detail. 
\begin{figure*}
    \centering
    \includegraphics[scale=1.0]{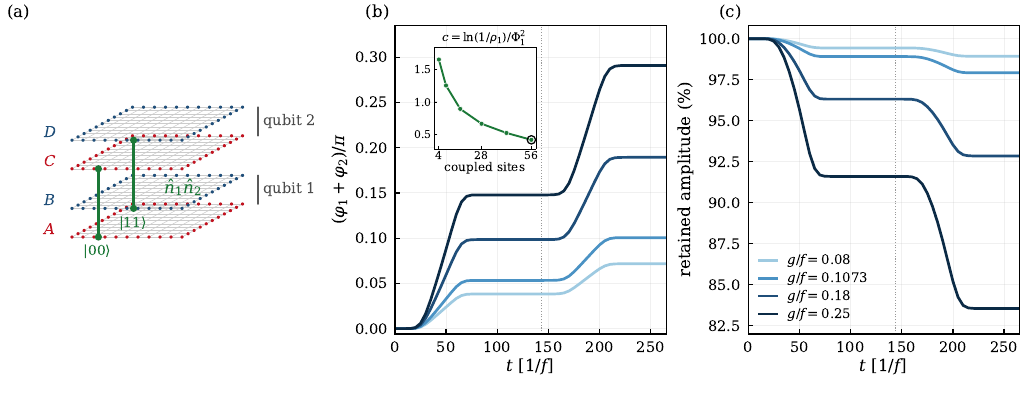}
    \caption{\label{fig:controlled-phase}%
Controlled-phase gate via a density--density nonlinearity.
(a) The four resonator arrays. Qubit $1$ is encoded in arrays $A$ and $B$ and
qubit $2$ in arrays $C$ and $D$. The
nonlinear coupler of Eq.~\eqref{eq:hamiltonian-cphase}, marked $\hat{n}_1\hat{n}_2$, joins the arrays carrying co-propagating
modes, $A$ with $C$ and $B$ with $D$.
(b) Phase $(\varphi_1+\varphi_2)/\pi$ accumulated by the two co-propagating
configurations, against time, for four values of the nonlinear coupling $g$, at the adopted
coupler, the whole edge row of $56$ sites. Each step is one passage of the packets through
the coupler; the dotted line marks one lap of the perimeter.
Inset: the coefficient $c$ of Eq.~\eqref{eq:cphase-cost} against the number of coupled sites at
$g/f=0.1288$; the circled point is the adopted coupler.
(c) Amplitude $\mathcal{E}_{\rm CP}$ retained in the computational subspace relative to its
initial value, for the same four couplings and with the same colors as in (b). }
\end{figure*}

We assume that the resonators of one edge row carry a Kerr
nonlinearity of density--density form between the two qubits, as could be implemented following
Ref.~\cite{Kounalakis2018}. The coupler acts at the same sites in both sectors: a single row,
the one at the edge, coupled along its entire length of $N_{\rm c}$$=L_x=56$ sites. The Hamiltonian is
\begin{align}
\label{eq:hamiltonian-cphase}
    H_{\rm CP}= g \sum_{x\in \mathcal{C}}\left(\hat{n}^{a}_{x,1}\hat{n}^{c}_{x,1}
    +\hat{n}^{b}_{x,1}\hat{n}^{d}_{x,1}\right),
\end{align}
where $\hat{n}^{a}_{x,y}=\hat{a}^{\dagger}_{x,y}\hat{a}_{x,y}$, and likewise for arrays $b$,
$c$ and $d$; $y=1$ is the row at the edge, and $g$ is the single coupling strength, varied in
Fig.~\ref{fig:controlled-phase}, $\mathcal{C}$ the set of the coupled sites. Each term joins a site of one qubit's array to the same
site of the other qubit's co-propagating array, $A$ with $C$ and $B$ with $D$. The length $N_{\rm c}$
and the choice of a single row are design decisions; both are justified below.

$H_{\rm CP}$ commutes with the photon number of every array and is diagonal in the site
basis. It neither moves nor creates photons, and only shifts the energy of configurations in
which the two photons sit on the same coupled site at the same time. Two consequences
follow. First, a state with one photon per qubit remains in the two-photon sector exactly, so
the dynamics can be simulated without truncating the Fock space. Second, the coupler acts
only in the two-qubit sectors in which both arrays it joins are occupied, $A$--$C$ and
$B$--$D$, which fixes which computational states acquire a phase; this is worked out below. The four arrays of Fig.~\ref{fig:controlled-phase}(a) need not
be four physical devices; the second qubit can instead occupy the lower-gap modes of the same
pair of arrays, as proposed in Sec.~\ref{sec:computational-basis-def}. What the simulation requires is that each qubit be represented on its
own copy of the lattice, so that the two-particle state is propagated in the tensor product
of the two single-particle spaces, one sector at a time. Both packets are injected exactly as in Sec.~\ref{sec:injection}, with the same Hann envelope of length $50\,f^{-1}$, and the
two-particle evolution begins once the drive is over.

Each qubit uses the encoding of Sec.~\ref{sec:computational-basis-def}:
$|0\rangle\equiv|\leftarrow\rangle$ is the clockwise mode of the first array of its pair and
$|1\rangle\equiv|\rightarrow\rangle$ the counterclockwise mode of the second, each a single
excitation delocalized over its array as in Eq.~\eqref{eq:basis-definition}. With qubit~1 in
arrays $a$ and $b$ and qubit~2 in arrays $c$ and $d$, at a fixed time we define
\begin{align}
	&|\leftarrow\rangle_1 = \sum_{x,y} \Psi^{a}_{x,y} \hat{a}_{x,y}^{\dagger} |A;0_{x,y}\rangle,\nn
	&|\rightarrow\rangle_1 = \sum_{x,y} \Psi^{b}_{x,y} \hat{b}_{x,y}^{\dagger} |B;0_{x,y}\rangle,\nn
	&|\leftarrow\rangle_2 = \sum_{x,y} \Psi^{c}_{x,y} \hat{c}_{x,y}^{\dagger} |C;0_{x,y}\rangle,\nn
	&|\rightarrow\rangle_2 = \sum_{x,y} \Psi^{d}_{x,y} \hat{d}_{x,y}^{\dagger} |D;0_{x,y}\rangle.
\end{align}
The gate that preserves the density of state and changes phases of individual components has the following form in computational basis $|00\rangle, |01\rangle, |10\rangle, |11\rangle$:
\begin{align}
    U_{CPhase}=\text{diag}(e^{i\varphi_1}, 1, 1, e^{i\varphi_2}).
\label{eq:cphase-matrix}
\end{align}

The coupler entangles only if $\vartheta=\varphi_1+\varphi_2\neq0$ modulo $2\pi$, equivalently if
the Schmidt rank of $U_{CPhase}$ exceeds one. The density--density term shifts both
co-propagating configurations with the same sign, so $\varphi_1$ and $\varphi_2$ add rather
than cancel, while the antiparallel sectors are annihilated and their entries in
Eq.~\eqref{eq:cphase-matrix} are exactly unity.
Fig.~\ref{fig:controlled-phase}(b) shows $\vartheta$ over a range of nonlinearity strengths.

Panel (c) of Fig.~\ref{fig:controlled-phase} tracks the amplitude the gate costs: the weight
remaining in the computational subspace relative to the same system without the gate,
\begin{align}
&\mathcal{E}_{\rm CP}(t)= \frac{\left|M_{A C}(t)\right|+\left|M_{B D}(t)\right|}
{\left|M_{A C}(0)\right|+\left|M_{B D}(0)\right|},\\
& M_{A C}(t)=\left\langle\psi_{1, A}^{\text {free }}(t) \otimes \psi_{1, C}^{\text {free }}(t) \mid \psi_{A C}^{\text {int }}(t)\right\rangle,
\end{align}
and correspondingly for $M_{BD}$. The two terms are the two co-propagating configurations on
which the nonlinearity acts, and each compares the interacting state with the freely
propagated tensor product of the two single-particle references. Built from the moduli of
these overlaps, $\mathcal{E}_{\rm CP}$ is a benchmark for an amplitude that does not take into an account the phase. This quantity is also normalized to the start of the interacting evolution rather than to the injected weight.

The coupler has only two parameters, the coupling strength $g$ and the number of coupled
sites $N_{\rm c}$, so we scan them directly rather than optimizing numerically as for the
single-qubit gates. Let the retention $\rho_1$ be the factor by which $\mathcal{E}_{\rm CP}$
is multiplied each time the packets cross the coupler, so that
$\mathcal{E}_{\rm CP}=\rho_1^{\,n}$ after $n$ passages, and let $\vartheta_1$ be the entangling
phase acquired in one passage, in radians. The phase is linear in the total coupling, as a
second-order process requires, and the amplitude lost is its square,
\begin{align}
\label{eq:cphase-cost}
    \ln(1/\rho_1)=c\,\vartheta_1^{2}.
\end{align}
The coefficient $c$ is dimensionless and measures how much amplitude a given phase costs. It
depends on the geometry alone: it saturates as $g\to0$ and falls as $c\sim N_{\rm c}^{-0.5}$, from
$1.65$ at four coupled sites to $0.42$ at fifty-six; see the inset of
Fig.~\ref{fig:controlled-phase}(b). A long weak coupler therefore reaches a given phase at
less loss than a short strong one. Since $N_\pi=\pi/\vartheta_1$ passages bring $\vartheta$ to $\pi$, the
amplitude surviving a controlled-$Z$ is $e^{-\pi c\vartheta_1}$, so the geometry enters only
through $c$ and the coupling only through the phase per passage.

We couple the edge row along its whole length, $N_{\rm c}$$=L_x$,
because $c$ falls monotonically with $N_{\rm c}$. We use a single row because the nonlinearity has
purchase only where the mode amplitude is large, which is the row at the edge: adding a second
row of the same sites doubles the number of nonlinear resonators but raises the phase per
passage by only $6\%$, leaves $c$ unchanged, and lowers the process fidelity of the completed
gate by $0.7$ percentage points.

Weakening the coupling always improves the gate, because the surviving amplitude
$e^{-\pi c\vartheta_1}$ grows as the phase per passage falls. But a smaller phase per passage needs
more passages, and each passage is one lap of the perimeter, so the count is bounded by
Eq.~\eqref{eq:laps-before-loss}: for $Q=10^6$ and $\omega_0/f=100$ the budget is $70$ laps on
the $56\times56$ arrays. We therefore adopt $g/f=0.1073$, which reaches $\vartheta=\pi$ in $20$
passages, well inside the budget, at a loss of about one percent in amplitude per passage. The count
assumes the per-passage phase and loss stay constant beyond the two passages we simulate.
Table~\ref{tab:cphase-sizes} collects the adopted parameters and the resulting fidelities.

At the adopted coupling the packets gain $\vartheta_1=0.0503\pi$ per passage and retain
$\rho_1=0.990$ of their amplitude, so that $c=0.42$. Twenty passages then bring $\vartheta$ to
within $0.6\%$ of $\pi$, and the gate matrix is
$M=\mathrm{diag}(m_{AC}^{N_\pi},1,1,m_{BD}^{N_\pi})$, up to the local $Z$ rotations that the platform
supplies as single-qubit gates. Evaluating Eq.~\eqref{eq:fidelity-1X-metrics} with $N=4$
gives $F_{\rm pro}=82.01\%$, $F_{\rm cond}=98.93\%$, $F_{\rm avg}=85.61\%$ and a leakage of
$17.09\%$. The leakage is the amplitude lost at the coupler: $81\%$ survives in the two
sectors it acts on, $66\%$ in probability. This is the key limiting factor for $F_{\rm cond}$. Unlike the single-qubit couplers, the
nonlinearity conserves only the total energy of the two photons, so one of them can leave the
gap. This is the smaller part of the loss: of the $2.2\%$ of the probability that leaves the
reference pair in the first passage, $2.0\%$ keeps both photons in the edge branch and $0.2\%$ has
at least one photon outside it. The two co-propagating sectors
are simulated separately and agree as the conjugation between the arrays requires. Resonator
loss at the same $Q$ and $\omega_0/f$ lowers $F_{\rm pro}$ to $61.6\%$. The gate takes
$2856\,f^{-1}$, or $9.1\,\mu$s at $f/2\pi=50$~MHz, and asks of the device a nonlinearity
$\chi/2\pi=5.4$~MHz, within reach of the tunable cross-Kerr couplers of
Ref.~\cite{Kounalakis2018}.

Table~\ref{tab:cphase-sizes} collects these numbers alongside the same gate on the
$42\times42$ arrays. Both rows are taken at twenty laps, so they differ only in the size of
the array, which separates the two things size does: the longer coupler lowers $c$ from
$0.48$ to $0.42$ and raises $F_{\rm pro}$ from $79.94\%$ to $82.01\%$, while the longer
perimeter makes the same twenty laps take a third more time, so the photon loss over the gate
grows in proportion. Since $c$ is still falling at $N_{\rm c}$$=56$, a larger array would potentially further improve the gate.

\begin{table}[tb]
\centering
\begin{tabular}{l c c c r r r}
\hline\hline
array & $N_{\rm c}$ & $g/f$ & $N_\pi$ & $F_{\rm pro}$ & $F_{\rm cond}$ & $F_{\rm pro}^{\rm loss}$\\
 & & & & (\%) & (\%) & (\%)\\
\hline
$56\times56$ & 56 & 0.1073 & 20 & \textbf{82.01} & 98.93 & 61.64\\
$42\times42$ & 42 & 0.132 & 20 & 79.94 & 98.62 & 64.48\\
\hline\hline
\end{tabular}
\caption{The controlled-phase gate on two array sizes, both at twenty laps. $N_{\rm c}$ is the
number of coupled sites, the whole edge row in each case, and $N_\pi$ the number of laps that
brings $\vartheta$ to $\pi$. $F_{\rm pro}$ and $F_{\rm cond}$ are unitary, as everywhere else in
this work; $F_{\rm pro}^{\rm loss}$ adds the resonator loss of
Eq.~\eqref{eq:laps-before-loss} at $Q=10^{6}$ and $\omega_0/f=100$, for which the budget is
$70$ laps on the $56\times56$ arrays and $93$ on the $42\times42$. Bold marks the adopted
gate.}
\label{tab:cphase-sizes}
\end{table}

\section{Protection against disorder}
\label{sec:disorder}

In this section we analyze the role of disorder. As established in Sec.~\ref{sec:model-shape}, disorder local to a single array cannot produce a logical bit flip, because that array carries only one chiral mode in the gap and the two arrays are physically disconnected. The leading residual error is therefore weight that remains in the edge mode of the same array but leaves the wave packet that encodes the qubit, which we refer to throughout as leakage. Meaning what we compute is not photon loss in the sense of absorption in the resonators, but a redistribution of the wave packet along the edge of the array. Resonator loss is a separate and independent error channel; we estimate its size from measured quality factors in Sec.~\ref{sec:computational-basis-def}. We first evaluate the leakage in the disordered system without gates, then show how disorder strength affects the fidelity of single-qubit gates, and finally propose possible error mitigation strategies.

Two kinds of imperfection are natural for couplers between resonators: a weak disorder affecting every coupling, as a finite fabrication tolerance would
produce, and a small number of strongly disordered couplings, as defects would. We simulate
the second, assuming fabrication uniform enough that the dominant imperfection is defects
introduced during fabrication and operation. All results in this section use that model.

Disorder is applied to the hoppings in both lattice directions: for each realization we draw
$n$ sites at random in each array and multiply the two hoppings emanating from each, one
along $x$ and one along $y$, by $1+\delta'$ with $\delta'$ uniform on $[-\delta,\delta]$.
Since $\delta\le0.3$ throughout, no coupling vanishes or changes sign. The two arrays are
disordered at independently chosen positions. We take $n$ from $50$ to $1000$ on the
$56\times56$ arrays, up to about $30\%$ of their $3136$ sites, and $\delta$ from $0.01$ to
$0.3$. Ten realizations are generated at every pair $(\delta,n)$, giving $550$ in all.

Rather than treating the disorder strength $\delta$ and the number of defects $n$ as two
independent controls, we note that they enter only through the product of
the number of scatterers and the variance of their strength. We therefore characterize a
realization by the single scattering parameter
\begin{align}
    \label{eq:scattering-parameter}
    x = \frac{n\delta^{2}}{3},
\end{align}
where the factor of three is the variance of a uniform distribution on $[-\delta,\delta]$.

First we answer a question what happens to a wave packet that simply circles a disordered array without any gate acting on it. We
inject a mode into a single disordered array with all gate couplings switched off and follow
it for eight laps of the perimeter, recording two quantities. The first is the fraction of
the weight that remains within the five outermost rows of resonators, which measures whether
the mode is still bound to the edge. The second is the squared overlap of the disordered
trajectory with the trajectory of the same packet in the clean system, which measures
whether it is still the same packet.

The two quantities behave completely differently, and the contrast is the main result of
this section. The edge confinement is untouched. Over the whole grid of $550$ realizations,
the fraction of the weight lying within five rows of an edge after eight laps stays
between $99.26\%$ and $99.36\%$, against a clean-system value of $99.33\%$. This is the quantitative verification of the topological protection present in this system: a single array carries one chiral mode in the gap and has no counter-propagating state at the same energy, so disorder within the array has
nothing to backscatter into and cannot detach the packet from the boundary.

\begin{figure}
    \centering
    \includegraphics[scale=1.0]{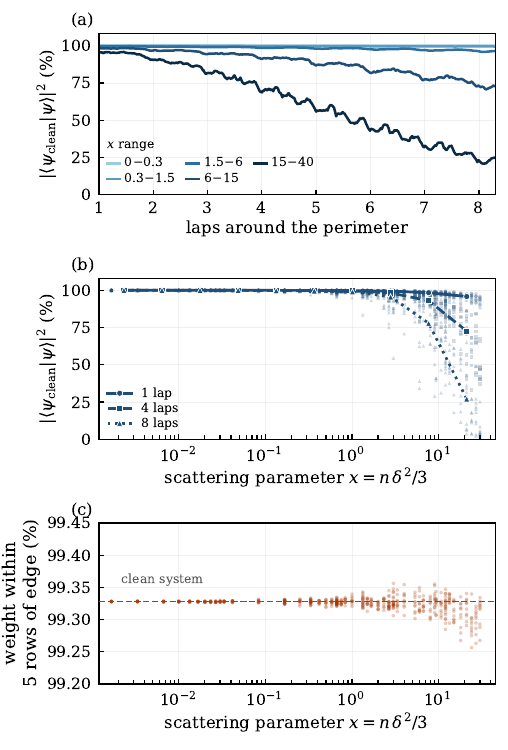}
    \caption{\label{fig:photon-loss-disorder-error}%
Behavior of an idle qubit in a disordered array, against the
scattering parameter $x$ of Eq.~\eqref{eq:scattering-parameter}.
(a) Squared overlap $|\langle\psi_{\rm clean}|\psi\rangle|^{2}$ of the
disordered trajectory with the trajectory of the same packet in the clean
system, as a function of the number of laps. Each curve is the median over all
realizations falling in one of five ranges of $x$, listed in the legend.
(b) The same quantity evaluated after one, four and eight laps, against $x$.
(c) Fraction of the weight in array $A$ lying within five rows of an edge,
evaluated after eight laps, against $x$; the dashed line is the clean-system value.
In (b) and (c) faint symbols are individual realizations and heavy symbols the
median within logarithmic bins of $x$. }
\end{figure}

What disorder does instead is dephase the packet as it travels. The overlap with the clean
trajectory decays lap by lap, at a rate set by the same scattering parameter $x$ of
Eq.~\eqref{eq:scattering-parameter} that organizes the gate data. Weak disorder leaves the
packet alone: for $x<0.3$ it is still $99.98\%$ intact after eight laps. The decay matters
only in the top two bands of Fig.~\ref{fig:photon-loss-disorder-error}(a), where the median
overlap after eight laps is $73\%$ around $x\simeq10$ and $23.0\%$ for $x\ge15$, the strongest
disorder on our grid. Even there it crosses $1/e$ only after $6.7$ laps. The packet
therefore survives roughly seven circuits of the array even at the strongest disorder
considered here, so the achievable circuit depth is set by how long the packet stays coherent
and not by any loss of topological protection.

\begin{figure}[t]
    \centering
    \includegraphics[scale=1.0]{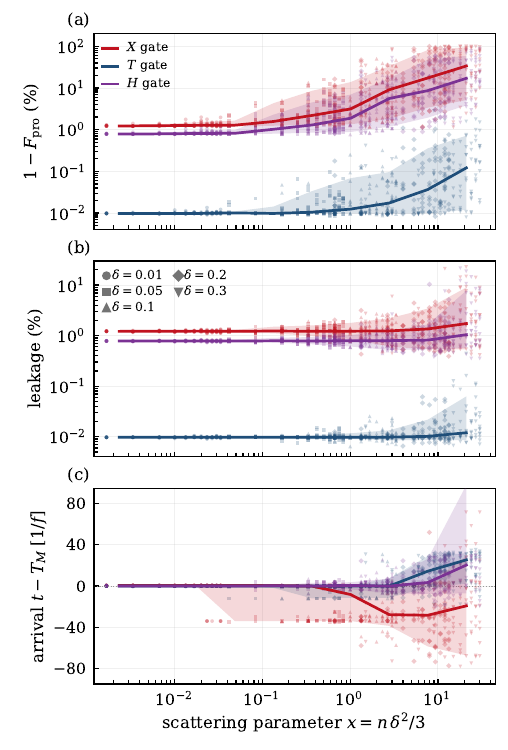}
    \caption{\label{fig:gate-fidelity-vs-disorder}%
Effect of coupling disorder on the three single-qubit gates, against the
scattering parameter $x$ of Eq.~\eqref{eq:scattering-parameter}. Faint symbols are individual realizations, with the symbol shape
indicating $\delta$; solid lines are the median and the shaded bands the range
between the tenth and ninetieth percentiles, both taken over all realizations
falling in logarithmic bins of $x$. Colors denote the gate: $X$ in red, $T$ in
blue and $H$ in purple.
(a) Deviation of the process fidelity from unity, $1-F_{\rm pro}$.
(b) Leakage out of the computational subspace.
(c) Arrival time of the wave packet, measured relative to the clean-system
measurement time $T_M$ of the corresponding gate. }
\end{figure}

We now apply disorder while the gates are switched on. We again organize the data by the
scattering parameter \eqref{eq:scattering-parameter}. We have verified the collapse directly by comparing pairs of $(\delta,n)$ that share a
value of $x$: all disagreements lie within twice the combined standard error, and most
within one. The results are shown in Fig.~\ref{fig:gate-fidelity-vs-disorder}.

The spread between disorder realizations grows from below
$0.1$ percentage points at weak disorder to tens of percentage points at the strongest, and
at the largest $x$ the outcome for the $X$ gate is close to bimodal: some realizations still
execute the gate almost perfectly while others fail outright. With ten realizations per grid
point the resulting uncertainty on any individual average is of order ten percentage points,
so the medians and percentile bands in Fig.~\ref{fig:gate-fidelity-vs-disorder} should be
read as indicating the trend and the width of the distribution.

The central observation is the contrast between panels (a) and (b). The process fidelity of
the $X$ gate degrades strongly with $x$, while its leakage moves hardly at all, staying
close to the clean-system value of about $1.2\%$ over the entire range. Disorder is
therefore not removing the wave packet from the edge, in agreement with our observations about topological protection mentioned above. As in the idle case, disorder perturbs the phase
accumulated along the edge and the time at which the packet reaches the coupler, and it is
that which spoils the gate. 

Panel (c) makes the mechanism explicit. It plots, for every realization, the time at which the
packet arrives, measured from the clean-system measurement time $T_M$ of that gate. At weak
disorder the arrival is within about $1\,f^{-1}$ of $T_M$. By $x\ge15$ it varies from one
realization to the next over a range of some $100\,f^{-1}$ for the $X$ gate.

The three gates are affected very differently according to the mechanism underlying each gate. The
$T$ gate is nearly immune to disorder since it modifies two bonds inside a single array and never transfers weight
between the arrays, so it does not depend on the inter-array transfer amplitude and phase,
which is precisely the quantity that is most sensitive to disorder. The $X$ and $H$ gates both open an
inter-array channel and both degrade, the Hadamard less than the $X$ gate, consistent with its transferring only half of the weight between the arrays. A practical consequence is that circuits should be
compiled to use phase gates freely and inter-array gates sparingly, and that calibration
effort is best spent on the arrival time.

The picture assembled above suggests two concrete mitigation strategies. The first concerns the measurement time. Figure~\ref{fig:gate-fidelity-vs-disorder}(c) shows
that what disorder principally does is move the arrival of the packet. Since a given device has a fixed realization of the disorder, that shift is systematic, and can be removed by calibrating $T_M$ on the device itself instead of taking it from the clean-system value. What cannot be removed this way is the spread between realizations, which is why the shift must be
measured per device.

The second concerns how circuits are compiled. Phase gates acting within one array are almost
untouched by disorder. A compiler
for this platform should therefore prefer $T$ gates over $X$ and $H$ gates wherever a
decomposition allows the choice, and should re-inject rather than extend a computation over
many laps.

\section{Conclusions}
\label{sec:conclusions}
In this work, we proposed a realization of a topologically protected qubit encoded in the
bosonic chiral edge modes of a pair of metamaterial Chern insulators, arranged such that the
pair reproduces the counter-propagating mode structure of the quantum spin Hall effect: bosonic resonators as
building blocks, tunable capacitive or inductive couplers as hoppings, and the Hofstadter model
as the underlying scheme. The logical states are the two directions of circulation. 

The qubit states are well defined and protected against bit flip by
construction, since the two arrays are physically disconnected and each carries a single chiral
mode in the gap. Injection is deterministic, by driving one edge resonator with a shaped-single-photon source. Behavior under disorder confirms
the topological protection and locates the limit of this platform: the circuits are limited by dephasing and resonator quality.

We built $X$, $T$, $Z$, $H$ and a controlled-phase gate, all by modifying
lattice bonds rather than by shaping pulses. Each single-qubit gate was optimized on the process
fidelity at a single application, $H$ with CMA-ES~\cite{Hansen2001,Hansen2016} and $X$, $T$
and $Z$ with Nelder-Mead, reaching $98.78\%$ for $X$, $99.99\%$ for $T$, $99.84\%$ for $Z$, composed of two passages, and
$99.22\%$ for $H$. Their residual error is almost entirely leakage, more than $97\%$ of the total for every gate. This leakage depends on how the coupler acts on the spectral content of the packet: for a fixed envelope, weaker and intra-array couplers leak least.

The controlled-phase gate uses a density--density nonlinearity along one
entire edge, and composing a controlled-$Z$ from $20$ passages reaches a process
fidelity of $82.0\%$, or $61.6\%$ once resonator loss is included. The two-qubit gate has the most room left for improvement, because it relies on the long trajectories along the device and is thus most affected by dephasing and leakage. Improving resonator and adding pulse shaping are the two natural trajectories for improvement here. As an alternative route to two-qubit operations, one could in the future work consider
the Knill--Laflamme--Milburn protocol~\cite{Knill2001Nature}, which builds a
probabilistic CNOT from linear elements alone. It requires beam splitters between the rails of
different qubits, together with auxiliary photons, whereas the $X$-gate coupler presented here
mixes only the two rails of one qubit; an inter-qubit coupler of the same kind would
therefore have to be designed first.

Finally, while we discussed possible routes for readout as well as reframing the leakage out of the computational subspace as a heralded erasure, we did not simulate them and we believe that these directions could constitute stimulating future work.

\begin{acknowledgements}
We acknowledge the fruitful discussions with Fabian Hassler, Anton Akhmerov, and Christian Kraglund Andersen. D. O. O. and E. G. acknowledge the support by Kavli Foundation. This work is also part of the project Engineered Topological Quantum Networks (Project No.VI.Veni.212.278) of the research program NWO Talent Programme Veni Science domain 2021, which is financed by the Dutch Research Council (NWO).
\end{acknowledgements}

\begin{center}
    {\bf Data availability statement}
\end{center}
The data that support the findings of this study were generated by numerical simulations. The source code and parameters used to generate the simulations is publicly available \cite{Supplement-code}. The generated data is available from Ref.\cite{Supplement-code-2}.

\appendix
\onecolumngrid
\clearpage %
\section{Table of optimized gate parameters}
\label{sec:table-of-optimized-params}
\begin{table}[ht]
\centering
\footnotesize
\setlength{\tabcolsep}{4pt}
\begin{tabular}{l c c p{5.9cm} r r r r r}
\hline\hline
\textbf{Gate} & $D$ & $d$ & \textbf{Couplings} $g/f$ &
$F_{\rm pro}$ & $F_{\rm avg}$ & $F_{\rm cond}$ & $L$ & $S_{\rm seep}$\\
 & & & & (\%) & (\%) & (\%) & (\%) & ($10^{-6}$)\\
\hline
\multicolumn{9}{l}{\textit{$X$ gate, inter-array coupler at $d=3$, couplings optimized on $F_{\rm pro}$}}\\
$X$  & 2  & 3 & \raggedright 1.7474, 0.6692
       & \textbf{98.780} & 99.187 & 99.994 & 1.214 & 3.9\\
$X$  & 4  & 3 & \raggedright 1.6979, 1.2416, 2.7807, 1.4801
       & 98.816 & 99.211 & 99.967 & 1.152 & 3.7\\
$X$  & 8  & 3 & \raggedright 1.5975, 0.8249, 0.0638, 1.8032, 0.9389, 1.3486, 1.9800,
                 1.3488
       & 98.784 & 99.190 & 99.958 & 1.174 & 3.7\\
$X$  & 12 & 3 & \raggedright 1.6992, 1.2602, 3.3319, 1.5800, 0.3268, 0.4550, 0.8322,
                 0.7197, 0.1470, 0.4411, 1.1182, 0.8710
       & 98.808 & 99.205 & 99.977 & 1.170 & 3.7\\
$X$  & 14 & 3 & \raggedright 1.7338, 1.1796, 1.8158, 1.1486, 0.4377, 0.7342, 1.4204,
                 0.4195, 0.3046, 0.1337, 1.9335, 1.8753, 0.6230, 1.0424
       & 98.805 & 99.203 & 99.970 & 1.166 & 3.7\\
\hline
\multicolumn{9}{l}{\textit{$X$ gate, the same optimization at wider separations}}\\
$X$  & 2  & 7 & \raggedright 1.7304, 1.2007
       & 98.727 & 99.151 & 99.980 & 1.253 & 4.0\\
$X$  & 12 & 7 & \raggedright 1.6867, 1.7223, 2.4460, 1.0759, 0.5012, 1.1518, 1.5948,
                 0.8328, 0.3228, 0.7517, 0.4738, 0.1511
       & 98.655 & 99.103 & 99.937 & 1.283 & 4.1\\
$X$  & 14 & 7 & \raggedright 1.7331, 1.6143, 1.3637, 0.8687, 1.1440, 1.1056, 0.2864,
                 0.2054, 1.5533, 1.4620, 0.7240, 1.6669, 2.5517, 2.2460
       & 98.661 & 99.107 & 99.955 & 1.294 & 4.1\\
$X$  & 2  & 5 & \raggedright 1.6932, 1.3459
       & 98.519 & 99.013 & 99.967 & 1.448 & 4.6\\
$X$  & 4  & 5 & \raggedright 1.7041, 1.4921, 0.0583, 0.1470
       & 98.516 & 99.011 & 99.974 & 1.458 & 4.7\\
$X$  & 8  & 5 & \raggedright 1.6718, 1.8694, 2.1260, 0.7002, 0.0274, 0.1529,
                 1.4029, 2.2374
       & 98.377 & 98.918 & 99.882 & 1.507 & 4.8\\
$X$  & 12 & 5 & \raggedright 1.7571, 1.7518, 1.0912, 0.2025, 1.4666, 0.4317,
                 0.6315, 1.1581, 2.3576, 1.0464, 0.4106, 1.9279
       & 98.387 & 98.924 & 99.949 & 1.563 & 5.0\\
$X$  & 14 & 5 & \raggedright 1.6882, 1.9351, 2.2758, 0.7582, 0.4599, 1.4007,
                 1.1816, 1.1981, 0.5666, 1.0627, 0.2666, 1.6226, 0.2510, 0.4299
       & 98.337 & 98.892 & 99.885 & 1.549 & 4.9\\
\hline
\multicolumn{9}{l}{\textit{$T$ and $Z$ gates, modified intra-array bonds in array $B$, optimized on $F_{\rm pro}$}}\\
$T$  & 2 & 5 & \raggedright 0.738, 3.991
       & \textbf{99.990} & 99.994 & 100.000 & 0.0097 & 0.03\\
$S$  & 2 & 5 & \raggedright 2.064, 0.863
       & 99.959 & 99.973 & 99.999 & 0.040 & 0.13\\
$Z$  & 2 & 5 & \raggedright two passages of $S$
       & 99.840 & 99.893 & 99.997 & 0.157 & 0.5\\
\hline
\multicolumn{9}{l}{\textit{$H$ gate: three-bond coupler at $d=5$, and the two-qubit controlled-$Z$}}\\
$H$ & 3\,/\,2 & 5\,/\,2 & \raggedright
   $g_{\rm inter}$: $-$1.9420, $-$0.3616, $-$0.0034\newline $g_{B}$: 1.2037, $-$0.4626\newline $g_{A}$: $-$0.1588, 1.5677
       & \textbf{99.217} & 99.478 & 99.996 & 0.779 & 2.5\\
$C\!-\!Z$ & 1 & $56$ & \raggedright $g$: 0.1073, uniform
       & \textbf{82.01} & 85.61 & 98.93 & 17.09 & 0.02\\
\hline\hline
\end{tabular}
\caption{Coupling parameters of the gates discussed in the text, together with their
clean-system performance. Every row is evaluated on the same frozen setup: a single
application of the gate to a freshly injected packet, Hann envelope of length
$T_{\rm drive}=50\,f^{-1}$, $56\times56$ arrays, measured at the first arrival. $D$ is the gate depth in rows inward from the edge and $d$ the separation along the
edge; for the inter-array coupler the attachment columns are $x_0$ and $x_0\pm d$, so the
structure spans $2d$, while for the intra-array bonds $d$ is the separation between the
modified bonds. Gates carrying both list them as inter\,/\,intra. The five metrics are defined in
Eqs.~\eqref{eq:gate-matrix-M} and \eqref{eq:fidelity-1X-metrics}; only two are independent. Bold marks the parameter set adopted for each gate. The $X$ blocks give the adopted
separation $d=3$ and the same optimization repeated at $d=5$ and $d=7$. The $H$ row lists its
intra-array couplings for array $B$ and then array $A$. For the $C\!-\!Z$ row, $D$ counts
coupled rows and the third column coupled sites along the edge.
The leakage $L$ is referenced to the gate-free trajectory and is therefore what
the coupler adds to the idle leakage of $1.5\times10^{-3}\,\%$,
Fig.~\ref{fig2:signal-vs-bandstructure}(d). The $Z$ row is the exception to the single application: it is two passages through the
coupler of the $S$ row (Sec.~\ref{sec:zgate}).
Coupling sets not listed here are given in the accompanying data
repository~\cite{Supplement-code}.}
\label{tab:params_fidelity}
\end{table}

\section{Optimization for repeated application on a single injected packet}
\label{sec:deep-circuit-appendix}

The gate parameters quoted in the main text are optimized for a single application of
the gate on a freshly injected wave packet. For completeness we record here what happens
when the objective is changed to the opposite regime, in which one packet is injected and
the same gate is applied repeatedly as the packet circles the array. The two objectives do
not have the same optimum, because the two error mechanisms scale differently: leakage
compounds, the surviving weight after $n$ passes being approximately $(1-L_1)^n$, whereas
the coherent error of a self-inverse gate partially cancels between successive
applications. A parameter set that is best at $n=1$ need not be best at large $n$, and
conversely.

For the Hadamard gate this can be seen directly. Table~\ref{tab:deep-H} compares the
adopted couplings of Sec.~\ref{sec:gates} with the same seven couplings re-optimized on
$F_{\rm pro}$ at seven applications. The adopted set is the better of the two at a single
application, $99.22\%$ against $96.09\%$, and the worse at $n=2$ and at every depth beyond the third: by
$n=8$ it has fallen to $75.06\%$ while the re-optimized set still holds $85.49\%$. The
re-optimization buys this by roughly halving the leakage, $10.82\%$ against $19.49\%$ at
$n=7$, at the price of a less accurate single application. It does so by weakening the couplings of the second row, the second inter-array row from $-0.36$ to
$-0.23$ and the second row of both intra-array bonds: once the objective penalizes
accumulated leakage, the optimizer weakens the second row of the coupler. The third inter-array row, on which the edge mode has no weight, is not withdrawn: it changes from $-0.003$ to $-0.146$.

\begin{table}[ht]
\centering
\begin{tabular}{c|cc|cc}
\hline
 & \multicolumn{2}{c|}{$F_{\rm pro}$ (\%)} & \multicolumn{2}{c}{leakage (\%)}\\
$n$ & adopted & $n{=}7$ opt & adopted & $n{=}7$ opt\\
\hline
1 & \textbf{99.22} & 96.09 &  0.78 &  0.76\\
2 & 98.29 & \textbf{99.17} &  1.71 &  0.82\\
3 & \textbf{95.87} & 94.52 &  4.10 &  2.35\\
4 & 93.33 & \textbf{96.71} &  6.63 &  3.24\\
5 & 89.49 & \textbf{91.37} & 10.43 &  5.52\\
6 & 85.53 & \textbf{92.52} & 14.39 &  7.35\\
7 & 80.37 & \textbf{86.13} & 19.49 & 10.82\\
8 & 75.06 & \textbf{85.49} & 24.85 & 14.32\\
\hline
\end{tabular}
\caption{Hadamard gate applied $n$ times to a single injected packet, clean system, Hann
envelope with $T_{\rm drive}=50\,f^{-1}$. Both columns are the three-bond coupler adopted in Sec.~\ref{sec:gates}. The first is the
adopted set, optimized on $F_{\rm pro}$ at a single application; the second is the same seven
couplings re-optimized at $n=7$, $g_{\rm inter}=\{-1.863,-0.226,-0.146\}$,
$g_{B}=\{1.223,-0.276\}$, $g_{A}=\{0.211,1.266\}$. Best value
in each row in bold. Only odd $n$ are meaningful as Hadamard gates, since $H^2=\mathbbm{1}$.}
\label{tab:deep-H}
\end{table}

We do not extend this re-optimization to the other gates, for two reasons. The first is
that the repeated-application regime cannot reach the fidelities of the single-application
regime: since leakage accumulates from one application to the next, the survival factor $(1-L_1)^n$
places a ceiling on $F_{\rm pro}$ that no choice of couplings can lift. Table~\ref{tab:deep-H}
shows $F_{\rm pro}$ down to $85\%$ by $n=8$ even for the set optimized at depth, against
$99.2\%$ for one application on a fresh packet. Re-injecting the
packet is simply a better use of the platform than optimizing against that ceiling.

The second reason is that the repeated-application objective admits solutions that are not
gates in the intended sense. Applying it to the $X$ gate returns couplings that are very
weak, for instance a single link of strength $g=0.197$ at $d=5$, or $g=\{0.227,0.028\}$ at
$d=3$ and depth $2$. These score well at $n=7$, reaching $97.9\%$ and $97.5\%$ against
$77.2\%$ and $81.1\%$ for the corresponding $n=1$ optima, but they transfer only $17\%$ and
$22\%$ of the mode between the arrays on each pass, accumulating a passage area of $\pi$
only after seven full laps. They score $16.5\%$ and $21.1\%$ at $n=1$. Such a solution is a
legitimate device, but it is a slow $X$ gate whose duration is seven laps, not a faster or
better one, and it is not what the objective was intended to find. Constraining the
per-pass transfer would be necessary to make the repeated-application objective meaningful
for the $X$ gate, and we leave this to future work.

\section{Spectral properties of Hofstadter Hamiltonian}
\label{app:spectral}
The Hofstadter Hamiltonian of a single array for an infinite lattice written for a single unit cell reads:
	\begin{align}
		H_{Hof}=\begin{pmatrix}
			f_x (e^{2i\pi/3+i k_x }+e^{-2i\pi/3-i k_x }) & f_y & f_y e^{i 3 k_y}\\
			f_y  &f_x (e^{i k_x}+e^{-i k_x }) & f_y\\
			f_y e^{-i 3 k_y}  & f_y & f_x (e^{-2i\pi/3+i k_x }+e^{2i\pi/3-i k_x })
		\end{pmatrix}.
	\end{align}
	From this we get an equation for its spectrum, where we omit the constant energy offset defined by a particular material implementation:
	\begin{align}
		E^3-3 E (f_x^2+f_y^2)-2 f_x^3\cos(3 k_x)-2 f_y^3 \cos(3 k_y)=0.
	\end{align}
	Using the trigonometric Cardano formula, we find the roots of spectral equation:
	\begin{align}
		E_{j=1,2,3}=2\sqrt{f_x^2+f_y^2}\cos\left[\frac{1}{3}\arccos\left[\frac{f_x^3\cos(3k_x)+f_y^3\cos(3k_y)}{\sqrt{(f_x^2+f_y^2)^3}}\right]+\frac{2j\pi}{3}\right].
	\end{align}
	For hopping parameters $f_x=f_y=f$ the gaps are centered around values $\pm (1+\sqrt{3})/2 f$ and have a size $(3-\sqrt{3})f$. For other ratios of $f_x/f_y$ parameters with the same value of spectrum width,  $f_x^2+f_y^2=2 f^2$, we find that the band gaps are smaller. That is a consequence of the symmetry of the spectrum with respect to interchange of $f_x$ and $f_y$ parameters at $k_x=k_y=0,\pi$ points.

 For the purpose of state signal optimization and gate construction, we analyze the localization of edge states for 1D infinite lattices. We take both orientations of such lattices with translation invariance along x and y directions. 

 Due to the maximization of the gap at $f_x=f_y$, we use these values in all numerical simulations. A particular material realization might allow the half-width of the Fourier transform to be narrowed, permitting an asymmetry between $f_x$ and $f_y$; that would in turn single out a preferred direction for gate implementation, since the two ribbon orientations then have different edge-mode decay lengths.

 In the present paper we have shown the possible implementation of gates when the mode propagates parallel to the $y$-direction of Eq.~\eqref{eq:model-no-gates}, for which a gate of depth two already covers most of the mode weight; the gate Hamiltonians of Sec.~\ref{sec:gates} label the position along this edge by $x$. 

\section{Injection envelope}
\label{app:envelope}
\begin{figure}[ht]
    \centering
    \includegraphics[scale=1.0]{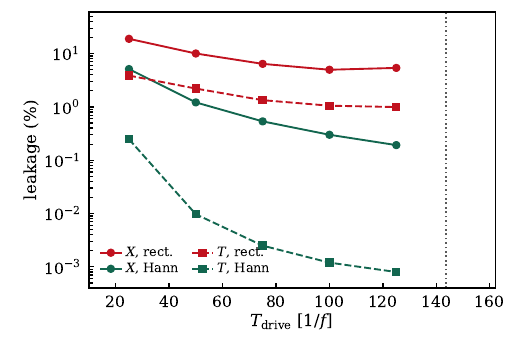}
    \caption{\label{fig:envelope-appendix}%
Leakage out of the computational subspace after a single gate application,
against the injection time $T_{\rm drive}$, for the $X$ gate (circles) and the
$T$ gate (squares) with a rectangular (red) and a Hann (green) injection
envelope. The dotted gray line marks $T_{\rm drive} = 2(L_x+L_y)/v = 143.6/f$, beyond
which the packet is longer than the perimeter and overlaps its own tail.}
\end{figure}

In this Appendix we compare the Hann envelope of Eq.~\eqref{eq:hann-window} with a rectangular one,
\begin{align}
    W(t)=\Theta(t)-\Theta(t-T_{\rm drive})
\end{align}
of the same length $T_{\rm drive}$. The expression for the Fourier transform of the function $W(t)$ is given by a $sinc$ function
\begin{align}
\label{eq:FFT-of-signal}
W(\omega)=T_{\rm drive} e^{-i(\omega-\omega_+) T_{\rm drive} / 2} \operatorname{sinc}\left(\frac{(\omega-\omega_+) T_{\rm drive}}{2}\right).
\end{align}
The sinc tails of Eq.~\eqref{eq:FFT-of-signal} are a direct consequence of the discontinuity of the rectangular window at $t=0$ and $t=T_{\rm drive}$, and for such an envelope they are the dominant source of leakage in the whole scheme. Their weight decays only as $(\omega-\omega_+)^{-2}$, so a fixed fraction of the injected signal is deposited outside the linear part of the edge branch no matter how the couplers are designed. Replacing the rectangular window by one that vanishes smoothly at both ends removes this floor at no cost in packet length. The Hann window is such a window. Fitting the measured spectra over the relevant range gives exponents $-2.02$ and $-6.04$ for the two windows, and the corresponding spectral weight falling outside the linear region of the edge dispersion is $2.6\%$ for the rectangular window against $0.026\%$ for the Hann window at $T_{\rm drive}=50\,f^{-1}$, both as a fraction of the weight lying within the bandwidth of the lattice. This difference propagates directly into the performance of the gates. These two numbers characterize the spectrum of the drive; the weight actually
deposited on the lattice outside the same window is given in Sec.~\ref{sec:injection}.
Figure~\ref{fig:envelope-appendix} shows the consequence for the gates: the leakage of the $X$
and $T$ gates after a single application, with the couplings held fixed, against the injection
time for the two envelopes.

\section{Transformation of the Hamiltonian to the real-valued coupling form}
\label{app:real-form}

In this Section we describe the transformation of a two-component model with Hofstadter Hamiltonians into the real-hopping representation. The main procedure was given in Ref. \cite{Huber2015Science} and was motivated by converting the model into classical mechanical analog. 
The unitary operation of transformation into the basis of circular polarizations for edge modes (Kramers pairs) has the form 
\begin{align}
    U=\frac{1}{\sqrt{2}}\left(\begin{array}{rr}
			1 & -i \\
			1 & i
		\end{array}\right)\otimes 1_{\text {lattice }}
\end{align}
 The bare model \eqref{eq:model-no-gates} becomes exactly real under this map. Here we show that the couplings of the symmetry-breaking gates do not.
Applying the transformation to the Hamiltonian with both the $X$ and the $Z$ (or $T$) gate present, we find:
\begin{align}
& U^{\dagger} \mathcal{H}_{X,Z} U=\frac{1}{\sqrt{2}}\left(\begin{array}{rr}
			1 & 1 \\
			i & -i
		\end{array}\right)\left(\begin{array}{cc}
			\mathcal{H}_{\Phi} & H_{X} \\
			H_{X}^{\dagger} & \mathcal{H}_{\Phi}^*+H_{Z}
		\end{array}\right) \frac{1}{\sqrt{2}}\left(\begin{array}{rr}
			1 & -i \\
			1 & i
		\end{array}\right)\nn
		&=\frac{1}{2}\left(\begin{array}{cc}
		\mathcal{H}_{\Phi} + H_{gate} + H_{gate}^{\dagger} + \mathcal{H}_{\Phi}^*+H_Z& -i\mathcal{H}_{\Phi} + iH_{X}-iH_{X}^{\dagger} + i\mathcal{H}_{\Phi}^*+iH_Z \\
		i(\mathcal{H}_{\Phi} + H_{gate})-i(H_{gate}^{\dagger} + \mathcal{H}_{\Phi}^*+H_Z) & i(-i\mathcal{H}_{\Phi} + iH_{gate})-i(-iH_{gate}^{\dagger} + i\mathcal{H}_{\Phi}^*+iH_Z)
		\end{array}\right)\nn
		&=\left(\begin{array}{ll}
	\operatorname{Re} \mathcal{H}_{\Phi}+	 (H_Z+H_{X}+H_{X}^{\dagger})/2 & \operatorname{Im} \mathcal{H}_{\Phi}+i(H_Z+H_{X}-H_{X}^{\dagger})/2 \\
			\operatorname{Im} \mathcal{H}_{\Phi}^*+i(-H_Z+H_{X}-H_{X}^{\dagger})/2 & \operatorname{Re} \mathcal{H}_{\Phi}+(H_Z-H_{X}-H_{X}^{\dagger})/2
		\end{array}\right)
\end{align}
The gate terms appear in all four blocks, some with a prefactor of $i$, so the transformation
does not leave them real. Two mechanisms are at work. The off-diagonal blocks contain
$iH_Z/2$, so an intra-array term such as the $T$ or $Z$ gate is imaginary at any strength.
They also contain $i(H_X-H_X^{\dagger})/2$, which vanishes only if the inter-array block is
symmetric; the coupler of Sec.~\ref{sec:gates} is not, because its long diagonal has no
mirrored partner, and for the parameters of Table~\ref{tab:params_fidelity}
$\max|H_X-H_X^{\mathsf{T}}|=1.75f$.

Counting bonds makes the trade explicit. Before the transformation the lattice has $4070$
complex bonds out of $12320$, and all gate couplings are real. After it every lattice bond is
real, but half of each gate coupler is complex: four of eight bonds for the $X$ gate, eight
of sixteen for $T$, and eighteen of thirty-six for $H$.

The complex coupling structures represent a hard challenge for implementation in modern metamaterial structures \cite{Splitthoff2024PRR}. But, a number of strategies have been developed for superconducting \cite{Peropadre2013PRB,Roushan2016,Rosen2024}  and nanomechanical \cite{Slim2025} platforms. These strategies are based on the same idea: two coupled resonators are detuned, and the coupling strength also has a time-dependent function. In the rotating wave approximation, one achieves a complex coupling with phase coming from the modulation phase shift, and the strength being defined by the ratio of frequencies of modulation and detuning \cite{Roushan2016}. In the Refs.\cite{Rosen2024,Slim2025} it was shown that it might be possible to implement the on-lattice version of single Hofstadter model Hamiltonian. Such single Hamiltonian describes integer quantum Hall system, and the two blocks would describe the quantum spin Hall device as in Fig.\ref{fig1:system-gates}. This exactly corresponds to the structure described in this paper, and does not require the unitary transformation shown above. In addition, while the relative strength of complex coupling cannot exceed 0.58 of initial coupling \cite{Rosen2024}, the gate coupling could be kept real and exceed this strength.  

\bibliography{metamaterial_bib}

\begin{thebibliography}{60}%
\makeatletter
\providecommand \@ifxundefined [1]{%
 \@ifx{#1\undefined}
}%
\providecommand \@ifnum [1]{%
 \ifnum #1\expandafter \@firstoftwo
 \else \expandafter \@secondoftwo
 \fi
}%
\providecommand \@ifx [1]{%
 \ifx #1\expandafter \@firstoftwo
 \else \expandafter \@secondoftwo
 \fi
}%
\providecommand \natexlab [1]{#1}%
\providecommand \enquote  [1]{``#1''}%
\providecommand \bibnamefont  [1]{#1}%
\providecommand \bibfnamefont [1]{#1}%
\providecommand \citenamefont [1]{#1}%
\providecommand \href@noop [0]{\@secondoftwo}%
\providecommand \href [0]{\begingroup \@sanitize@url \@href}%
\providecommand \@href[1]{\@@startlink{#1}\@@href}%
\providecommand \@@href[1]{\endgroup#1\@@endlink}%
\providecommand \@sanitize@url [0]{\catcode `\\12\catcode `\$12\catcode `\&12\catcode `\#12\catcode `\^12\catcode `\_12\catcode `\%12\relax}%
\providecommand \@@startlink[1]{}%
\providecommand \@@endlink[0]{}%
\providecommand \url  [0]{\begingroup\@sanitize@url \@url }%
\providecommand \@url [1]{\endgroup\@href {#1}{\urlprefix }}%
\providecommand \urlprefix  [0]{URL }%
\providecommand \Eprint [0]{\href }%
\providecommand \doibase [0]{https://doi.org/}%
\providecommand \selectlanguage [0]{\@gobble}%
\providecommand \bibinfo  [0]{\@secondoftwo}%
\providecommand \bibfield  [0]{\@secondoftwo}%
\providecommand \translation [1]{[#1]}%
\providecommand \BibitemOpen [0]{}%
\providecommand \bibitemStop [0]{}%
\providecommand \bibitemNoStop [0]{.\EOS\space}%
\providecommand \EOS [0]{\spacefactor3000\relax}%
\providecommand \BibitemShut  [1]{\csname bibitem#1\endcsname}%
\let\auto@bib@innerbib\@empty
%</preamble>
\bibitem [{\citenamefont {Koch}\ \emph {et~al.}(2007)\citenamefont {Koch}, \citenamefont {Yu}, \citenamefont {Gambetta}, \citenamefont {Houck}, \citenamefont {Schuster}, \citenamefont {Majer}, \citenamefont {Blais}, \citenamefont {Devoret}, \citenamefont {Girvin},\ and\ \citenamefont {Schoelkopf}}]{Koch2007transmon}%
  \BibitemOpen
  \bibfield  {author} {\bibinfo {author} {\bibfnamefont {J.}~\bibnamefont {Koch}}, \bibinfo {author} {\bibfnamefont {T.~M.}\ \bibnamefont {Yu}}, \bibinfo {author} {\bibfnamefont {J.}~\bibnamefont {Gambetta}}, \bibinfo {author} {\bibfnamefont {A.~A.}\ \bibnamefont {Houck}}, \bibinfo {author} {\bibfnamefont {D.~I.}\ \bibnamefont {Schuster}}, \bibinfo {author} {\bibfnamefont {J.}~\bibnamefont {Majer}}, \bibinfo {author} {\bibfnamefont {A.}~\bibnamefont {Blais}}, \bibinfo {author} {\bibfnamefont {M.~H.}\ \bibnamefont {Devoret}}, \bibinfo {author} {\bibfnamefont {S.~M.}\ \bibnamefont {Girvin}},\ and\ \bibinfo {author} {\bibfnamefont {R.~J.}\ \bibnamefont {Schoelkopf}},\ }\bibfield  {title} {\bibinfo {title} {Charge-insensitive qubit design derived from the cooper pair box},\ }\href {https://doi.org/10.1103/physreva.76.042319} {\bibfield  {journal} {\bibinfo  {journal} {Physical Review A}\ }\textbf {\bibinfo {volume} {76}},\ \bibinfo {pages} {042319} (\bibinfo {year} {2007})}\BibitemShut {NoStop}%
\bibitem [{\citenamefont {Loss}\ and\ \citenamefont {DiVincenzo}(1998)}]{spin-qubits}%
  \BibitemOpen
  \bibfield  {author} {\bibinfo {author} {\bibfnamefont {D.}~\bibnamefont {Loss}}\ and\ \bibinfo {author} {\bibfnamefont {D.~P.}\ \bibnamefont {DiVincenzo}},\ }\bibfield  {title} {\bibinfo {title} {Quantum computation with quantum dots},\ }\href {https://doi.org/10.1103/PhysRevA.57.120} {\bibfield  {journal} {\bibinfo  {journal} {Phys. Rev. A}\ }\textbf {\bibinfo {volume} {57}},\ \bibinfo {pages} {120} (\bibinfo {year} {1998})}\BibitemShut {NoStop}%
\bibitem [{\citenamefont {Kitaev}(2000)}]{majorana}%
  \BibitemOpen
  \bibfield  {author} {\bibinfo {author} {\bibfnamefont {A.}~\bibnamefont {Kitaev}},\ }\bibfield  {title} {\bibinfo {title} {Unpaired {M}ajorana fermions in quantum wires},\ }\href {https://doi.org/10.1070/1063-7869/44/10S/S29} {\bibfield  {journal} {\bibinfo  {journal} {Uspekhi Fizicheskikh Nauk}\ }\textbf {\bibinfo {volume} {171}},\ \bibinfo {pages} {131} (\bibinfo {year} {2000})},\ \Eprint {https://arxiv.org/abs/cond-mat/0010440} {arXiv:cond-mat/0010440} \BibitemShut {NoStop}%
\bibitem [{\citenamefont {Wang}\ \emph {et~al.}(2022)\citenamefont {Wang}, \citenamefont {Dvir}, \citenamefont {Mazur}, \citenamefont {Liu}, \citenamefont {van Loo}, \citenamefont {ten Haaf}, \citenamefont {Bordin}, \citenamefont {Gazibegovic}, \citenamefont {Badawy}, \citenamefont {Bakkers}, \citenamefont {Wimmer},\ and\ \citenamefont {Kouwenhoven}}]{Wang2022poormans}%
  \BibitemOpen
  \bibfield  {author} {\bibinfo {author} {\bibfnamefont {G.}~\bibnamefont {Wang}}, \bibinfo {author} {\bibfnamefont {T.}~\bibnamefont {Dvir}}, \bibinfo {author} {\bibfnamefont {G.~P.}\ \bibnamefont {Mazur}}, \bibinfo {author} {\bibfnamefont {C.-X.}\ \bibnamefont {Liu}}, \bibinfo {author} {\bibfnamefont {N.}~\bibnamefont {van Loo}}, \bibinfo {author} {\bibfnamefont {S.~L.~D.}\ \bibnamefont {ten Haaf}}, \bibinfo {author} {\bibfnamefont {A.}~\bibnamefont {Bordin}}, \bibinfo {author} {\bibfnamefont {S.}~\bibnamefont {Gazibegovic}}, \bibinfo {author} {\bibfnamefont {G.}~\bibnamefont {Badawy}}, \bibinfo {author} {\bibfnamefont {E.~P. A.~M.}\ \bibnamefont {Bakkers}}, \bibinfo {author} {\bibfnamefont {M.}~\bibnamefont {Wimmer}},\ and\ \bibinfo {author} {\bibfnamefont {L.~P.}\ \bibnamefont {Kouwenhoven}},\ }\bibfield  {title} {\bibinfo {title} {Singlet and triplet cooper pair splitting in hybrid superconducting nanowires},\ }\href {https://doi.org/10.1038/s41586-022-05352-2} {\bibfield  {journal} {\bibinfo  {journal}
  {Nature}\ }\textbf {\bibinfo {volume} {612}},\ \bibinfo {pages} {448} (\bibinfo {year} {2022})}\BibitemShut {NoStop}%
\bibitem [{\citenamefont {Dvir}\ \emph {et~al.}(2023)\citenamefont {Dvir}, \citenamefont {Wang}, \citenamefont {van Loo}, \citenamefont {Liu}, \citenamefont {Mazur}, \citenamefont {Bordin}, \citenamefont {ten Haaf}, \citenamefont {Wang}, \citenamefont {van Driel}, \citenamefont {Zatelli}, \citenamefont {Li}, \citenamefont {Malinowski}, \citenamefont {Gazibegovic}, \citenamefont {Badawy}, \citenamefont {Bakkers}, \citenamefont {Wimmer},\ and\ \citenamefont {Kouwenhoven}}]{Dvir2023Kitaev}%
  \BibitemOpen
  \bibfield  {author} {\bibinfo {author} {\bibfnamefont {T.}~\bibnamefont {Dvir}}, \bibinfo {author} {\bibfnamefont {G.}~\bibnamefont {Wang}}, \bibinfo {author} {\bibfnamefont {N.}~\bibnamefont {van Loo}}, \bibinfo {author} {\bibfnamefont {C.-X.}\ \bibnamefont {Liu}}, \bibinfo {author} {\bibfnamefont {G.~P.}\ \bibnamefont {Mazur}}, \bibinfo {author} {\bibfnamefont {A.}~\bibnamefont {Bordin}}, \bibinfo {author} {\bibfnamefont {S.~L.~D.}\ \bibnamefont {ten Haaf}}, \bibinfo {author} {\bibfnamefont {J.-Y.}\ \bibnamefont {Wang}}, \bibinfo {author} {\bibfnamefont {D.}~\bibnamefont {van Driel}}, \bibinfo {author} {\bibfnamefont {F.}~\bibnamefont {Zatelli}}, \bibinfo {author} {\bibfnamefont {X.}~\bibnamefont {Li}}, \bibinfo {author} {\bibfnamefont {F.~K.}\ \bibnamefont {Malinowski}}, \bibinfo {author} {\bibfnamefont {S.}~\bibnamefont {Gazibegovic}}, \bibinfo {author} {\bibfnamefont {G.}~\bibnamefont {Badawy}}, \bibinfo {author} {\bibfnamefont {E.~P. A.~M.}\ \bibnamefont {Bakkers}}, \bibinfo {author} {\bibfnamefont
  {M.}~\bibnamefont {Wimmer}},\ and\ \bibinfo {author} {\bibfnamefont {L.~P.}\ \bibnamefont {Kouwenhoven}},\ }\bibfield  {title} {\bibinfo {title} {Realization of a minimal kitaev chain in coupled quantum dots},\ }\href {https://doi.org/10.1038/s41586-022-05585-1} {\bibfield  {journal} {\bibinfo  {journal} {Nature}\ }\textbf {\bibinfo {volume} {614}},\ \bibinfo {pages} {445} (\bibinfo {year} {2023})}\BibitemShut {NoStop}%
\bibitem [{\citenamefont {Zatelli}\ \emph {et~al.}(2024)\citenamefont {Zatelli}, \citenamefont {van Driel}, \citenamefont {Xu}, \citenamefont {Wang}, \citenamefont {Liu}, \citenamefont {Bordin}, \citenamefont {Roovers}, \citenamefont {Mazur}, \citenamefont {van Loo}, \citenamefont {Wolff}, \citenamefont {Bozkurt}, \citenamefont {Badawy}, \citenamefont {Gazibegovic}, \citenamefont {Bakkers}, \citenamefont {Wimmer}, \citenamefont {Kouwenhoven},\ and\ \citenamefont {Dvir}}]{Zatelli2024}%
  \BibitemOpen
  \bibfield  {author} {\bibinfo {author} {\bibfnamefont {F.}~\bibnamefont {Zatelli}}, \bibinfo {author} {\bibfnamefont {D.}~\bibnamefont {van Driel}}, \bibinfo {author} {\bibfnamefont {D.}~\bibnamefont {Xu}}, \bibinfo {author} {\bibfnamefont {G.}~\bibnamefont {Wang}}, \bibinfo {author} {\bibfnamefont {C.-X.}\ \bibnamefont {Liu}}, \bibinfo {author} {\bibfnamefont {A.}~\bibnamefont {Bordin}}, \bibinfo {author} {\bibfnamefont {B.}~\bibnamefont {Roovers}}, \bibinfo {author} {\bibfnamefont {G.~P.}\ \bibnamefont {Mazur}}, \bibinfo {author} {\bibfnamefont {N.}~\bibnamefont {van Loo}}, \bibinfo {author} {\bibfnamefont {J.~C.}\ \bibnamefont {Wolff}}, \bibinfo {author} {\bibfnamefont {A.~M.}\ \bibnamefont {Bozkurt}}, \bibinfo {author} {\bibfnamefont {G.}~\bibnamefont {Badawy}}, \bibinfo {author} {\bibfnamefont {S.}~\bibnamefont {Gazibegovic}}, \bibinfo {author} {\bibfnamefont {E.~P. A.~M.}\ \bibnamefont {Bakkers}}, \bibinfo {author} {\bibfnamefont {M.}~\bibnamefont {Wimmer}}, \bibinfo {author} {\bibfnamefont {L.~P.}\
  \bibnamefont {Kouwenhoven}},\ and\ \bibinfo {author} {\bibfnamefont {T.}~\bibnamefont {Dvir}},\ }\bibfield  {title} {\bibinfo {title} {Robust poor man’s majorana zero modes using yu-shiba-rusinov states},\ }\href {https://doi.org/10.1038/s41467-024-52066-2} {\bibfield  {journal} {\bibinfo  {journal} {Nature Communications}\ }\textbf {\bibinfo {volume} {15}},\ \bibinfo {pages} {7933} (\bibinfo {year} {2024})}\BibitemShut {NoStop}%
\bibitem [{\citenamefont {Bordin}\ \emph {et~al.}(2025)\citenamefont {Bordin}, \citenamefont {Liu}, \citenamefont {Dvir}, \citenamefont {Zatelli}, \citenamefont {ten Haaf}, \citenamefont {van Driel}, \citenamefont {Wang}, \citenamefont {van Loo}, \citenamefont {Zhang}, \citenamefont {Wolff}, \citenamefont {Van~Caekenberghe}, \citenamefont {Badawy}, \citenamefont {Gazibegovic}, \citenamefont {Bakkers}, \citenamefont {Wimmer}, \citenamefont {Kouwenhoven},\ and\ \citenamefont {Mazur}}]{Bordin2025}%
  \BibitemOpen
  \bibfield  {author} {\bibinfo {author} {\bibfnamefont {A.}~\bibnamefont {Bordin}}, \bibinfo {author} {\bibfnamefont {C.-X.}\ \bibnamefont {Liu}}, \bibinfo {author} {\bibfnamefont {T.}~\bibnamefont {Dvir}}, \bibinfo {author} {\bibfnamefont {F.}~\bibnamefont {Zatelli}}, \bibinfo {author} {\bibfnamefont {S.~L.~D.}\ \bibnamefont {ten Haaf}}, \bibinfo {author} {\bibfnamefont {D.}~\bibnamefont {van Driel}}, \bibinfo {author} {\bibfnamefont {G.}~\bibnamefont {Wang}}, \bibinfo {author} {\bibfnamefont {N.}~\bibnamefont {van Loo}}, \bibinfo {author} {\bibfnamefont {Y.}~\bibnamefont {Zhang}}, \bibinfo {author} {\bibfnamefont {J.~C.}\ \bibnamefont {Wolff}}, \bibinfo {author} {\bibfnamefont {T.}~\bibnamefont {Van~Caekenberghe}}, \bibinfo {author} {\bibfnamefont {G.}~\bibnamefont {Badawy}}, \bibinfo {author} {\bibfnamefont {S.}~\bibnamefont {Gazibegovic}}, \bibinfo {author} {\bibfnamefont {E.~P. A.~M.}\ \bibnamefont {Bakkers}}, \bibinfo {author} {\bibfnamefont {M.}~\bibnamefont {Wimmer}}, \bibinfo {author} {\bibfnamefont
  {L.~P.}\ \bibnamefont {Kouwenhoven}},\ and\ \bibinfo {author} {\bibfnamefont {G.~P.}\ \bibnamefont {Mazur}},\ }\bibfield  {title} {\bibinfo {title} {Enhanced majorana stability in a three-site kitaev chain},\ }\href {https://doi.org/10.1038/s41565-025-01894-4} {\bibfield  {journal} {\bibinfo  {journal} {Nature Nanotechnology}\ }\textbf {\bibinfo {volume} {20}},\ \bibinfo {pages} {726} (\bibinfo {year} {2025})}\BibitemShut {NoStop}%
\bibitem [{\citenamefont {ten Haaf}\ \emph {et~al.}(2025)\citenamefont {ten Haaf}, \citenamefont {Zhang}, \citenamefont {Wang}, \citenamefont {Bordin}, \citenamefont {Liu}, \citenamefont {Kulesh}, \citenamefont {Sietses}, \citenamefont {Prosko}, \citenamefont {Xiao}, \citenamefont {Thomas}, \citenamefont {Manfra}, \citenamefont {Wimmer},\ and\ \citenamefont {Goswami}}]{Haaf2025}%
  \BibitemOpen
  \bibfield  {author} {\bibinfo {author} {\bibfnamefont {S.~L.~D.}\ \bibnamefont {ten Haaf}}, \bibinfo {author} {\bibfnamefont {Y.}~\bibnamefont {Zhang}}, \bibinfo {author} {\bibfnamefont {Q.}~\bibnamefont {Wang}}, \bibinfo {author} {\bibfnamefont {A.}~\bibnamefont {Bordin}}, \bibinfo {author} {\bibfnamefont {C.-X.}\ \bibnamefont {Liu}}, \bibinfo {author} {\bibfnamefont {I.}~\bibnamefont {Kulesh}}, \bibinfo {author} {\bibfnamefont {V.~P.~M.}\ \bibnamefont {Sietses}}, \bibinfo {author} {\bibfnamefont {C.~G.}\ \bibnamefont {Prosko}}, \bibinfo {author} {\bibfnamefont {D.}~\bibnamefont {Xiao}}, \bibinfo {author} {\bibfnamefont {C.}~\bibnamefont {Thomas}}, \bibinfo {author} {\bibfnamefont {M.~J.}\ \bibnamefont {Manfra}}, \bibinfo {author} {\bibfnamefont {M.}~\bibnamefont {Wimmer}},\ and\ \bibinfo {author} {\bibfnamefont {S.}~\bibnamefont {Goswami}},\ }\bibfield  {title} {\bibinfo {title} {Observation of edge and bulk states in a three-site kitaev chain},\ }\href {https://doi.org/10.1038/s41586-025-08892-5}
  {\bibfield  {journal} {\bibinfo  {journal} {Nature}\ }\textbf {\bibinfo {volume} {641}},\ \bibinfo {pages} {890} (\bibinfo {year} {2025})}\BibitemShut {NoStop}%
\bibitem [{\citenamefont {van Loo}\ \emph {et~al.}(2026)\citenamefont {van Loo}, \citenamefont {Zatelli}, \citenamefont {Steffensen}, \citenamefont {Roovers}, \citenamefont {Wang}, \citenamefont {Van~Caekenberghe}, \citenamefont {Bordin}, \citenamefont {van Driel}, \citenamefont {Zhang}, \citenamefont {Huisman}, \citenamefont {Badawy}, \citenamefont {Bakkers}, \citenamefont {Mazur}, \citenamefont {Aguado},\ and\ \citenamefont {Kouwenhoven}}]{Loo2026}%
  \BibitemOpen
  \bibfield  {author} {\bibinfo {author} {\bibfnamefont {N.}~\bibnamefont {van Loo}}, \bibinfo {author} {\bibfnamefont {F.}~\bibnamefont {Zatelli}}, \bibinfo {author} {\bibfnamefont {G.~O.}\ \bibnamefont {Steffensen}}, \bibinfo {author} {\bibfnamefont {B.}~\bibnamefont {Roovers}}, \bibinfo {author} {\bibfnamefont {G.}~\bibnamefont {Wang}}, \bibinfo {author} {\bibfnamefont {T.}~\bibnamefont {Van~Caekenberghe}}, \bibinfo {author} {\bibfnamefont {A.}~\bibnamefont {Bordin}}, \bibinfo {author} {\bibfnamefont {D.}~\bibnamefont {van Driel}}, \bibinfo {author} {\bibfnamefont {Y.}~\bibnamefont {Zhang}}, \bibinfo {author} {\bibfnamefont {W.~D.}\ \bibnamefont {Huisman}}, \bibinfo {author} {\bibfnamefont {G.}~\bibnamefont {Badawy}}, \bibinfo {author} {\bibfnamefont {E.~P. A.~M.}\ \bibnamefont {Bakkers}}, \bibinfo {author} {\bibfnamefont {G.~P.}\ \bibnamefont {Mazur}}, \bibinfo {author} {\bibfnamefont {R.}~\bibnamefont {Aguado}},\ and\ \bibinfo {author} {\bibfnamefont {L.~P.}\ \bibnamefont {Kouwenhoven}},\ }\bibfield
  {title} {\bibinfo {title} {Single-shot parity readout of a minimal kitaev chain},\ }\href {https://doi.org/10.1038/s41586-025-09927-7} {\bibfield  {journal} {\bibinfo  {journal} {Nature}\ }\textbf {\bibinfo {volume} {650}},\ \bibinfo {pages} {334} (\bibinfo {year} {2026})}\BibitemShut {NoStop}%
\bibitem [{\citenamefont {R\'{e}glade}\ \emph {et~al.}(2024)\citenamefont {R\'{e}glade}, \citenamefont {Bocquet}, \citenamefont {Gautier}, \citenamefont {Cohen}, \citenamefont {Marquet}, \citenamefont {Albertinale}, \citenamefont {Pankratova}, \citenamefont {Hall\'{e}n}, \citenamefont {Rautschke}, \citenamefont {Sellem}, \citenamefont {Rouchon}, \citenamefont {Sarlette}, \citenamefont {Mirrahimi}, \citenamefont {Campagne-Ibarcq}, \citenamefont {Lescanne}, \citenamefont {Jezouin},\ and\ \citenamefont {Leghtas}}]{Reglade2024}%
  \BibitemOpen
  \bibfield  {author} {\bibinfo {author} {\bibfnamefont {U.}~\bibnamefont {R\'{e}glade}}, \bibinfo {author} {\bibfnamefont {A.}~\bibnamefont {Bocquet}}, \bibinfo {author} {\bibfnamefont {R.}~\bibnamefont {Gautier}}, \bibinfo {author} {\bibfnamefont {J.}~\bibnamefont {Cohen}}, \bibinfo {author} {\bibfnamefont {A.}~\bibnamefont {Marquet}}, \bibinfo {author} {\bibfnamefont {E.}~\bibnamefont {Albertinale}}, \bibinfo {author} {\bibfnamefont {N.}~\bibnamefont {Pankratova}}, \bibinfo {author} {\bibfnamefont {M.}~\bibnamefont {Hall\'{e}n}}, \bibinfo {author} {\bibfnamefont {F.}~\bibnamefont {Rautschke}}, \bibinfo {author} {\bibfnamefont {L.-A.}\ \bibnamefont {Sellem}}, \bibinfo {author} {\bibfnamefont {P.}~\bibnamefont {Rouchon}}, \bibinfo {author} {\bibfnamefont {A.}~\bibnamefont {Sarlette}}, \bibinfo {author} {\bibfnamefont {M.}~\bibnamefont {Mirrahimi}}, \bibinfo {author} {\bibfnamefont {P.}~\bibnamefont {Campagne-Ibarcq}}, \bibinfo {author} {\bibfnamefont {R.}~\bibnamefont {Lescanne}}, \bibinfo {author}
  {\bibfnamefont {S.}~\bibnamefont {Jezouin}},\ and\ \bibinfo {author} {\bibfnamefont {Z.}~\bibnamefont {Leghtas}},\ }\bibfield  {title} {\bibinfo {title} {Quantum control of a cat qubit with bit-flip times exceeding ten seconds},\ }\href {https://doi.org/10.1038/s41586-024-07294-3} {\bibfield  {journal} {\bibinfo  {journal} {Nature}\ }\textbf {\bibinfo {volume} {629}},\ \bibinfo {pages} {778} (\bibinfo {year} {2024})}\BibitemShut {NoStop}%
\bibitem [{\citenamefont {Guillaud}\ \emph {et~al.}(2023)\citenamefont {Guillaud}, \citenamefont {Cohen},\ and\ \citenamefont {Mirrahimi}}]{Guillaud2023}%
  \BibitemOpen
  \bibfield  {author} {\bibinfo {author} {\bibfnamefont {J.}~\bibnamefont {Guillaud}}, \bibinfo {author} {\bibfnamefont {J.}~\bibnamefont {Cohen}},\ and\ \bibinfo {author} {\bibfnamefont {M.}~\bibnamefont {Mirrahimi}},\ }\bibfield  {title} {\bibinfo {title} {Quantum computation with cat qubits},\ }\href {https://doi.org/10.21468/scipostphyslectnotes.72} {\bibfield  {journal} {\bibinfo  {journal} {SciPost Physics Lecture Notes}\ }\textbf {\bibinfo {volume} {72}} (\bibinfo {year} {2023})}\BibitemShut {NoStop}%
\bibitem [{\citenamefont {Manucharyan}\ \emph {et~al.}(2009)\citenamefont {Manucharyan}, \citenamefont {Koch}, \citenamefont {Glazman},\ and\ \citenamefont {Devoret}}]{Manucharyan2009science}%
  \BibitemOpen
  \bibfield  {author} {\bibinfo {author} {\bibfnamefont {V.~E.}\ \bibnamefont {Manucharyan}}, \bibinfo {author} {\bibfnamefont {J.}~\bibnamefont {Koch}}, \bibinfo {author} {\bibfnamefont {L.~I.}\ \bibnamefont {Glazman}},\ and\ \bibinfo {author} {\bibfnamefont {M.~H.}\ \bibnamefont {Devoret}},\ }\bibfield  {title} {\bibinfo {title} {Fluxonium: Single {C}ooper-pair circuit free of charge offsets},\ }\href {https://doi.org/10.1126/science.1175552} {\bibfield  {journal} {\bibinfo  {journal} {Science}\ }\textbf {\bibinfo {volume} {326}},\ \bibinfo {pages} {113} (\bibinfo {year} {2009})}\BibitemShut {NoStop}%
\bibitem [{\citenamefont {Su}\ \emph {et~al.}(1979)\citenamefont {Su}, \citenamefont {Schrieffer},\ and\ \citenamefont {Heeger}}]{Su1979}%
  \BibitemOpen
  \bibfield  {author} {\bibinfo {author} {\bibfnamefont {W.~P.}\ \bibnamefont {Su}}, \bibinfo {author} {\bibfnamefont {J.~R.}\ \bibnamefont {Schrieffer}},\ and\ \bibinfo {author} {\bibfnamefont {A.~J.}\ \bibnamefont {Heeger}},\ }\bibfield  {title} {\bibinfo {title} {Solitons in {P}olyacetylene},\ }\href {https://doi.org/10.1103/physrevlett.42.1698} {\bibfield  {journal} {\bibinfo  {journal} {Physical Review Letters}\ }\textbf {\bibinfo {volume} {42}},\ \bibinfo {pages} {1698} (\bibinfo {year} {1979})}\BibitemShut {NoStop}%
\bibitem [{\citenamefont {Splitthoff}\ \emph {et~al.}(2024)\citenamefont {Splitthoff}, \citenamefont {Belo}, \citenamefont {Jin}, \citenamefont {Liu}, \citenamefont {Greplova},\ and\ \citenamefont {Andersen}}]{Splitthoff2024PRR}%
  \BibitemOpen
  \bibfield  {author} {\bibinfo {author} {\bibfnamefont {L.~J.}\ \bibnamefont {Splitthoff}}, \bibinfo {author} {\bibfnamefont {M.~C.}\ \bibnamefont {Belo}}, \bibinfo {author} {\bibfnamefont {G.}~\bibnamefont {Jin}}, \bibinfo {author} {\bibfnamefont {Y.}~\bibnamefont {Liu}}, \bibinfo {author} {\bibfnamefont {E.}~\bibnamefont {Greplova}},\ and\ \bibinfo {author} {\bibfnamefont {C.~K.}\ \bibnamefont {Andersen}},\ }\bibfield  {title} {\bibinfo {title} {Gate-tunable phase transition in a resonator-based {S}u-{S}chrieffer-{H}eeger chain},\ }\href {https://doi.org/10.1103/PhysRevResearch.6.043286} {\bibfield  {journal} {\bibinfo  {journal} {Phys. Rev. Res.}\ }\textbf {\bibinfo {volume} {6}},\ \bibinfo {pages} {043286} (\bibinfo {year} {2024})}\BibitemShut {NoStop}%
\bibitem [{\citenamefont {Jouanny}\ \emph {et~al.}(2025)\citenamefont {Jouanny}, \citenamefont {Frasca}, \citenamefont {Weibel}, \citenamefont {Peyruchat}, \citenamefont {Scigliuzzo}, \citenamefont {Oppliger}, \citenamefont {De~Palma}, \citenamefont {Sbroggiò}, \citenamefont {Beaulieu}, \citenamefont {Zilberberg},\ and\ \citenamefont {Scarlino}}]{Jouanny2025natcomm}%
  \BibitemOpen
  \bibfield  {author} {\bibinfo {author} {\bibfnamefont {V.}~\bibnamefont {Jouanny}}, \bibinfo {author} {\bibfnamefont {S.}~\bibnamefont {Frasca}}, \bibinfo {author} {\bibfnamefont {V.~J.}\ \bibnamefont {Weibel}}, \bibinfo {author} {\bibfnamefont {l.}~\bibnamefont {Peyruchat}}, \bibinfo {author} {\bibfnamefont {M.}~\bibnamefont {Scigliuzzo}}, \bibinfo {author} {\bibfnamefont {F.}~\bibnamefont {Oppliger}}, \bibinfo {author} {\bibfnamefont {F.}~\bibnamefont {De~Palma}}, \bibinfo {author} {\bibfnamefont {D.}~\bibnamefont {Sbroggiò}}, \bibinfo {author} {\bibfnamefont {G.}~\bibnamefont {Beaulieu}}, \bibinfo {author} {\bibfnamefont {O.}~\bibnamefont {Zilberberg}},\ and\ \bibinfo {author} {\bibfnamefont {P.}~\bibnamefont {Scarlino}},\ }\bibfield  {title} {\bibinfo {title} {High kinetic inductance cavity arrays for compact band engineering and topology-based disorder meters},\ }\href {https://doi.org/10.1038/s41467-025-58595-8} {\bibfield  {journal} {\bibinfo  {journal} {Nature Communications}\ }\textbf {\bibinfo
  {volume} {16}},\ \bibinfo {pages} {3396} (\bibinfo {year} {2025})}\BibitemShut {NoStop}%
\bibitem [{\citenamefont {Jin}\ and\ \citenamefont {Greplova}(2023)}]{JinGreplova2023}%
  \BibitemOpen
  \bibfield  {author} {\bibinfo {author} {\bibfnamefont {G.}~\bibnamefont {Jin}}\ and\ \bibinfo {author} {\bibfnamefont {E.}~\bibnamefont {Greplova}},\ }\bibfield  {title} {\bibinfo {title} {Topological entanglement stabilization in superconducting quantum circuits},\ }\href {https://doi.org/10.1103/PhysRevResearch.5.023088} {\bibfield  {journal} {\bibinfo  {journal} {Phys. Rev. Research}\ }\textbf {\bibinfo {volume} {5}},\ \bibinfo {pages} {023088} (\bibinfo {year} {2023})}\BibitemShut {NoStop}%
\bibitem [{\citenamefont {Bello}\ \emph {et~al.}(2016)\citenamefont {Bello}, \citenamefont {Creffield},\ and\ \citenamefont {Platero}}]{Bello2016}%
  \BibitemOpen
  \bibfield  {author} {\bibinfo {author} {\bibfnamefont {M.}~\bibnamefont {Bello}}, \bibinfo {author} {\bibfnamefont {C.~E.}\ \bibnamefont {Creffield}},\ and\ \bibinfo {author} {\bibfnamefont {G.}~\bibnamefont {Platero}},\ }\bibfield  {title} {\bibinfo {title} {Long-range doublon transfer in a dimer chain induced by topology and ac fields},\ }\href {https://doi.org/10.1038/srep22562} {\bibfield  {journal} {\bibinfo  {journal} {Sci. Rep.}\ }\textbf {\bibinfo {volume} {6}},\ \bibinfo {pages} {22562} (\bibinfo {year} {2016})}\BibitemShut {NoStop}%
\bibitem [{\citenamefont {Halperin}(1982)}]{Halperin1982PhysicalReviewQuantizedHallConductance}%
  \BibitemOpen
  \bibfield  {author} {\bibinfo {author} {\bibfnamefont {B.~I.}\ \bibnamefont {Halperin}},\ }\bibfield  {title} {\bibinfo {title} {Quantized {H}all conductance, current-carrying edge states, and the existence of extended states in a two-dimensional disordered potential},\ }\href {https://doi.org/10.1103/physrevb.25.2185} {\bibfield  {journal} {\bibinfo  {journal} {Physical Review B}\ }\textbf {\bibinfo {volume} {25}},\ \bibinfo {pages} {2185–2190} (\bibinfo {year} {1982})}\BibitemShut {NoStop}%
\bibitem [{\citenamefont {Büttiker}(1988)}]{BTtiker1988PhysicalReviewAbsenceBackscatteringQuantum}%
  \BibitemOpen
  \bibfield  {author} {\bibinfo {author} {\bibfnamefont {M.}~\bibnamefont {Büttiker}},\ }\bibfield  {title} {\bibinfo {title} {Absence of backscattering in the quantum {H}all effect in multiprobe conductors},\ }\href {https://doi.org/10.1103/physrevb.38.9375} {\bibfield  {journal} {\bibinfo  {journal} {Physical Review B}\ }\textbf {\bibinfo {volume} {38}},\ \bibinfo {pages} {9375–9389} (\bibinfo {year} {1988})}\BibitemShut {NoStop}%
\bibitem [{\citenamefont {Bordone}\ \emph {et~al.}(2019)\citenamefont {Bordone}, \citenamefont {Bellentani},\ and\ \citenamefont {Bertoni}}]{Bordone2019SemiconductorScienceQuantumComputingQuantum}%
  \BibitemOpen
  \bibfield  {author} {\bibinfo {author} {\bibfnamefont {P.}~\bibnamefont {Bordone}}, \bibinfo {author} {\bibfnamefont {L.}~\bibnamefont {Bellentani}},\ and\ \bibinfo {author} {\bibfnamefont {A.}~\bibnamefont {Bertoni}},\ }\bibfield  {title} {\bibinfo {title} {Quantum computing with quantum-{H}all edge state interferometry},\ }\href {https://doi.org/10.1088/1361-6641/ab3be6} {\bibfield  {journal} {\bibinfo  {journal} {Semiconductor Science and Technology}\ }\textbf {\bibinfo {volume} {34}},\ \bibinfo {pages} {103001} (\bibinfo {year} {2019})}\BibitemShut {NoStop}%
\bibitem [{\citenamefont {Ko\"{o}nig}\ \emph {et~al.}(2007)\citenamefont {Ko\"{o}nig}, \citenamefont {Wiedmann}, \citenamefont {Bru\"{u}ne}, \citenamefont {Roth}, \citenamefont {Buhmann}, \citenamefont {Molenkamp}, \citenamefont {Qi},\ and\ \citenamefont {Zhang}}]{Koenig2007science}%
  \BibitemOpen
  \bibfield  {author} {\bibinfo {author} {\bibfnamefont {M.}~\bibnamefont {Ko\"{o}nig}}, \bibinfo {author} {\bibfnamefont {S.}~\bibnamefont {Wiedmann}}, \bibinfo {author} {\bibfnamefont {C.}~\bibnamefont {Bru\"{u}ne}}, \bibinfo {author} {\bibfnamefont {A.}~\bibnamefont {Roth}}, \bibinfo {author} {\bibfnamefont {H.}~\bibnamefont {Buhmann}}, \bibinfo {author} {\bibfnamefont {L.~W.}\ \bibnamefont {Molenkamp}}, \bibinfo {author} {\bibfnamefont {X.-L.}\ \bibnamefont {Qi}},\ and\ \bibinfo {author} {\bibfnamefont {S.-C.}\ \bibnamefont {Zhang}},\ }\bibfield  {title} {\bibinfo {title} {Quantum spin {H}all insulator state in {H}g{T}e quantum wells},\ }\href {https://doi.org/10.1126/science.1148047} {\bibfield  {journal} {\bibinfo  {journal} {Science}\ }\textbf {\bibinfo {volume} {318}},\ \bibinfo {pages} {766} (\bibinfo {year} {2007})}\BibitemShut {NoStop}%
\bibitem [{\citenamefont {Ghiasi}\ \emph {et~al.}(2025)\citenamefont {Ghiasi}, \citenamefont {Petrosyan}, \citenamefont {Ingla-Ayn\'{e}s}, \citenamefont {Bras}, \citenamefont {Watanabe}, \citenamefont {Taniguchi}, \citenamefont {Ma\~{n}as Valero}, \citenamefont {Coronado}, \citenamefont {Zollner}, \citenamefont {Fabian}, \citenamefont {Kim},\ and\ \citenamefont {van~der Zant}}]{Ghiasi2025}%
  \BibitemOpen
  \bibfield  {author} {\bibinfo {author} {\bibfnamefont {T.~S.}\ \bibnamefont {Ghiasi}}, \bibinfo {author} {\bibfnamefont {D.}~\bibnamefont {Petrosyan}}, \bibinfo {author} {\bibfnamefont {J.}~\bibnamefont {Ingla-Ayn\'{e}s}}, \bibinfo {author} {\bibfnamefont {T.}~\bibnamefont {Bras}}, \bibinfo {author} {\bibfnamefont {K.}~\bibnamefont {Watanabe}}, \bibinfo {author} {\bibfnamefont {T.}~\bibnamefont {Taniguchi}}, \bibinfo {author} {\bibfnamefont {S.}~\bibnamefont {Ma\~{n}as Valero}}, \bibinfo {author} {\bibfnamefont {E.}~\bibnamefont {Coronado}}, \bibinfo {author} {\bibfnamefont {K.}~\bibnamefont {Zollner}}, \bibinfo {author} {\bibfnamefont {J.}~\bibnamefont {Fabian}}, \bibinfo {author} {\bibfnamefont {P.}~\bibnamefont {Kim}},\ and\ \bibinfo {author} {\bibfnamefont {H.~S.~J.}\ \bibnamefont {van~der Zant}},\ }\bibfield  {title} {\bibinfo {title} {Quantum spin {H}all effect in magnetic graphene},\ }\href {https://doi.org/10.1038/s41467-025-60377-1} {\bibfield  {journal} {\bibinfo  {journal} {Nat. Commun.}\
  }\textbf {\bibinfo {volume} {16}},\ \bibinfo {pages} {5336} (\bibinfo {year} {2025})}\BibitemShut {NoStop}%
\bibitem [{\citenamefont {Kollár}\ \emph {et~al.}(2019)\citenamefont {Kollár}, \citenamefont {Fitzpatrick},\ and\ \citenamefont {Houck}}]{Kollar2019Nature}%
  \BibitemOpen
  \bibfield  {author} {\bibinfo {author} {\bibfnamefont {A.~J.}\ \bibnamefont {Kollár}}, \bibinfo {author} {\bibfnamefont {M.}~\bibnamefont {Fitzpatrick}},\ and\ \bibinfo {author} {\bibfnamefont {A.~A.}\ \bibnamefont {Houck}},\ }\bibfield  {title} {\bibinfo {title} {Hyperbolic lattices in circuit quantum electrodynamics},\ }\href {https://doi.org/10.1038/s41586-019-1348-3} {\bibfield  {journal} {\bibinfo  {journal} {Nature}\ }\textbf {\bibinfo {volume} {571}},\ \bibinfo {pages} {45} (\bibinfo {year} {2019})}\BibitemShut {NoStop}%
\bibitem [{\citenamefont {Süsstrunk}\ and\ \citenamefont {Huber}(2015)}]{Huber2015Science}%
  \BibitemOpen
  \bibfield  {author} {\bibinfo {author} {\bibfnamefont {R.}~\bibnamefont {Süsstrunk}}\ and\ \bibinfo {author} {\bibfnamefont {S.~D.}\ \bibnamefont {Huber}},\ }\bibfield  {title} {\bibinfo {title} {Observation of phononic helical edge states in a mechanical topological insulator},\ }\href {https://doi.org/10.1126/science.aab0239} {\bibfield  {journal} {\bibinfo  {journal} {Science}\ }\textbf {\bibinfo {volume} {349}},\ \bibinfo {pages} {47} (\bibinfo {year} {2015})}\BibitemShut {NoStop}%
\bibitem [{\citenamefont {Serra-Garcia}\ \emph {et~al.}(2018)\citenamefont {Serra-Garcia}, \citenamefont {Peri}, \citenamefont {S\"{u}sstrunk}, \citenamefont {Bilal}, \citenamefont {Larsen}, \citenamefont {Villanueva},\ and\ \citenamefont {Huber}}]{SerraGarcia2018}%
  \BibitemOpen
  \bibfield  {author} {\bibinfo {author} {\bibfnamefont {M.}~\bibnamefont {Serra-Garcia}}, \bibinfo {author} {\bibfnamefont {V.}~\bibnamefont {Peri}}, \bibinfo {author} {\bibfnamefont {R.}~\bibnamefont {S\"{u}sstrunk}}, \bibinfo {author} {\bibfnamefont {O.~R.}\ \bibnamefont {Bilal}}, \bibinfo {author} {\bibfnamefont {T.}~\bibnamefont {Larsen}}, \bibinfo {author} {\bibfnamefont {L.~G.}\ \bibnamefont {Villanueva}},\ and\ \bibinfo {author} {\bibfnamefont {S.~D.}\ \bibnamefont {Huber}},\ }\bibfield  {title} {\bibinfo {title} {Observation of a phononic quadrupole topological insulator},\ }\href {https://doi.org/10.1038/nature25156} {\bibfield  {journal} {\bibinfo  {journal} {Nature}\ }\textbf {\bibinfo {volume} {555}},\ \bibinfo {pages} {342} (\bibinfo {year} {2018})}\BibitemShut {NoStop}%
\bibitem [{\citenamefont {Serra-Garcia}\ \emph {et~al.}(2019)\citenamefont {Serra-Garcia}, \citenamefont {S\"{u}sstrunk},\ and\ \citenamefont {Huber}}]{SerraGarcia2019}%
  \BibitemOpen
  \bibfield  {author} {\bibinfo {author} {\bibfnamefont {M.}~\bibnamefont {Serra-Garcia}}, \bibinfo {author} {\bibfnamefont {R.}~\bibnamefont {S\"{u}sstrunk}},\ and\ \bibinfo {author} {\bibfnamefont {S.~D.}\ \bibnamefont {Huber}},\ }\bibfield  {title} {\bibinfo {title} {Observation of quadrupole transitions and edge mode topology in an {LC} circuit network},\ }\href {https://doi.org/10.1103/PhysRevB.99.020304} {\bibfield  {journal} {\bibinfo  {journal} {Phys. Rev. B}\ }\textbf {\bibinfo {volume} {99}},\ \bibinfo {pages} {020304} (\bibinfo {year} {2019})}\BibitemShut {NoStop}%
\bibitem [{\citenamefont {Slot}\ \emph {et~al.}(2017)\citenamefont {Slot}, \citenamefont {Gardenier}, \citenamefont {Jacobse}, \citenamefont {van Miert}, \citenamefont {Kempkes}, \citenamefont {Zevenhuizen}, \citenamefont {Smith}, \citenamefont {Vanmaekelbergh},\ and\ \citenamefont {Swart}}]{Slot2017}%
  \BibitemOpen
  \bibfield  {author} {\bibinfo {author} {\bibfnamefont {M.~R.}\ \bibnamefont {Slot}}, \bibinfo {author} {\bibfnamefont {T.~S.}\ \bibnamefont {Gardenier}}, \bibinfo {author} {\bibfnamefont {P.~H.}\ \bibnamefont {Jacobse}}, \bibinfo {author} {\bibfnamefont {G.~C.~P.}\ \bibnamefont {van Miert}}, \bibinfo {author} {\bibfnamefont {S.~N.}\ \bibnamefont {Kempkes}}, \bibinfo {author} {\bibfnamefont {S.~J.~M.}\ \bibnamefont {Zevenhuizen}}, \bibinfo {author} {\bibfnamefont {C.~M.}\ \bibnamefont {Smith}}, \bibinfo {author} {\bibfnamefont {D.}~\bibnamefont {Vanmaekelbergh}},\ and\ \bibinfo {author} {\bibfnamefont {I.}~\bibnamefont {Swart}},\ }\bibfield  {title} {\bibinfo {title} {Experimental realization and characterization of an electronic lieb lattice},\ }\href {https://doi.org/10.1038/nphys4105} {\bibfield  {journal} {\bibinfo  {journal} {Nature Physics}\ }\textbf {\bibinfo {volume} {13}},\ \bibinfo {pages} {672} (\bibinfo {year} {2017})}\BibitemShut {NoStop}%
\bibitem [{\citenamefont {Kempkes}\ \emph {et~al.}(2018)\citenamefont {Kempkes}, \citenamefont {Slot}, \citenamefont {Freeney}, \citenamefont {Zevenhuizen}, \citenamefont {Vanmaekelbergh}, \citenamefont {Swart},\ and\ \citenamefont {Smith}}]{Kempkes2018}%
  \BibitemOpen
  \bibfield  {author} {\bibinfo {author} {\bibfnamefont {S.~N.}\ \bibnamefont {Kempkes}}, \bibinfo {author} {\bibfnamefont {M.~R.}\ \bibnamefont {Slot}}, \bibinfo {author} {\bibfnamefont {S.~E.}\ \bibnamefont {Freeney}}, \bibinfo {author} {\bibfnamefont {S.~J.~M.}\ \bibnamefont {Zevenhuizen}}, \bibinfo {author} {\bibfnamefont {D.}~\bibnamefont {Vanmaekelbergh}}, \bibinfo {author} {\bibfnamefont {I.}~\bibnamefont {Swart}},\ and\ \bibinfo {author} {\bibfnamefont {C.~M.}\ \bibnamefont {Smith}},\ }\bibfield  {title} {\bibinfo {title} {Design and characterization of electrons in a fractal geometry},\ }\href {https://doi.org/10.1038/s41567-018-0328-0} {\bibfield  {journal} {\bibinfo  {journal} {Nature Physics}\ }\textbf {\bibinfo {volume} {15}},\ \bibinfo {pages} {127} (\bibinfo {year} {2018})}\BibitemShut {NoStop}%
\bibitem [{\citenamefont {Hofstadter}(1976)}]{Hofstadter1976}%
  \BibitemOpen
  \bibfield  {author} {\bibinfo {author} {\bibfnamefont {D.~R.}\ \bibnamefont {Hofstadter}},\ }\bibfield  {title} {\bibinfo {title} {Energy levels and wave functions of bloch electrons in rational and irrational magnetic fields},\ }\href {https://doi.org/10.1103/physrevb.14.2239} {\bibfield  {journal} {\bibinfo  {journal} {Physical Review B}\ }\textbf {\bibinfo {volume} {14}},\ \bibinfo {pages} {2239} (\bibinfo {year} {1976})}\BibitemShut {NoStop}%
\bibitem [{\citenamefont {Thouless}\ \emph {et~al.}(1982)\citenamefont {Thouless}, \citenamefont {Kohmoto}, \citenamefont {Nightingale},\ and\ \citenamefont {den Nijs}}]{Thouless1982}%
  \BibitemOpen
  \bibfield  {author} {\bibinfo {author} {\bibfnamefont {D.~J.}\ \bibnamefont {Thouless}}, \bibinfo {author} {\bibfnamefont {M.}~\bibnamefont {Kohmoto}}, \bibinfo {author} {\bibfnamefont {M.~P.}\ \bibnamefont {Nightingale}},\ and\ \bibinfo {author} {\bibfnamefont {M.}~\bibnamefont {den Nijs}},\ }\bibfield  {title} {\bibinfo {title} {Quantized {H}all conductance in a two-dimensional periodic potential},\ }\href {https://doi.org/10.1103/physrevlett.49.405} {\bibfield  {journal} {\bibinfo  {journal} {Physical Review Letters}\ }\textbf {\bibinfo {volume} {49}},\ \bibinfo {pages} {405} (\bibinfo {year} {1982})}\BibitemShut {NoStop}%
\bibitem [{\citenamefont {Dana}\ \emph {et~al.}(1985)\citenamefont {Dana}, \citenamefont {Avron},\ and\ \citenamefont {Zak}}]{Dana1985}%
  \BibitemOpen
  \bibfield  {author} {\bibinfo {author} {\bibfnamefont {I.}~\bibnamefont {Dana}}, \bibinfo {author} {\bibfnamefont {Y.}~\bibnamefont {Avron}},\ and\ \bibinfo {author} {\bibfnamefont {J.}~\bibnamefont {Zak}},\ }\bibfield  {title} {\bibinfo {title} {Quantised hall conductance in a perfect crystal},\ }\href {https://doi.org/10.1088/0022-3719/18/22/004} {\bibfield  {journal} {\bibinfo  {journal} {Journal of Physics C: Solid State Physics}\ }\textbf {\bibinfo {volume} {18}},\ \bibinfo {pages} {L679} (\bibinfo {year} {1985})}\BibitemShut {NoStop}%
\bibitem [{\citenamefont {Rosen}\ \emph {et~al.}(2024)\citenamefont {Rosen}, \citenamefont {Muschinske}, \citenamefont {Barrett}, \citenamefont {Chatterjee}, \citenamefont {Hays}, \citenamefont {DeMarco}, \citenamefont {Karamlou}, \citenamefont {Rower}, \citenamefont {Das}, \citenamefont {Kim}, \citenamefont {Niedzielski}, \citenamefont {Schuldt}, \citenamefont {Serniak}, \citenamefont {Schwartz}, \citenamefont {Yoder}, \citenamefont {Grover},\ and\ \citenamefont {Oliver}}]{Rosen2024}%
  \BibitemOpen
  \bibfield  {author} {\bibinfo {author} {\bibfnamefont {I.~T.}\ \bibnamefont {Rosen}}, \bibinfo {author} {\bibfnamefont {S.}~\bibnamefont {Muschinske}}, \bibinfo {author} {\bibfnamefont {C.~N.}\ \bibnamefont {Barrett}}, \bibinfo {author} {\bibfnamefont {A.}~\bibnamefont {Chatterjee}}, \bibinfo {author} {\bibfnamefont {M.}~\bibnamefont {Hays}}, \bibinfo {author} {\bibfnamefont {M.~A.}\ \bibnamefont {DeMarco}}, \bibinfo {author} {\bibfnamefont {A.~H.}\ \bibnamefont {Karamlou}}, \bibinfo {author} {\bibfnamefont {D.~A.}\ \bibnamefont {Rower}}, \bibinfo {author} {\bibfnamefont {R.}~\bibnamefont {Das}}, \bibinfo {author} {\bibfnamefont {D.~K.}\ \bibnamefont {Kim}}, \bibinfo {author} {\bibfnamefont {B.~M.}\ \bibnamefont {Niedzielski}}, \bibinfo {author} {\bibfnamefont {M.}~\bibnamefont {Schuldt}}, \bibinfo {author} {\bibfnamefont {K.}~\bibnamefont {Serniak}}, \bibinfo {author} {\bibfnamefont {M.~E.}\ \bibnamefont {Schwartz}}, \bibinfo {author} {\bibfnamefont {J.~L.}\ \bibnamefont {Yoder}}, \bibinfo {author}
  {\bibfnamefont {J.~A.}\ \bibnamefont {Grover}},\ and\ \bibinfo {author} {\bibfnamefont {W.~D.}\ \bibnamefont {Oliver}},\ }\bibfield  {title} {\bibinfo {title} {A synthetic magnetic vector potential in a 2{D} superconducting qubit array},\ }\href {https://doi.org/10.1038/s41567-024-02661-3} {\bibfield  {journal} {\bibinfo  {journal} {Nature Physics}\ }\textbf {\bibinfo {volume} {20}},\ \bibinfo {pages} {1881} (\bibinfo {year} {2024})}\BibitemShut {NoStop}%
\bibitem [{\citenamefont {Kane}\ and\ \citenamefont {Mele}(2005)}]{Kane2005PRL}%
  \BibitemOpen
  \bibfield  {author} {\bibinfo {author} {\bibfnamefont {C.~L.}\ \bibnamefont {Kane}}\ and\ \bibinfo {author} {\bibfnamefont {E.~J.}\ \bibnamefont {Mele}},\ }\bibfield  {title} {\bibinfo {title} {Quantum spin {H}all effect in graphene},\ }\href {https://doi.org/10.1103/PhysRevLett.95.226801} {\bibfield  {journal} {\bibinfo  {journal} {Phys. Rev. Lett.}\ }\textbf {\bibinfo {volume} {95}},\ \bibinfo {pages} {226801} (\bibinfo {year} {2005})}\BibitemShut {NoStop}%
\bibitem [{\citenamefont {Bernevig}\ \emph {et~al.}(2006)\citenamefont {Bernevig}, \citenamefont {Hughes},\ and\ \citenamefont {Zhang}}]{Bernevig2006Science}%
  \BibitemOpen
  \bibfield  {author} {\bibinfo {author} {\bibfnamefont {B.~A.}\ \bibnamefont {Bernevig}}, \bibinfo {author} {\bibfnamefont {T.~L.}\ \bibnamefont {Hughes}},\ and\ \bibinfo {author} {\bibfnamefont {S.-C.}\ \bibnamefont {Zhang}},\ }\bibfield  {title} {\bibinfo {title} {Quantum spin {H}all effect and topological phase transition in {HgTe} quantum wells},\ }\href {https://doi.org/10.1126/science.1133734} {\bibfield  {journal} {\bibinfo  {journal} {Science}\ }\textbf {\bibinfo {volume} {314}},\ \bibinfo {pages} {1757} (\bibinfo {year} {2006})}\BibitemShut {NoStop}%
\bibitem [{\citenamefont {Hatsugai}(1993)}]{Hatsugai1993}%
  \BibitemOpen
  \bibfield  {author} {\bibinfo {author} {\bibfnamefont {Y.}~\bibnamefont {Hatsugai}},\ }\bibfield  {title} {\bibinfo {title} {Chern number and edge states in the integer quantum {H}all effect},\ }\href {https://doi.org/10.1103/physrevlett.71.3697} {\bibfield  {journal} {\bibinfo  {journal} {Physical Review Letters}\ }\textbf {\bibinfo {volume} {71}},\ \bibinfo {pages} {3697} (\bibinfo {year} {1993})}\BibitemShut {NoStop}%
\bibitem [{\citenamefont {Kubica}\ \emph {et~al.}(2023)\citenamefont {Kubica}, \citenamefont {Haim}, \citenamefont {Vaknin}, \citenamefont {Levine}, \citenamefont {Brand\~{a}o},\ and\ \citenamefont {Retzker}}]{Kubica2023erasure}%
  \BibitemOpen
  \bibfield  {author} {\bibinfo {author} {\bibfnamefont {A.}~\bibnamefont {Kubica}}, \bibinfo {author} {\bibfnamefont {A.}~\bibnamefont {Haim}}, \bibinfo {author} {\bibfnamefont {Y.}~\bibnamefont {Vaknin}}, \bibinfo {author} {\bibfnamefont {H.}~\bibnamefont {Levine}}, \bibinfo {author} {\bibfnamefont {F.}~\bibnamefont {Brand\~{a}o}},\ and\ \bibinfo {author} {\bibfnamefont {A.}~\bibnamefont {Retzker}},\ }\bibfield  {title} {\bibinfo {title} {Erasure qubits: Overcoming the $t_1$ limit in superconducting circuits},\ }\href {https://doi.org/10.1103/PhysRevX.13.041022} {\bibfield  {journal} {\bibinfo  {journal} {Phys. Rev. X}\ }\textbf {\bibinfo {volume} {13}},\ \bibinfo {pages} {041022} (\bibinfo {year} {2023})}\BibitemShut {NoStop}%
\bibitem [{\citenamefont {Levine}\ \emph {et~al.}(2024)\citenamefont {Levine}, \citenamefont {Haim}, \citenamefont {Hung} \emph {et~al.}}]{Levine2024dualrail}%
  \BibitemOpen
  \bibfield  {author} {\bibinfo {author} {\bibfnamefont {H.}~\bibnamefont {Levine}}, \bibinfo {author} {\bibfnamefont {A.}~\bibnamefont {Haim}}, \bibinfo {author} {\bibfnamefont {J.~S.~C.}\ \bibnamefont {Hung}}, \emph {et~al.},\ }\bibfield  {title} {\bibinfo {title} {Demonstrating a long-coherence dual-rail erasure qubit using tunable transmons},\ }\href {https://doi.org/10.1103/PhysRevX.14.011051} {\bibfield  {journal} {\bibinfo  {journal} {Phys. Rev. X}\ }\textbf {\bibinfo {volume} {14}},\ \bibinfo {pages} {011051} (\bibinfo {year} {2024})}\BibitemShut {NoStop}%
\bibitem [{\citenamefont {Romanenko}\ \emph {et~al.}(2020)\citenamefont {Romanenko}, \citenamefont {Pilipenko}, \citenamefont {Zorzetti}, \citenamefont {Frolov}, \citenamefont {Awida}, \citenamefont {Belomestnykh}, \citenamefont {Posen},\ and\ \citenamefont {Grassellino}}]{Romanenko2020PRA}%
  \BibitemOpen
  \bibfield  {author} {\bibinfo {author} {\bibfnamefont {A.}~\bibnamefont {Romanenko}}, \bibinfo {author} {\bibfnamefont {R.}~\bibnamefont {Pilipenko}}, \bibinfo {author} {\bibfnamefont {S.}~\bibnamefont {Zorzetti}}, \bibinfo {author} {\bibfnamefont {D.}~\bibnamefont {Frolov}}, \bibinfo {author} {\bibfnamefont {M.}~\bibnamefont {Awida}}, \bibinfo {author} {\bibfnamefont {S.}~\bibnamefont {Belomestnykh}}, \bibinfo {author} {\bibfnamefont {S.}~\bibnamefont {Posen}},\ and\ \bibinfo {author} {\bibfnamefont {A.}~\bibnamefont {Grassellino}},\ }\bibfield  {title} {\bibinfo {title} {Three-dimensional superconducting resonators at $t<20$ mk with photon lifetimes up to $\tau=2$ s},\ }\href {https://doi.org/10.1103/PhysRevApplied.13.034032} {\bibfield  {journal} {\bibinfo  {journal} {Phys. Rev. Appl.}\ }\textbf {\bibinfo {volume} {13}},\ \bibinfo {pages} {034032} (\bibinfo {year} {2020})}\BibitemShut {NoStop}%
\bibitem [{\citenamefont {de~Ory}\ \emph {et~al.}(2025)\citenamefont {de~Ory}, \citenamefont {Rollano}, \citenamefont {Rodriguez}, \citenamefont {Magaz}, \citenamefont {Granados},\ and\ \citenamefont {Gomez}}]{Ory2025}%
  \BibitemOpen
  \bibfield  {author} {\bibinfo {author} {\bibfnamefont {M.~C.}\ \bibnamefont {de~Ory}}, \bibinfo {author} {\bibfnamefont {V.}~\bibnamefont {Rollano}}, \bibinfo {author} {\bibfnamefont {D.}~\bibnamefont {Rodriguez}}, \bibinfo {author} {\bibfnamefont {M.~T.}\ \bibnamefont {Magaz}}, \bibinfo {author} {\bibfnamefont {D.}~\bibnamefont {Granados}},\ and\ \bibinfo {author} {\bibfnamefont {A.}~\bibnamefont {Gomez}},\ }\bibfield  {title} {\bibinfo {title} {Low loss hybrid {N}b/{A}u superconducting resonators for quantum circuit applications},\ }\href {https://doi.org/10.1002/qute.202400699} {\bibfield  {journal} {\bibinfo  {journal} {Advanced Quantum Technologies}\ }\textbf {\bibinfo {volume} {8}},\ \bibinfo {pages} {2400699} (\bibinfo {year} {2025})}\BibitemShut {NoStop}%
\bibitem [{\citenamefont {Pechal}\ \emph {et~al.}(2014)\citenamefont {Pechal}, \citenamefont {Huthmacher}, \citenamefont {Eichler}, \citenamefont {Zeytino\u{g}lu}, \citenamefont {Abdumalikov}, \citenamefont {Berger}, \citenamefont {Wallraff},\ and\ \citenamefont {Filipp}}]{Pechal2014PRX}%
  \BibitemOpen
  \bibfield  {author} {\bibinfo {author} {\bibfnamefont {M.}~\bibnamefont {Pechal}}, \bibinfo {author} {\bibfnamefont {L.}~\bibnamefont {Huthmacher}}, \bibinfo {author} {\bibfnamefont {C.}~\bibnamefont {Eichler}}, \bibinfo {author} {\bibfnamefont {S.}~\bibnamefont {Zeytino\u{g}lu}}, \bibinfo {author} {\bibfnamefont {A.~A.}\ \bibnamefont {Abdumalikov}, \bibfnamefont {Jr.}}, \bibinfo {author} {\bibfnamefont {S.}~\bibnamefont {Berger}}, \bibinfo {author} {\bibfnamefont {A.}~\bibnamefont {Wallraff}},\ and\ \bibinfo {author} {\bibfnamefont {S.}~\bibnamefont {Filipp}},\ }\bibfield  {title} {\bibinfo {title} {Microwave-controlled generation of shaped single photons in circuit quantum electrodynamics},\ }\href {https://doi.org/10.1103/PhysRevX.4.041010} {\bibfield  {journal} {\bibinfo  {journal} {Phys. Rev. X}\ }\textbf {\bibinfo {volume} {4}},\ \bibinfo {pages} {041010} (\bibinfo {year} {2014})}\BibitemShut {NoStop}%
\bibitem [{\citenamefont {Harris}(1978)}]{Harris1978}%
  \BibitemOpen
  \bibfield  {author} {\bibinfo {author} {\bibfnamefont {F.~J.}\ \bibnamefont {Harris}},\ }\bibfield  {title} {\bibinfo {title} {On the use of windows for harmonic analysis with the discrete {F}ourier transform},\ }\href {https://doi.org/10.1109/PROC.1978.10837} {\bibfield  {journal} {\bibinfo  {journal} {Proc. IEEE}\ }\textbf {\bibinfo {volume} {66}},\ \bibinfo {pages} {51} (\bibinfo {year} {1978})}\BibitemShut {NoStop}%
\bibitem [{\citenamefont {Khaneja}\ \emph {et~al.}(2005)\citenamefont {Khaneja}, \citenamefont {Reiss}, \citenamefont {Kehlet}, \citenamefont {Schulte-Herbrüggen},\ and\ \citenamefont {Glaser}}]{Khaneja2005grape}%
  \BibitemOpen
  \bibfield  {author} {\bibinfo {author} {\bibfnamefont {N.}~\bibnamefont {Khaneja}}, \bibinfo {author} {\bibfnamefont {T.}~\bibnamefont {Reiss}}, \bibinfo {author} {\bibfnamefont {C.}~\bibnamefont {Kehlet}}, \bibinfo {author} {\bibfnamefont {T.}~\bibnamefont {Schulte-Herbrüggen}},\ and\ \bibinfo {author} {\bibfnamefont {S.~J.}\ \bibnamefont {Glaser}},\ }\bibfield  {title} {\bibinfo {title} {Optimal control of coupled spin dynamics: design of {NMR} pulse sequences by gradient ascent algorithms},\ }\href {https://doi.org/10.1016/j.jmr.2004.11.004} {\bibfield  {journal} {\bibinfo  {journal} {Journal of Magnetic Resonance}\ }\textbf {\bibinfo {volume} {172}},\ \bibinfo {pages} {296} (\bibinfo {year} {2005})}\BibitemShut {NoStop}%
\bibitem [{\citenamefont {Motzoi}\ \emph {et~al.}(2009)\citenamefont {Motzoi}, \citenamefont {Gambetta}, \citenamefont {Rebentrost},\ and\ \citenamefont {Wilhelm}}]{Motzoi2009}%
  \BibitemOpen
  \bibfield  {author} {\bibinfo {author} {\bibfnamefont {F.}~\bibnamefont {Motzoi}}, \bibinfo {author} {\bibfnamefont {J.~M.}\ \bibnamefont {Gambetta}}, \bibinfo {author} {\bibfnamefont {P.}~\bibnamefont {Rebentrost}},\ and\ \bibinfo {author} {\bibfnamefont {F.~K.}\ \bibnamefont {Wilhelm}},\ }\bibfield  {title} {\bibinfo {title} {Simple pulses for elimination of leakage in weakly nonlinear qubits},\ }\href {https://doi.org/10.1103/physrevlett.103.110501} {\bibfield  {journal} {\bibinfo  {journal} {Physical Review Letters}\ }\textbf {\bibinfo {volume} {103}},\ \bibinfo {pages} {110501} (\bibinfo {year} {2009})}\BibitemShut {NoStop}%
\bibitem [{\citenamefont {Barenco}\ \emph {et~al.}(1995)\citenamefont {Barenco}, \citenamefont {Bennett}, \citenamefont {Cleve}, \citenamefont {DiVincenzo}, \citenamefont {Margolus}, \citenamefont {Shor}, \citenamefont {Sleator}, \citenamefont {Smolin},\ and\ \citenamefont {Weinfurter}}]{Barenco1995}%
  \BibitemOpen
  \bibfield  {author} {\bibinfo {author} {\bibfnamefont {A.}~\bibnamefont {Barenco}}, \bibinfo {author} {\bibfnamefont {C.~H.}\ \bibnamefont {Bennett}}, \bibinfo {author} {\bibfnamefont {R.}~\bibnamefont {Cleve}}, \bibinfo {author} {\bibfnamefont {D.~P.}\ \bibnamefont {DiVincenzo}}, \bibinfo {author} {\bibfnamefont {N.}~\bibnamefont {Margolus}}, \bibinfo {author} {\bibfnamefont {P.}~\bibnamefont {Shor}}, \bibinfo {author} {\bibfnamefont {T.}~\bibnamefont {Sleator}}, \bibinfo {author} {\bibfnamefont {J.~A.}\ \bibnamefont {Smolin}},\ and\ \bibinfo {author} {\bibfnamefont {H.}~\bibnamefont {Weinfurter}},\ }\bibfield  {title} {\bibinfo {title} {Elementary gates for quantum computation},\ }\href {https://doi.org/10.1103/physreva.52.3457} {\bibfield  {journal} {\bibinfo  {journal} {Physical Review A}\ }\textbf {\bibinfo {volume} {52}},\ \bibinfo {pages} {3457} (\bibinfo {year} {1995})}\BibitemShut {NoStop}%
\bibitem [{\citenamefont {Boykin}\ \emph {et~al.}(2000)\citenamefont {Boykin}, \citenamefont {Mor}, \citenamefont {Pulver}, \citenamefont {Roychowdhury},\ and\ \citenamefont {Vatan}}]{Boykin2000}%
  \BibitemOpen
  \bibfield  {author} {\bibinfo {author} {\bibfnamefont {P.}~\bibnamefont {Boykin}}, \bibinfo {author} {\bibfnamefont {T.}~\bibnamefont {Mor}}, \bibinfo {author} {\bibfnamefont {M.}~\bibnamefont {Pulver}}, \bibinfo {author} {\bibfnamefont {V.}~\bibnamefont {Roychowdhury}},\ and\ \bibinfo {author} {\bibfnamefont {F.}~\bibnamefont {Vatan}},\ }\bibfield  {title} {\bibinfo {title} {A new universal and fault-tolerant quantum basis},\ }\href {https://doi.org/10.1016/s0020-0190(00)00084-3} {\bibfield  {journal} {\bibinfo  {journal} {Information Processing Letters}\ }\textbf {\bibinfo {volume} {75}},\ \bibinfo {pages} {101} (\bibinfo {year} {2000})}\BibitemShut {NoStop}%
\bibitem [{\citenamefont {Kitaev}(1997)}]{Kitaev1997}%
  \BibitemOpen
  \bibfield  {author} {\bibinfo {author} {\bibfnamefont {A.~Y.}\ \bibnamefont {Kitaev}},\ }\bibfield  {title} {\bibinfo {title} {Quantum computations: algorithms and error correction},\ }\href {https://doi.org/10.1070/rm1997v052n06abeh002155} {\bibfield  {journal} {\bibinfo  {journal} {Russian Mathematical Surveys}\ }\textbf {\bibinfo {volume} {52}},\ \bibinfo {pages} {1191} (\bibinfo {year} {1997})}\BibitemShut {NoStop}%
\bibitem [{\citenamefont {Wood}\ and\ \citenamefont {Gambetta}(2018)}]{Wood2018PRA}%
  \BibitemOpen
  \bibfield  {author} {\bibinfo {author} {\bibfnamefont {C.~J.}\ \bibnamefont {Wood}}\ and\ \bibinfo {author} {\bibfnamefont {J.~M.}\ \bibnamefont {Gambetta}},\ }\bibfield  {title} {\bibinfo {title} {Quantification and characterization of leakage errors},\ }\href {https://doi.org/10.1103/PhysRevA.97.032306} {\bibfield  {journal} {\bibinfo  {journal} {Phys. Rev. A}\ }\textbf {\bibinfo {volume} {97}},\ \bibinfo {pages} {032306} (\bibinfo {year} {2018})}\BibitemShut {NoStop}%
\bibitem [{\citenamefont {Horodecki}\ \emph {et~al.}(1999)\citenamefont {Horodecki}, \citenamefont {Horodecki},\ and\ \citenamefont {Horodecki}}]{Horodecki1999PRA}%
  \BibitemOpen
  \bibfield  {author} {\bibinfo {author} {\bibfnamefont {M.}~\bibnamefont {Horodecki}}, \bibinfo {author} {\bibfnamefont {P.}~\bibnamefont {Horodecki}},\ and\ \bibinfo {author} {\bibfnamefont {R.}~\bibnamefont {Horodecki}},\ }\bibfield  {title} {\bibinfo {title} {General teleportation channel, singlet fraction, and quasidistillation},\ }\href {https://doi.org/10.1103/PhysRevA.60.1888} {\bibfield  {journal} {\bibinfo  {journal} {Phys. Rev. A}\ }\textbf {\bibinfo {volume} {60}},\ \bibinfo {pages} {1888} (\bibinfo {year} {1999})}\BibitemShut {NoStop}%
\bibitem [{\citenamefont {Nielsen}(2002)}]{Nielsen2002PLA}%
  \BibitemOpen
  \bibfield  {author} {\bibinfo {author} {\bibfnamefont {M.~A.}\ \bibnamefont {Nielsen}},\ }\bibfield  {title} {\bibinfo {title} {A simple formula for the average gate fidelity of a quantum dynamical operation},\ }\href {https://doi.org/10.1016/S0375-9601(02)01272-0} {\bibfield  {journal} {\bibinfo  {journal} {Phys. Lett. A}\ }\textbf {\bibinfo {volume} {303}},\ \bibinfo {pages} {249} (\bibinfo {year} {2002})}\BibitemShut {NoStop}%
\bibitem [{\citenamefont {Hansen}\ and\ \citenamefont {Ostermeier}(2001)}]{Hansen2001}%
  \BibitemOpen
  \bibfield  {author} {\bibinfo {author} {\bibfnamefont {N.}~\bibnamefont {Hansen}}\ and\ \bibinfo {author} {\bibfnamefont {A.}~\bibnamefont {Ostermeier}},\ }\bibfield  {title} {\bibinfo {title} {Completely derandomized self-adaptation in evolution strategies},\ }\href {https://doi.org/10.1162/106365601750190398} {\bibfield  {journal} {\bibinfo  {journal} {Evolutionary Computation}\ }\textbf {\bibinfo {volume} {9}},\ \bibinfo {pages} {159} (\bibinfo {year} {2001})}\BibitemShut {NoStop}%
\bibitem [{\citenamefont {Hansen}(2016)}]{Hansen2016}%
  \BibitemOpen
  \bibfield  {author} {\bibinfo {author} {\bibfnamefont {N.}~\bibnamefont {Hansen}},\ }\bibfield  {title} {\bibinfo {title} {The cma evolution strategy: A tutorial},\ }\href {https://doi.org/10.48550/ARXIV.1604.00772} {\bibfield  {journal} {\bibinfo  {journal} {arxiv:1604.00772}\ } (\bibinfo {year} {2016})},\ \Eprint {https://arxiv.org/abs/1604.00772} {arXiv:1604.00772 [cs.LG]} \BibitemShut {NoStop}%
\bibitem [{\citenamefont {Virtanen}\ \emph {et~al.}(2020)\citenamefont {Virtanen}, \citenamefont {Gommers}, \citenamefont {Oliphant}, \citenamefont {Haberland}, \citenamefont {Reddy}, \citenamefont {Cournapeau}, \citenamefont {Burovski}, \citenamefont {Peterson}, \citenamefont {Weckesser}, \citenamefont {Bright}, \citenamefont {van~der Walt}, \citenamefont {Brett}, \citenamefont {Wilson}, \citenamefont {Millman}, \citenamefont {Mayorov}, \citenamefont {Nelson}, \citenamefont {Jones}, \citenamefont {Kern}, \citenamefont {Larson}, \citenamefont {Carey}, \citenamefont {Polat}, \citenamefont {Feng}, \citenamefont {Moore}, \citenamefont {VanderPlas}, \citenamefont {Laxalde}, \citenamefont {Perktold}, \citenamefont {Cimrman}, \citenamefont {Henriksen}, \citenamefont {Quintero}, \citenamefont {Harris}, \citenamefont {Archibald}, \citenamefont {Ribeiro}, \citenamefont {Pedregosa}, \citenamefont {van Mulbregt}, \citenamefont {Vijaykumar}, \citenamefont {Bardelli}, \citenamefont {Rothberg}, \citenamefont {Hilboll},
  \citenamefont {Kloeckner}, \citenamefont {Scopatz}, \citenamefont {Lee}, \citenamefont {Rokem}, \citenamefont {Woods}, \citenamefont {Fulton}, \citenamefont {Masson}, \citenamefont {Häggstrom}, \citenamefont {Fitzgerald}, \citenamefont {Nicholson}, \citenamefont {Hagen}, \citenamefont {Pasechnik}, \citenamefont {Olivetti}, \citenamefont {Martin}, \citenamefont {Wieser}, \citenamefont {Silva}, \citenamefont {Lenders}, \citenamefont {Wilhelm}, \citenamefont {Young}, \citenamefont {Price}, \citenamefont {Ingold}, \citenamefont {Allen}, \citenamefont {Lee}, \citenamefont {Audren}, \citenamefont {Probst}, \citenamefont {Dietrich}, \citenamefont {Silterra}, \citenamefont {Webber}, \citenamefont {Slavic}, \citenamefont {Nothman}, \citenamefont {Buchner}, \citenamefont {Kulick}, \citenamefont {Schönberger}, \citenamefont {de~Miranda~Cardoso}, \citenamefont {Reimer}, \citenamefont {Harrington}, \citenamefont {Rodríguez}, \citenamefont {Nunez-Iglesias}, \citenamefont {Kuczynski}, \citenamefont {Tritz},
  \citenamefont {Thoma}, \citenamefont {Newville}, \citenamefont {Kümmerer}, \citenamefont {Bolingbroke}, \citenamefont {Tartre}, \citenamefont {Pak}, \citenamefont {Smith}, \citenamefont {Nowaczyk}, \citenamefont {Shebanov}, \citenamefont {Pavlyk}, \citenamefont {Brodtkorb}, \citenamefont {Lee}, \citenamefont {McGibbon}, \citenamefont {Feldbauer}, \citenamefont {Lewis}, \citenamefont {Tygier}, \citenamefont {Sievert}, \citenamefont {Vigna}, \citenamefont {Peterson}, \citenamefont {More}, \citenamefont {Pudlik}, \citenamefont {Oshima}, \citenamefont {Pingel}, \citenamefont {Robitaille}, \citenamefont {Spura}, \citenamefont {Jones}, \citenamefont {Cera}, \citenamefont {Leslie}, \citenamefont {Zito}, \citenamefont {Krauss}, \citenamefont {Upadhyay}, \citenamefont {Halchenko},\ and\ \citenamefont {Vazquez-Baeza}}]{Virtanen2020}%
  \BibitemOpen
  \bibfield  {author} {\bibinfo {author} {\bibfnamefont {P.}~\bibnamefont {Virtanen}}, \bibinfo {author} {\bibfnamefont {R.}~\bibnamefont {Gommers}}, \bibinfo {author} {\bibfnamefont {T.~E.}\ \bibnamefont {Oliphant}}, \bibinfo {author} {\bibfnamefont {M.}~\bibnamefont {Haberland}}, \bibinfo {author} {\bibfnamefont {T.}~\bibnamefont {Reddy}}, \bibinfo {author} {\bibfnamefont {D.}~\bibnamefont {Cournapeau}}, \bibinfo {author} {\bibfnamefont {E.}~\bibnamefont {Burovski}}, \bibinfo {author} {\bibfnamefont {P.}~\bibnamefont {Peterson}}, \bibinfo {author} {\bibfnamefont {W.}~\bibnamefont {Weckesser}}, \bibinfo {author} {\bibfnamefont {J.}~\bibnamefont {Bright}}, \bibinfo {author} {\bibfnamefont {S.~J.}\ \bibnamefont {van~der Walt}}, \bibinfo {author} {\bibfnamefont {M.}~\bibnamefont {Brett}}, \bibinfo {author} {\bibfnamefont {J.}~\bibnamefont {Wilson}}, \bibinfo {author} {\bibfnamefont {K.~J.}\ \bibnamefont {Millman}}, \bibinfo {author} {\bibfnamefont {N.}~\bibnamefont {Mayorov}}, \bibinfo {author} {\bibfnamefont
  {A.~R.~J.}\ \bibnamefont {Nelson}}, \bibinfo {author} {\bibfnamefont {E.}~\bibnamefont {Jones}}, \bibinfo {author} {\bibfnamefont {R.}~\bibnamefont {Kern}}, \bibinfo {author} {\bibfnamefont {E.}~\bibnamefont {Larson}}, \bibinfo {author} {\bibfnamefont {C.~J.}\ \bibnamefont {Carey}}, \bibinfo {author} {\bibfnamefont {I.}~\bibnamefont {Polat}}, \bibinfo {author} {\bibfnamefont {Y.}~\bibnamefont {Feng}}, \bibinfo {author} {\bibfnamefont {E.~W.}\ \bibnamefont {Moore}}, \bibinfo {author} {\bibfnamefont {J.}~\bibnamefont {VanderPlas}}, \bibinfo {author} {\bibfnamefont {D.}~\bibnamefont {Laxalde}}, \bibinfo {author} {\bibfnamefont {J.}~\bibnamefont {Perktold}}, \bibinfo {author} {\bibfnamefont {R.}~\bibnamefont {Cimrman}}, \bibinfo {author} {\bibfnamefont {I.}~\bibnamefont {Henriksen}}, \bibinfo {author} {\bibfnamefont {E.~A.}\ \bibnamefont {Quintero}}, \bibinfo {author} {\bibfnamefont {C.~R.}\ \bibnamefont {Harris}}, \bibinfo {author} {\bibfnamefont {A.~M.}\ \bibnamefont {Archibald}}, \bibinfo {author}
  {\bibfnamefont {A.~H.}\ \bibnamefont {Ribeiro}}, \bibinfo {author} {\bibfnamefont {F.}~\bibnamefont {Pedregosa}}, \bibinfo {author} {\bibfnamefont {P.}~\bibnamefont {van Mulbregt}}, \bibinfo {author} {\bibfnamefont {A.}~\bibnamefont {Vijaykumar}}, \bibinfo {author} {\bibfnamefont {A.~P.}\ \bibnamefont {Bardelli}}, \bibinfo {author} {\bibfnamefont {A.}~\bibnamefont {Rothberg}}, \bibinfo {author} {\bibfnamefont {A.}~\bibnamefont {Hilboll}}, \bibinfo {author} {\bibfnamefont {A.}~\bibnamefont {Kloeckner}}, \bibinfo {author} {\bibfnamefont {A.}~\bibnamefont {Scopatz}}, \bibinfo {author} {\bibfnamefont {A.}~\bibnamefont {Lee}}, \bibinfo {author} {\bibfnamefont {A.}~\bibnamefont {Rokem}}, \bibinfo {author} {\bibfnamefont {C.~N.}\ \bibnamefont {Woods}}, \bibinfo {author} {\bibfnamefont {C.}~\bibnamefont {Fulton}}, \bibinfo {author} {\bibfnamefont {C.}~\bibnamefont {Masson}}, \bibinfo {author} {\bibfnamefont {C.}~\bibnamefont {Häggstrom}}, \bibinfo {author} {\bibfnamefont {C.}~\bibnamefont {Fitzgerald}}, \bibinfo
  {author} {\bibfnamefont {D.~A.}\ \bibnamefont {Nicholson}}, \bibinfo {author} {\bibfnamefont {D.~R.}\ \bibnamefont {Hagen}}, \bibinfo {author} {\bibfnamefont {D.~V.}\ \bibnamefont {Pasechnik}}, \bibinfo {author} {\bibfnamefont {E.}~\bibnamefont {Olivetti}}, \bibinfo {author} {\bibfnamefont {E.}~\bibnamefont {Martin}}, \bibinfo {author} {\bibfnamefont {E.}~\bibnamefont {Wieser}}, \bibinfo {author} {\bibfnamefont {F.}~\bibnamefont {Silva}}, \bibinfo {author} {\bibfnamefont {F.}~\bibnamefont {Lenders}}, \bibinfo {author} {\bibfnamefont {F.}~\bibnamefont {Wilhelm}}, \bibinfo {author} {\bibfnamefont {G.}~\bibnamefont {Young}}, \bibinfo {author} {\bibfnamefont {G.~A.}\ \bibnamefont {Price}}, \bibinfo {author} {\bibfnamefont {G.-L.}\ \bibnamefont {Ingold}}, \bibinfo {author} {\bibfnamefont {G.~E.}\ \bibnamefont {Allen}}, \bibinfo {author} {\bibfnamefont {G.~R.}\ \bibnamefont {Lee}}, \bibinfo {author} {\bibfnamefont {H.}~\bibnamefont {Audren}}, \bibinfo {author} {\bibfnamefont {I.}~\bibnamefont {Probst}}, \bibinfo
  {author} {\bibfnamefont {J.~P.}\ \bibnamefont {Dietrich}}, \bibinfo {author} {\bibfnamefont {J.}~\bibnamefont {Silterra}}, \bibinfo {author} {\bibfnamefont {J.~T.}\ \bibnamefont {Webber}}, \bibinfo {author} {\bibfnamefont {J.}~\bibnamefont {Slavic}}, \bibinfo {author} {\bibfnamefont {J.}~\bibnamefont {Nothman}}, \bibinfo {author} {\bibfnamefont {J.}~\bibnamefont {Buchner}}, \bibinfo {author} {\bibfnamefont {J.}~\bibnamefont {Kulick}}, \bibinfo {author} {\bibfnamefont {J.~L.}\ \bibnamefont {Schönberger}}, \bibinfo {author} {\bibfnamefont {J.~V.}\ \bibnamefont {de~Miranda~Cardoso}}, \bibinfo {author} {\bibfnamefont {J.}~\bibnamefont {Reimer}}, \bibinfo {author} {\bibfnamefont {J.}~\bibnamefont {Harrington}}, \bibinfo {author} {\bibfnamefont {J.~L.~C.}\ \bibnamefont {Rodríguez}}, \bibinfo {author} {\bibfnamefont {J.}~\bibnamefont {Nunez-Iglesias}}, \bibinfo {author} {\bibfnamefont {J.}~\bibnamefont {Kuczynski}}, \bibinfo {author} {\bibfnamefont {K.}~\bibnamefont {Tritz}}, \bibinfo {author} {\bibfnamefont
  {M.}~\bibnamefont {Thoma}}, \bibinfo {author} {\bibfnamefont {M.}~\bibnamefont {Newville}}, \bibinfo {author} {\bibfnamefont {M.}~\bibnamefont {Kümmerer}}, \bibinfo {author} {\bibfnamefont {M.}~\bibnamefont {Bolingbroke}}, \bibinfo {author} {\bibfnamefont {M.}~\bibnamefont {Tartre}}, \bibinfo {author} {\bibfnamefont {M.}~\bibnamefont {Pak}}, \bibinfo {author} {\bibfnamefont {N.~J.}\ \bibnamefont {Smith}}, \bibinfo {author} {\bibfnamefont {N.}~\bibnamefont {Nowaczyk}}, \bibinfo {author} {\bibfnamefont {N.}~\bibnamefont {Shebanov}}, \bibinfo {author} {\bibfnamefont {O.}~\bibnamefont {Pavlyk}}, \bibinfo {author} {\bibfnamefont {P.~A.}\ \bibnamefont {Brodtkorb}}, \bibinfo {author} {\bibfnamefont {P.}~\bibnamefont {Lee}}, \bibinfo {author} {\bibfnamefont {R.~T.}\ \bibnamefont {McGibbon}}, \bibinfo {author} {\bibfnamefont {R.}~\bibnamefont {Feldbauer}}, \bibinfo {author} {\bibfnamefont {S.}~\bibnamefont {Lewis}}, \bibinfo {author} {\bibfnamefont {S.}~\bibnamefont {Tygier}}, \bibinfo {author} {\bibfnamefont
  {S.}~\bibnamefont {Sievert}}, \bibinfo {author} {\bibfnamefont {S.}~\bibnamefont {Vigna}}, \bibinfo {author} {\bibfnamefont {S.}~\bibnamefont {Peterson}}, \bibinfo {author} {\bibfnamefont {S.}~\bibnamefont {More}}, \bibinfo {author} {\bibfnamefont {T.}~\bibnamefont {Pudlik}}, \bibinfo {author} {\bibfnamefont {T.}~\bibnamefont {Oshima}}, \bibinfo {author} {\bibfnamefont {T.~J.}\ \bibnamefont {Pingel}}, \bibinfo {author} {\bibfnamefont {T.~P.}\ \bibnamefont {Robitaille}}, \bibinfo {author} {\bibfnamefont {T.}~\bibnamefont {Spura}}, \bibinfo {author} {\bibfnamefont {T.~R.}\ \bibnamefont {Jones}}, \bibinfo {author} {\bibfnamefont {T.}~\bibnamefont {Cera}}, \bibinfo {author} {\bibfnamefont {T.}~\bibnamefont {Leslie}}, \bibinfo {author} {\bibfnamefont {T.}~\bibnamefont {Zito}}, \bibinfo {author} {\bibfnamefont {T.}~\bibnamefont {Krauss}}, \bibinfo {author} {\bibfnamefont {U.}~\bibnamefont {Upadhyay}}, \bibinfo {author} {\bibfnamefont {Y.~O.}\ \bibnamefont {Halchenko}},\ and\ \bibinfo {author} {\bibfnamefont
  {Y.}~\bibnamefont {Vazquez-Baeza}},\ }\bibfield  {title} {\bibinfo {title} {Scipy 1.0: fundamental algorithms for scientific computing in python},\ }\href {https://doi.org/10.1038/s41592-019-0686-2} {\bibfield  {journal} {\bibinfo  {journal} {Nature Methods}\ }\textbf {\bibinfo {volume} {17}},\ \bibinfo {pages} {261} (\bibinfo {year} {2020})}\BibitemShut {NoStop}%
\bibitem [{\citenamefont {Katiraee-Far}\ \emph {et~al.}(2025)\citenamefont {Katiraee-Far}, \citenamefont {Matsumoto}, \citenamefont {Undseth}, \citenamefont {Smet}, \citenamefont {Gualtieri}, \citenamefont {Meinersen}, \citenamefont {de~Fuentes}, \citenamefont {Capannelli}, \citenamefont {Rimbach-Russ}, \citenamefont {Scappucci}, \citenamefont {Vandersypen},\ and\ \citenamefont {Greplova}}]{katiraeefar2025cmaes}%
  \BibitemOpen
  \bibfield  {author} {\bibinfo {author} {\bibfnamefont {S.~R.}\ \bibnamefont {Katiraee-Far}}, \bibinfo {author} {\bibfnamefont {Y.}~\bibnamefont {Matsumoto}}, \bibinfo {author} {\bibfnamefont {B.}~\bibnamefont {Undseth}}, \bibinfo {author} {\bibfnamefont {M.~D.}\ \bibnamefont {Smet}}, \bibinfo {author} {\bibfnamefont {V.}~\bibnamefont {Gualtieri}}, \bibinfo {author} {\bibfnamefont {C.~V.}\ \bibnamefont {Meinersen}}, \bibinfo {author} {\bibfnamefont {I.~F.}\ \bibnamefont {de~Fuentes}}, \bibinfo {author} {\bibfnamefont {K.}~\bibnamefont {Capannelli}}, \bibinfo {author} {\bibfnamefont {M.}~\bibnamefont {Rimbach-Russ}}, \bibinfo {author} {\bibfnamefont {G.}~\bibnamefont {Scappucci}}, \bibinfo {author} {\bibfnamefont {L.~M.~K.}\ \bibnamefont {Vandersypen}},\ and\ \bibinfo {author} {\bibfnamefont {E.}~\bibnamefont {Greplova}},\ }\href {https://arxiv.org/abs/2503.12256} {\bibinfo {title} {Unified evolutionary optimization for high-fidelity spin qubit operations}} (\bibinfo {year} {2025}),\ \Eprint
  {https://arxiv.org/abs/2503.12256} {arXiv:2503.12256 [quant-ph]} \BibitemShut {NoStop}%
\bibitem [{\citenamefont {Oriekhov}\ \emph {et~al.}(2026{\natexlab{a}})\citenamefont {Oriekhov}, \citenamefont {Luhanko}, \citenamefont {Crossman}, \citenamefont {Jin}, \citenamefont {Dra\v{s}ki\'{c}},\ and\ \citenamefont {Greplova}}]{Supplement-code}%
  \BibitemOpen
  \bibfield  {author} {\bibinfo {author} {\bibfnamefont {D.~O.}\ \bibnamefont {Oriekhov}}, \bibinfo {author} {\bibfnamefont {M.}~\bibnamefont {Luhanko}}, \bibinfo {author} {\bibfnamefont {G.}~\bibnamefont {Crossman}}, \bibinfo {author} {\bibfnamefont {G.}~\bibnamefont {Jin}}, \bibinfo {author} {\bibfnamefont {R.}~\bibnamefont {Dra\v{s}ki\'{c}}},\ and\ \bibinfo {author} {\bibfnamefont {E.}~\bibnamefont {Greplova}},\ }\href@noop {} {} (\bibinfo {year} {2026}{\natexlab{a}}),\ \bibinfo {note} {{T}he code containing the implementation and optimization of dynamics of chiral modes in metamaterial array, all gates, and the calculation of fidelities in the clean and disordered systems can be found at gitlab repository: \url{https://gitlab.com/QMAI/papers/metamaterial-qubit-platform}}\BibitemShut {NoStop}%
\bibitem [{\citenamefont {Kounalakis}\ \emph {et~al.}(2018)\citenamefont {Kounalakis}, \citenamefont {Dickel}, \citenamefont {Bruno}, \citenamefont {Langford},\ and\ \citenamefont {Steele}}]{Kounalakis2018}%
  \BibitemOpen
  \bibfield  {author} {\bibinfo {author} {\bibfnamefont {M.}~\bibnamefont {Kounalakis}}, \bibinfo {author} {\bibfnamefont {C.}~\bibnamefont {Dickel}}, \bibinfo {author} {\bibfnamefont {A.}~\bibnamefont {Bruno}}, \bibinfo {author} {\bibfnamefont {N.~K.}\ \bibnamefont {Langford}},\ and\ \bibinfo {author} {\bibfnamefont {G.~A.}\ \bibnamefont {Steele}},\ }\bibfield  {title} {\bibinfo {title} {Tuneable hopping and nonlinear cross-{K}err interactions in a high-coherence superconducting circuit},\ }\href {https://doi.org/10.1038/s41534-018-0088-9} {\bibfield  {journal} {\bibinfo  {journal} {npj Quantum Information}\ }\textbf {\bibinfo {volume} {4}},\ \bibinfo {pages} {38} (\bibinfo {year} {2018})}\BibitemShut {NoStop}%
\bibitem [{\citenamefont {Knill}\ \emph {et~al.}(2001)\citenamefont {Knill}, \citenamefont {Laflamme},\ and\ \citenamefont {Milburn}}]{Knill2001Nature}%
  \BibitemOpen
  \bibfield  {author} {\bibinfo {author} {\bibfnamefont {E.}~\bibnamefont {Knill}}, \bibinfo {author} {\bibfnamefont {R.}~\bibnamefont {Laflamme}},\ and\ \bibinfo {author} {\bibfnamefont {G.~J.}\ \bibnamefont {Milburn}},\ }\bibfield  {title} {\bibinfo {title} {A scheme for efficient quantum computation with linear optics},\ }\href {https://doi.org/10.1038/35051009} {\bibfield  {journal} {\bibinfo  {journal} {Nature}\ }\textbf {\bibinfo {volume} {409}},\ \bibinfo {pages} {46} (\bibinfo {year} {2001})}\BibitemShut {NoStop}%
\bibitem [{\citenamefont {Oriekhov}\ \emph {et~al.}(2026{\natexlab{b}})\citenamefont {Oriekhov}, \citenamefont {Luhanko}, \citenamefont {Crossman}, \citenamefont {Jin}, \citenamefont {Dra\v{s}ki\'{c}},\ and\ \citenamefont {Greplova}}]{Supplement-code-2}%
  \BibitemOpen
  \bibfield  {author} {\bibinfo {author} {\bibfnamefont {D.~O.}\ \bibnamefont {Oriekhov}}, \bibinfo {author} {\bibfnamefont {M.}~\bibnamefont {Luhanko}}, \bibinfo {author} {\bibfnamefont {G.}~\bibnamefont {Crossman}}, \bibinfo {author} {\bibfnamefont {G.}~\bibnamefont {Jin}}, \bibinfo {author} {\bibfnamefont {R.}~\bibnamefont {Dra\v{s}ki\'{c}}},\ and\ \bibinfo {author} {\bibfnamefont {E.}~\bibnamefont {Greplova}},\ }\href@noop {} {} (\bibinfo {year} {2026}{\natexlab{b}}),\ \bibinfo {note} {{T}he data generated by the code \cite{Supplement-code} can be found at zenodo repository: \doi{10.5281/zenodo.23327306}}\BibitemShut {NoStop}%
\bibitem [{\citenamefont {Peropadre}\ \emph {et~al.}(2013)\citenamefont {Peropadre}, \citenamefont {Zueco}, \citenamefont {Wulschner}, \citenamefont {Deppe}, \citenamefont {Marx}, \citenamefont {Gross},\ and\ \citenamefont {García-Ripoll}}]{Peropadre2013PRB}%
  \BibitemOpen
  \bibfield  {author} {\bibinfo {author} {\bibfnamefont {B.}~\bibnamefont {Peropadre}}, \bibinfo {author} {\bibfnamefont {D.}~\bibnamefont {Zueco}}, \bibinfo {author} {\bibfnamefont {F.}~\bibnamefont {Wulschner}}, \bibinfo {author} {\bibfnamefont {F.}~\bibnamefont {Deppe}}, \bibinfo {author} {\bibfnamefont {A.}~\bibnamefont {Marx}}, \bibinfo {author} {\bibfnamefont {R.}~\bibnamefont {Gross}},\ and\ \bibinfo {author} {\bibfnamefont {J.~J.}\ \bibnamefont {García-Ripoll}},\ }\bibfield  {title} {\bibinfo {title} {Tunable coupling engineering between superconducting resonators: From sidebands to effective gauge fields},\ }\href {https://doi.org/10.1103/physrevb.87.134504} {\bibfield  {journal} {\bibinfo  {journal} {Physical Review B}\ }\textbf {\bibinfo {volume} {87}},\ \bibinfo {pages} {134504} (\bibinfo {year} {2013})}\BibitemShut {NoStop}%
\bibitem [{\citenamefont {Roushan}\ \emph {et~al.}(2016)\citenamefont {Roushan}, \citenamefont {Neill}, \citenamefont {Megrant}, \citenamefont {Chen}, \citenamefont {Babbush}, \citenamefont {Barends}, \citenamefont {Campbell}, \citenamefont {Chen}, \citenamefont {Chiaro}, \citenamefont {Dunsworth}, \citenamefont {Fowler}, \citenamefont {Jeffrey}, \citenamefont {Kelly}, \citenamefont {Lucero}, \citenamefont {Mutus}, \citenamefont {O’Malley}, \citenamefont {Neeley}, \citenamefont {Quintana}, \citenamefont {Sank}, \citenamefont {Vainsencher}, \citenamefont {Wenner}, \citenamefont {White}, \citenamefont {Kapit}, \citenamefont {Neven},\ and\ \citenamefont {Martinis}}]{Roushan2016}%
  \BibitemOpen
  \bibfield  {author} {\bibinfo {author} {\bibfnamefont {P.}~\bibnamefont {Roushan}}, \bibinfo {author} {\bibfnamefont {C.}~\bibnamefont {Neill}}, \bibinfo {author} {\bibfnamefont {A.}~\bibnamefont {Megrant}}, \bibinfo {author} {\bibfnamefont {Y.}~\bibnamefont {Chen}}, \bibinfo {author} {\bibfnamefont {R.}~\bibnamefont {Babbush}}, \bibinfo {author} {\bibfnamefont {R.}~\bibnamefont {Barends}}, \bibinfo {author} {\bibfnamefont {B.}~\bibnamefont {Campbell}}, \bibinfo {author} {\bibfnamefont {Z.}~\bibnamefont {Chen}}, \bibinfo {author} {\bibfnamefont {B.}~\bibnamefont {Chiaro}}, \bibinfo {author} {\bibfnamefont {A.}~\bibnamefont {Dunsworth}}, \bibinfo {author} {\bibfnamefont {A.}~\bibnamefont {Fowler}}, \bibinfo {author} {\bibfnamefont {E.}~\bibnamefont {Jeffrey}}, \bibinfo {author} {\bibfnamefont {J.}~\bibnamefont {Kelly}}, \bibinfo {author} {\bibfnamefont {E.}~\bibnamefont {Lucero}}, \bibinfo {author} {\bibfnamefont {J.}~\bibnamefont {Mutus}}, \bibinfo {author} {\bibfnamefont {P.~J.~J.}\ \bibnamefont
  {O’Malley}}, \bibinfo {author} {\bibfnamefont {M.}~\bibnamefont {Neeley}}, \bibinfo {author} {\bibfnamefont {C.}~\bibnamefont {Quintana}}, \bibinfo {author} {\bibfnamefont {D.}~\bibnamefont {Sank}}, \bibinfo {author} {\bibfnamefont {A.}~\bibnamefont {Vainsencher}}, \bibinfo {author} {\bibfnamefont {J.}~\bibnamefont {Wenner}}, \bibinfo {author} {\bibfnamefont {T.}~\bibnamefont {White}}, \bibinfo {author} {\bibfnamefont {E.}~\bibnamefont {Kapit}}, \bibinfo {author} {\bibfnamefont {H.}~\bibnamefont {Neven}},\ and\ \bibinfo {author} {\bibfnamefont {J.}~\bibnamefont {Martinis}},\ }\bibfield  {title} {\bibinfo {title} {Chiral ground-state currents of interacting photons in a synthetic magnetic field},\ }\href {https://doi.org/10.1038/nphys3930} {\bibfield  {journal} {\bibinfo  {journal} {Nature Physics}\ }\textbf {\bibinfo {volume} {13}},\ \bibinfo {pages} {146} (\bibinfo {year} {2016})}\BibitemShut {NoStop}%
\bibitem [{\citenamefont {Slim}\ \emph {et~al.}(2025)\citenamefont {Slim}, \citenamefont {del Pino},\ and\ \citenamefont {Verhagen}}]{Slim2025}%
  \BibitemOpen
  \bibfield  {author} {\bibinfo {author} {\bibfnamefont {J.~J.}\ \bibnamefont {Slim}}, \bibinfo {author} {\bibfnamefont {J.}~\bibnamefont {del Pino}},\ and\ \bibinfo {author} {\bibfnamefont {E.}~\bibnamefont {Verhagen}},\ }\bibfield  {title} {\bibinfo {title} {Programmable synthetic magnetism and chiral edge states in nano-optomechanical quantum {H}all networks},\ }\href {https://doi.org/10.1038/s41467-025-62541-z} {\bibfield  {journal} {\bibinfo  {journal} {Nature Communications}\ }\textbf {\bibinfo {volume} {16}},\ \bibinfo {pages} {7471} (\bibinfo {year} {2025})}\BibitemShut {NoStop}%
\end{thebibliography}%

\end{document}